\documentclass{SciPost}

\usepackage{graphicx}
\usepackage{dcolumn}
\usepackage{bm}

\usepackage{hyperref}
\usepackage{graphicx} 
\usepackage{url} 
\usepackage{subcaption}
\usepackage{amsmath}
\usepackage{xcolor}
\usepackage{cleveref}
\usepackage{physics}
\usepackage{booktabs}
\usepackage{comment}
\usepackage[normalem]{ulem}
\usepackage{slashed}

\newcommand{\met}{\mathrm{MET}}
\newcommand{\SU}{\mathrm{SU}}
\newcommand{\SO}{\mathrm{SO}}
\newcommand{\Sp}{\mathrm{Sp}}
\newcommand{\U}{\mathrm{U}}
\newcommand{\hc}{\mathrm{h.c.}}

\newcommand{\blue}[1]{{\color{blue} #1}}
\newcommand{\red}[1]{{\color{red} #1}}

\hypersetup{
    colorlinks,
    linkcolor={red!50!black},
    citecolor={blue!50!black},
    urlcolor={blue!80!black}
}

\usepackage[bitstream-charter]{mathdesign}
\DeclareSymbolFont{usualmathcal}{OMS}{cmsy}{m}{n}
\DeclareSymbolFontAlphabet{\mathcal}{usualmathcal}

\fancypagestyle{SPstyle}{
\fancyhf{}
\lhead{\colorbox{scipostblue}{\bf \color{white} ~SciPost Physics }}
\rhead{{\bf \color{scipostdeepblue} ~Submission }}

\fancyfoot[C]{\textbf{\thepage}}
}

\begin{document}

\pagestyle{SPstyle}

\begin{center}{\Large \textbf{\color{scipostdeepblue}{
The recasting bound atlas\\on pair production of vector-like quarks
}}}\end{center}

\begin{center}\textbf{
Thomas Flacke\textsuperscript{$\star$} and
Manuel Kunkel\textsuperscript{$\dagger$}
}\end{center}

\begin{center}
$\star$ Quantum Universe Center, KIAS, Seoul, Republic of Korea \\
$\dagger$ Institute of Theoretical Physics and Astrophysics, University of Würzburg, Germany
\\[\baselineskip]
$\star$ \href{mailto:flacke@kias.re.kr}{\small flacke@kias.re.kr}\,,
$\dagger$ \href{mailto:manuel.kunkel@uni-wuerzburg.de}{\small manuel.kunkel@uni-wuerzburg.de}
\end{center}

\noindent Report number: KIAS-Q26020

\section*{\color{scipostdeepblue}{Abstract}}
\textbf{\boldmath{%
We present a comprehensive recasting study of pair production of vector-like quarks.
We consider decays of vector-like quarks into third generation quarks and Standard Model bosons or exotic top-philic scalars. 
We derive recasting bounds on all 43 combinations of decay channels.
Using the efficiencies of each channel, we obtain combined limits across decay channels by adding up the contributions to each signal region.
A data base of recasting results along with example codes is published, allowing for application to different models without the need for rerunning recasting tools.
We demonstrate the utility of the data base by setting bounds on several sample models.
}}
\vspace{\baselineskip}

\vspace{10pt}
\noindent\rule{\textwidth}{1pt}
\tableofcontents
\noindent\rule{\textwidth}{1pt}
\vspace{10pt}

\section{Introduction}

Vector-like quarks (VLQs) are amongst the prime search targets for Beyond the Standard Model (BSM) searches at the LHC. 
ATLAS and CMS so far mainly focused on VLQs with charge $5/3$, $2/3$, and $-1/3$ which are assumed to decay into one 3rd generation quark $(t, b)$ and a Standard Model boson $(h, W, Z)$, covering both pair \cite{ATLAS:2022tla,ATLAS:2024gyc,ATLAS:2026ojf,ATLAS:2022hnn,CMS:2024xbc,CMS:2020ttz,CMS:2022fck,CMS:2026ggz} and single production \cite{ATLAS:2022ozf,ATLAS:2023pja,ATLAS:2023bfh,ATLAS:2023ixh,ATLAS:2024xne,ATLAS:2024xdc,CMS:2022yxp,CMS:2023agg,CMS:2024qdd,CMS:2026vwc,CMS:2026iyo}. 
By now, more exotic searches also became available targeting VLQs with decays into first or second generation quarks \cite{ATLAS:2024zlo} or into 3rd generation quarks and BSM bosons \cite{CMS:2026fky,CMS:2026ggz,ATLAS:2026rnb}.

On the theoretical side, VLQs are predicted to be present at the TeV scale by many BSM models which aim to address the hierarchy problem, including Composite Higgs Models \cite{Kaplan:1983fs,Kaplan:1983sm,Dugan:1984hq,Kaplan:1991dc}, little Higgs models \cite{Arkani-Hamed:2002ikv}, models with extra dimension(s) \cite{Contino:2003ve,Contino:2004vy}, or extensions of supersymmetric models \cite{Aguilar-Saavedra:2017giu}.
Underlying models typically predict not only one charge 2/3 VLQ which serves as a top partner, i.e.\ mixes with the SM-like chiral top, but multiple $\SU(2)_L$ multiplets of VLQs. 
In addition, many underlying models predict additional BSM states such as an extended scalar sector, which may be lighter than the VLQs.
This provides additional decay channels for the VLQs, most of which have not yet been searched for.

The phenomenology of VLQs has been studied extensively in the literature.
Simplified models for standard decay channels were developed in \cite{DeSimone:2012fs,Buchkremer:2013bha} and later extended to include decays into BSM scalars \cite{Banerjee:2022xmu}.
Exotic decays of VLQs have been investigated in \cite{Matsedonskyi:2014lla,Banerjee:2016wls,Chala:2017xgc,Bizot:2018tds,Chala:2018qdf,Han:2018hcu,Kim:2018mks,Aguilar-Saavedra:2019ghg,Benbrik:2019zdp,Cacciapaglia:2019zmj,Ramos:2019qqa,Xie:2019gya,Buckley:2020wzk,Dermisek:2020gbr,Wang:2020ips,Corcella:2021mdl,Dasgupta:2021fzw,Dermisek:2021zjd,Banerjee:2022izw,Bardhan:2022sif,Belyaev:2022shr,Bhardwaj:2022nko,Bhardwaj:2022wfz,Banerjee:2022xmu,Cornell:2022nky,Ghosh:2022rta,Verma:2022nyd,Banerjee:2023upj,Arhrib:2024mbq,Arhrib:2024nbj,Qureshi:2024naw}, see \cite{Banerjee:2024zvg} for an overview.
However, these works study either specific models or selected decay channels.
While this work was in progress, Ref.~\cite{Arhrib:2026coy} appeared which provides a tool utilizing experimental bounds on standard decay channels.
What is still absent in the literature is a comprehensive overview and current bounds on all standard and exotic decay channels.

In this article, we aim to partially close this gap. 
We consider the VLQs $X_{8/3}$, $X$, $T$, $B$, $Y$ of charge $8/3$, $5/3$, $2/3$, $-1/3$, and $-4/3$, covering all states that can couple to quarks and the Higgs \cite{Aguilar-Saavedra:2013qpa}.
Besides decays to a third generation quark and a SM boson, we also allow for exotic decays into a third generation quark and a BSM scalar.
For the scalars we consider colour singlets $S^{\pm \pm}$, $S^{\pm}$, $S^0$ of charge $\pm 2$, $\pm 1$, and $0$, and assume dominant decays to third generation quarks.
Using a simplified model \cite{Banerjee:2022xmu}, we derive recasting bounds on VLQ pair production for all decay channels.
We provide a data base of bounds on production cross section times branching ratio in all 43 possible decay channel combinations in VLQ pair production. 
The data base also includes the efficiency of each signal region, which allows for combination of bounds across decay channels.
This allows obtaining correct bounds even when the decays are split over multiple channels with similar branching ratio.
We demonstrate how this data base can be used in order to determine bounds on models with single VLQs as well as on models with several VLQ multiplets.

This article is structured as follows. 
In \cref{sec:model} we briefly discuss Composite Higgs models as one example of a motivation for studying VLQs with additional top-philic BSM scalars.
We then describe the simplified model used in this article and provide a list of the channels we study.
In \cref{sec:recast} we summarise the technical aspects of this work, starting with our toolchain for event generation and recasting.
We then detail how a combined limit can be obtained from multiple channels.
We briefly describe the data base available at \cite{zenodo} where the recasting results have been published, and finally we critically assess its limitations.
We turn to the physics results in \cref{sec:bounds}. 
First, we present bounds on selected individual channels and compare them to experimental limits where available. 
We then show several examples for combining limits across channels, demonstrating that this significantly increases the obtained bound.
In \cref{sec:conclusion} we draw our conclusions and provide an outlook for how this work can be extended.
We collect figures of the upper limits on all channels in \cref{app:allplots}.

\section{Model, implementation and channels}\label{sec:model}

In this section we motivate our choice of decay channels of the VLQs and scalars by reviewing the situation in composite Higgs models.
We then discuss the pair production channels under study and the simplified model we use to describe them.

\subsection{Vector-like quarks in Composite Higgs Models}\label{sec:Model:CH}

Vector-like quarks and extended scalar sectors are a staple of many BSM models.
Here, we briefly review a class of composite Higgs models with top partial compositeness following \cite{Banerjee:2022xmu}.
This serves as motivation for our choice in VLQ and BSM scalar charges, as well as the focus on top-philic decays of the scalars.

In composite Higgs models with an underlying fermionic description \cite{Barnard:2013zea,Ferretti:2013kya}, the Higgs field is realized as a pseudo Nambu-Goldstone boson (pNGB).
It emerges as a bound state of ``hyper'' fermions charged under a new ``hypercolor'' gauge group, which becomes strongly coupled at the multi-TeV scale.
The confinement breaks the global flavour symmetry $G$ of the hyper fermions to a subgroup $H$ which contains the SM gauge group.
The $G/H$ breaking patterns are dictated by the multiplicity and representations of the hyper fermions.
The minimal cosets $\SU(4)/\Sp(4)$, $\SU(4)^2/\SU(4)$ and $\SU(5)/\SO(5)$ each contain at least one neutral BSM scalar beyond the Higgs.
The last two contain additional neutral and singly charged scalars, and $\SU(5)/\SO(5)$ also features a doubly charged state.
Thus, pNGBs beyond $H$ are a generic prediction of these models, and their electric charges range from 0 to 2.
We therefore consider scalars $S^{\pm\pm}$, $S^\pm$, and $S^0$ in this article. 

Composite Higgs models generate the top Yukawa coupling via partial compositeness \cite{Kaplan:1991dc}, wherein the elementary top quark has to mix with VLQs (``top partners'') with the same SM quantum numbers as $t_L$ and $t_R$.
In order to form coloured fermionic resonances, a second set of hyper fermions is introduced which carries QCD colour.
Top partners are then realized as 3 particle bound states containing both types of hyper fermions. 
Minimal colour embeddings which yield top partners have been classified in \cite{Ferretti:2013kya,Ferretti:2016upr,Belyaev:2016ftv}, giving rise to 12 ``minimal'' models. 
These models always feature an $X$ and $B$ as well as several $T$ states, with some of them also featuring an $X_{8/3}$ and $Y$ \cite{Kunkel:2025qld}.
This covers all possible VLQs since forming a Yukawa-type coupling with the Higgs restricts VLQs to only seven $\SU(2)_L \times \U(1)_Y$ representations \cite{Aguilar-Saavedra:2013qpa}:
singlets $T$ and $B$, doublets $(X,T)$, $(T,B)$ and $(B,Y)$, and triplets $(X_{8/3}, X, T)$ and $(T,B,Y)$, where $X_{8/3}$, $X$, $T$, $B$, $Y$ have electric charge $8/3$, $5/3$, $2/3$, $-1/3$, and $-4/3$.
In this article, we take all five charges of VLQs into account.

The partial compositeness interactions include couplings of the pNGBs to third generation quarks and top partners.
With the mixing of the top partners with the quarks, this leads to top-philic decays of the BSM scalars into third generation quarks, which we assume to dominate in this article.
A competing decay channel is into two electroweak vector bosons, mediated by the topological anomaly \cite{Wess:1971yu,Witten:1983tw,Witten:1983tx} analogously to $\pi^0 \to \gamma\gamma$ in the SM.
Being loop induced, these decays are typically subdominant however\footnote{In each model, the couplings can be calculated from the hyper fermion multiplicity and their electroweak quantum numbers \cite{Banerjee:2022izw}.}.
If $S$ acquires a vacuum expectation value, that can also allow for the scalar to decay to $W$ and $Z$ bosons. 
This VEV would however contribute to the masses of electroweak gauge bosons and is thus strongly constrained by electroweak precision tests. 
Finally, the kinetic term of the pNGBs induces cascade decays $S\rightarrow S' V$, which however requires a considerable mass splitting between $m_S$ and $m_{S'}$. 

The class of Composite Higgs models described above contains a rich spectrum beyond VLQs and colour singlet pNGBs, including vector-like fermions in different colour representations \cite{Cacciapaglia:2021uqh,Cacciapaglia:2026jlv}, coloured pNGBs \cite{Cacciapaglia:2015eqa,Flacke:2025xwl} and spin-1 resonances \cite{BuarqueFranzosi:2016ooy,Cacciapaglia:2024wdn,Caliri:2024jdk}. 
We do not cover these states in this work.
Despite this motivation focusing on composite Higgs models, we stress that the results of this work are applicable to any model that features VLQs and uncoloured scalars.

\subsection{Model}\label{sec:model:imp}

\begin{figure}
    \centering
    \includegraphics[height=3.5cm]{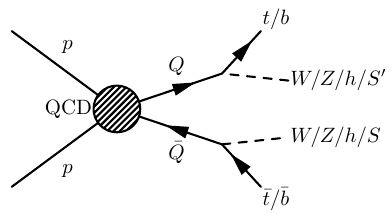} 
    \caption{Feynman diagram of VLQ pair production with decay to a third generation quark and a SM boson or an exotic scalar.}
    \label{fig:VLQpp}
\end{figure}

In this article, we study QCD pair production of VLQs 
\begin{align}
    Q=\left(X_{8/3},\,X,\,T,\,B,\,Y\right)   
\end{align}
of charge $8/3$, $5/3$, $2/3$, $-1/3$, and $-4/3$.
We consider decays into third generation quarks $(t,b)$ and Standard Model bosons $(h,W,Z)$ or BSM scalars $(S^{++},\,S^+,S^0)$ with charge $2$, $1$, and $0$,
\begin{align}
    X_{8/3} &\to t S^{++}, \label{eq:decays_x83}\\
    X &\to tW^+ ,\, bS^{++}, \, tS^+, \label{eq:decays_x}\\
    T &\to th,\, tZ,\, bW^+,\, bS^+,\, tS^0 , \label{eq:decays_tp} \\
    B &\to bh,\, bZ,\, tW^-, \, tS^- ,\, bS^0, \label{eq:decays_bp} \\
    Y &\to bW^- ,\, tS^{--}, \, bS^-. \label{eq:decays_y}
\end{align}
For the BSM scalars we assume decays into third generation quarks,
\begin{align}    
    S^{++}\rightarrow W^+t\bar{b}, \quad S^{+}\rightarrow t\bar{b}, \quad S^0\rightarrow t\bar{t}. \label{eq:Sdec}
\end{align}
The decay of $S^{++}$ is mediated by an off-shell $S^{+}$, where we assume that they come from a common multiplet and are close in mass.
From these VLQs and decays, QCD pair production yields 43 ``channels'' which are listed in \cref{tab:channels} and generically illustrated in \cref{fig:VLQpp}. 
By ``channel'', we mean a $Q\bar Q$ pair and its decays into third generation quarks and SM bosons.\footnote{We refrain from calling it a final state as we classify channels at the level \emph{before} decaying SM particles $(t,h,W,Z)$.} 
Note that for channels where $Q$ and $\bar Q$ decay differently, we only take one combination into account as we do not employ charge sensitive searches.
For the convenience of the reader, we show one VLQ and its decay products in blue and the other one in red. 
We also pre-emptively provide links to the figures showing the constraints from recasts in each channel. 
Details on the recasting procedure are provided in the following section.

\begin{table}[]
    \centering
    \small
    \begin{tabular}{|l|c||l|c|}
      \hline
        Channel & Fig. &   Channel & Fig. \\
      \hline\hline
      $\blue{X_{8/3}}\red{\bar{X}_{8/3}}\rightarrow\blue{tS^{++}}\red{\bar{t}S^{--}}\rightarrow \blue{tt\bar{b}W+}\red{\bar{t}\bar{t}bW^-}$ &\ref{fig:x83x83}  &  &\\
      \hline
      $\blue{X}\red{\bar{X}} \rightarrow\blue{tW^+}\red{\bar{t}W^-}$ & \ref{sfig:xxtwtw} 
      & $\blue{Y}\red{\bar{Y}}\rightarrow \blue{bW^{-}}\red{\bar{b}W^{+}}$ & \ref{sfig:yybwbw}\\
      $\blue{X}\red{\bar{X}} \rightarrow\blue{bS^{++}}\red{\bar{t}W^-}\rightarrow \blue{b\bar{b}tW^+}\red{\bar{t}W^-}(+\mbox{cc})$ & \ref{sfig:xxbstw}  
      & $\blue{Y}\red{\bar{Y}}\rightarrow \blue{bW^{-}}\red{\bar{t}S^{--}}\rightarrow \blue{bW^-}\red{\bar{t}b\bar{t}W^-}(+\mbox{cc})$ & \ref{sfig:yybwts} \\
      $\blue{X}\red{\bar{X}} \rightarrow\blue{tS^+}\red{\bar{t}W^{-}} \rightarrow \blue{t\bar{b}t}\red{\bar{t}W^-}(+\mbox{cc})$  & \ref{sfig:xxtstw}    
      & $\blue{Y}\red{\bar{Y}}\rightarrow \blue{bW^{-}}\red{\bar{b}S^{+}}\rightarrow \blue{bW^-}\red{\bar{b}\bar{b}t}(+\mbox{cc})$  & \ref{sfig:yybwbs} \\
      $\blue{X}\red{\bar{X}}\rightarrow \blue{bS^{++}}\red{\bar{b}S^{--}}\rightarrow\blue{b\bar{b}tW^+}\red{\bar{b}b\bar{t}W^-}$   & \ref{sfig:xxbsbs}  
      & $\blue{Y}\red{\bar{Y}}\rightarrow \blue{tS^{--}}\red{\bar{t}S^{++}}\rightarrow \blue{tb\bar{t}W^-}\red{\bar{t}\bar{b}tW^+}$  & \ref{sfig:yytsts} \\
      $\blue{X}\red{\bar{X}}\rightarrow \blue{bS^{++}}\red{\bar{t}S^{-}}\rightarrow\blue{b\bar{b}tW^+}\red{\bar{t}b\bar{t}}(+\mbox{cc})$   & \ref{sfig:xxbsts}       
      & $\blue{Y}\red{\bar{Y}}\rightarrow \blue{tS^{--}}\red{\bar{b}S^{+}}\rightarrow \blue{tb\bar{t}W^-}\red{\bar{b}\bar{b}t}(+\mbox{cc}) $  & \ref{sfig:yytsbs} \\
      $\blue{X}\red{\bar{X}}\rightarrow \blue{bS^{+}}\red{\bar{t}S^{-}}\rightarrow\blue{b\bar{b}t}\red{\bar{t}b\bar{t}}$   & \ref{sfig:xxtsts} 
      & $\blue{Y}\red{\bar{Y}}\rightarrow \blue{bS^{-}}\red{\bar{b}S^{+}}\rightarrow \blue{bb\bar{t}}\red{\bar{b}\bar{b}t}$  & \ref{sfig:yybsbs} \\
      \hline
      $\blue{T}\red{\bar{T}}\rightarrow\blue{bW^+}\red{\bar{b}W^-}$   & \ref{sfig:ttstd} 
      & $\blue{B}\red{\bar{B}}\rightarrow\blue{bh}\red{\bar{b}h}$  & \ref{sfig:bbstd}\\ 
      $\blue{T}\red{\bar{T}}\rightarrow\blue{bW^+}\red{\bar{t}h}(+\mbox{cc})$   & \ref{sfig:ttstd}
      & $\blue{B}\red{\bar{B}}\rightarrow\blue{bh}\red{\bar{b}Z}(+\mbox{cc})$  & \ref{sfig:bbstd} \\ 
      $\blue{T}\red{\bar{T}}\rightarrow\blue{bW^+}\red{\bar{t}Z}(+\mbox{cc})$   & \ref{sfig:ttstd}
      & $\blue{B}\red{\bar{B}}\rightarrow\blue{bh}\red{\bar{t}W^+}(+\mbox{cc})$  & \ref{sfig:bbstd} \\ 
      $\blue{T}\red{\bar{T}}\rightarrow\blue{th}\red{\bar{t}h}$   & \ref{sfig:ttstd}
      & $\blue{B}\red{\bar{B}}\rightarrow\blue{bZ}\red{\bar{b}Z}$  & \ref{sfig:bbstd} \\ 
      $\blue{T}\red{\bar{T}}\rightarrow\blue{th}\red{\bar{t}Z}(+\mbox{cc})$   & \ref{sfig:ttstd}
      & $\blue{B}\red{\bar{B}}\rightarrow\blue{bZ}\red{\bar{t}W^+}(+\mbox{cc})$  & \ref{sfig:bbstd} \\ 
      $\blue{T}\red{\bar{T}}\rightarrow\blue{tZ}\red{\bar{t}Z}$   & \ref{sfig:ttstd}
      & $\blue{B}\red{\bar{B}}\rightarrow\blue{tW^-}\red{\bar{t}W^+}$  & \ref{sfig:bbstd} \\ 
      $\blue{T}\red{\bar{T}}\rightarrow\blue{bS^{+}}\red{\bar{b}W^-}\rightarrow \blue{b\bar{b}t}\red{\bar{b}W^-}(+\mbox{cc})$   & \ref{sfig:ttbsbw}
      & $\blue{B}\red{\bar{B}}\rightarrow\blue{bh}\red{\bar{b}S^0}\rightarrow \blue{bh}\red{\bar{b}t\bar{b}}(+\mbox{cc})$  & \ref{sfig:bbbhbs} \\ 
      $\blue{T}\red{\bar{T}}\rightarrow\blue{bS^{+}}\red{\bar{t}Z}\rightarrow \blue{b\bar{b}t}\red{\bar{t}Z}(+\mbox{cc})$   & \ref{sfig:ttbstz}
      & $\blue{B}\red{\bar{B}}\rightarrow\blue{bS^{0}}\red{\bar{b}Z}\rightarrow \blue{bt\bar{t}}\red{\bar{b}Z}(+\mbox{cc})$  & \ref{sfig:bbbsbz} \\ 
      $\blue{T}\red{\bar{T}}\rightarrow\blue{bS^{+}}\red{\bar{t}h}\rightarrow \blue{b\bar{b}t}\red{\bar{t}h}(+\mbox{cc})$   & \ref{sfig:ttbsth} 
      & $\blue{B}\red{\bar{B}}\rightarrow\blue{bS^{0}}\red{\bar{t}W^+}\rightarrow \blue{bt\bar{t}}\red{\bar{t}W^+}(+\mbox{cc})$  & \ref{sfig:bbbstw} \\ 
      $\blue{T}\red{\bar{T}}\rightarrow\blue{bW^{+}}\red{\bar{t}S^0}\rightarrow \blue{bW^+}\red{\bar{t}t\bar{t}}(+\mbox{cc})$   & \ref{sfig:ttbwts} 
      & $\blue{B}\red{\bar{B}}\rightarrow\blue{bh}\red{\bar{t}S^+}\rightarrow \blue{bh}\red{\bar{t}\bar{b}t}(+\mbox{cc})$  & \ref{sfig:bbbhts} \\ 
      $\blue{T}\red{\bar{T}}\rightarrow\blue{tS^0}\red{\bar{t}Z}\rightarrow \blue{tt\bar{t}}\red{\bar{t}Z}(+\mbox{cc})$   & \ref{sfig:tttstz} 
      & $\blue{B}\red{\bar{B}}\rightarrow\blue{bZ}\red{\bar{t}S^+}\rightarrow \blue{bZ}\red{\bar{t}\bar{b}t}(+\mbox{cc})$  & \ref{sfig:bbbzts} \\ 
      $\blue{T}\red{\bar{T}}\rightarrow\blue{th}\red{\bar{t}S^0}\rightarrow \blue{th\red{\bar{t}t\bar{t}}}(+\mbox{cc})$   & \ref{sfig:ttthts} 
      & $\blue{B}\red{\bar{B}}\rightarrow\blue{tS^{-}}\red{\bar{t}W^+}\rightarrow \blue{tb\bar{t}}\red{\bar{t}W^+}(+\mbox{cc})$  & \ref{sfig:bbtstw} \\ 
      $\blue{T}\red{\bar{T}}\rightarrow\blue{bS^{+}}\red{\bar{b}S^-}\rightarrow \blue{b\bar{b}t}\red{\bar{b}b\bar{t}}$   & \ref{sfig:ttbsbs} 
      & $\blue{B}\red{\bar{B}}\rightarrow\blue{bS^{0}}\red{\bar{b}S^0}\rightarrow \blue{bt\bar{t}}\red{\bar{b}t\bar{t}}$  & \ref{sfig:bbbsbs} \\ 
      $\blue{T}\red{\bar{T}}\rightarrow\blue{bS^{+}}\red{\bar{t}S^0}\rightarrow \blue{b\bar{b}t}\red{\bar{t}t\bar{t}}(+\mbox{cc})$   & \ref{sfig:ttbsts} 
      & $\blue{B}\red{\bar{B}}\rightarrow\blue{bS^{0}}\red{\bar{t}S^+}\rightarrow \blue{bt\bar{t}}\red{\bar{t}\bar{b}t}(+\mbox{cc})$  & \ref{sfig:bbbsts} \\ 
      $\blue{T}\red{\bar{T}}\rightarrow\blue{tS^{0}}\red{\bar{t}S^0}\rightarrow \blue{tt\bar{t}}\red{\bar{t}t\bar{t}}$   & \ref{sfig:tttsts} 
      & $\blue{B}\red{\bar{B}}\rightarrow\blue{tS^{+}}\red{\bar{t}S^-}\rightarrow \blue{t\bar{b}t}\red{\bar{t}b\bar{t}}$  & \ref{sfig:bbtsts} \\ 
      \hline
      \end{tabular}
    \caption{Channels of VLQ pair production. The colours distinguish between the decays of the VLQ and its conjugate. }
    \label{tab:channels}
\end{table}

We parametrise the VLQs with the simplified model introduced in Ref.~\cite{Banerjee:2022xmu}. The part of the Lagrangian relevant to this article reads
\begin{equation}
    \mathcal{L}_\mathrm{int}=\mathcal{L}_{Q,QCD}+\mathcal{L}_{Q,EW}+\mathcal{L}_{Q f S}+\mathcal{L}_{ffS}+\mathcal{L}_{SSV}.
\end{equation}
The QCD interactions of the VLQs 
\begin{equation}
\mathcal{L}_{Q,QCD}=\frac{g_s}{2}\bar{Q}\slashed G^a\lambda^a Q
\end{equation}
determine the QCD pair production.
The couplings to electroweak bosons
\begin{align}
\mathcal{L}_{Q,EW} =~&\frac{e}{\sqrt{2}s_W}\kappa^W_{T,L}\bar{T}\slashed W^+ P_L b+\frac{e}{2c_Ws_W}\kappa^Z_{T,L}\bar{T}\slashed Z P_L t+\frac{e}{\sqrt{2}s_W}\kappa^W_{B,L}\bar{B}\slashed W^- P_L t + (L \leftrightarrow R) + \hc \nonumber\\
&+\frac{e}{2c_Ws_W}\kappa^Z_{B,L}\bar{B}\slashed Z P_L b+\frac{e}{\sqrt{2}s_W}\kappa^W_{X,L}\bar{X}\slashed W^+ P_L t + (L \leftrightarrow R) + \hc \nonumber \\
&+ \kappa_{T,L}^h \bar T P_L t \, h + \kappa_{B,L}^h \bar B P_L b \, h + (L \leftrightarrow R) + \hc 
\label{L_psifV}
\end{align}
yield the ``standard'' VLQ decays into third generation quarks and electroweak bosons.
The ``exotic'' decay channels into third generation quarks and BSM scalars are described by
\begin{align}
\mathcal{L}_{Q f  S} =~&  S^+\left[\kappa^{ S^+}_{T,L} \bar{T} P_L b + \kappa^{ S^+}_{X,L} \bar{X} P_L t+ L\leftrightarrow R\right]+ {\rm h.c.} +  S^-\left[\kappa^{ S^-}_{B,L} \bar{B} P_L t + L\leftrightarrow R\right]+ {\rm h.c.} \nonumber\\
&+ S^0\left[\kappa^{ S^0}_{T,L}\bar{T} P_L t+\kappa^{ S^0}_{B,L}\bar{B} P_L b+ L\leftrightarrow R\right]+ {\rm h.c.}\nonumber \\
&+  S^{++}\left[\kappa^{ S^{++}}_{X_{8/3},L} \bar{X} P_L t +\kappa^{ S^{++}}_{X,L} \bar{X} P_L b + L\leftrightarrow R\right]+ {\rm h.c.} .
\label{L_psifpi}
\end{align}
Finally, the scalar decays to third generation quarks are mediated by
\begin{align}
\mathcal{L}_{f f  S} & =S^0 \bar{t}\left[\kappa^{ S^0}_{t}+i\tilde{\kappa}^{ S^0}_{t}\gamma_5\right]t
 + S^+\left[\kappa^{ S^+}_{tb,L} \bar{t} P_L b+ L\leftrightarrow R\right]+ {\rm h.c.},
\label{L_ffpi}
\end{align}
and 
\begin{equation}
\mathcal{L}_{SSV}\supset\frac{ie}{s_W}W^{-\mu} K^{S^-_i S^{++}_j}_{W}S^-_i\overleftrightarrow{\partial_\mu}S^{++}_j+{\mathrm{h.c.}}
\end{equation}
is relevant for the decay $S^{++}\rightarrow (W^+S^+)^* \rightarrow W^+t\bar{b}$. 
The simplified model in Ref.~\cite{Banerjee:2022xmu} contains additional scalar-vector interactions in $\mathcal{L}_{SVV}$ (and also in $\mathcal{L}_{SSVV}$ and $\mathcal{L}_{SV\tilde{V}}$) which however do not affect VLQ production or decay. They have an impact on scalar decays into electroweak bosons, but as we assume a top-philic scenario with dominant decays into third generation quarks, we consistently neglect these additional couplings.

\section{Technical aspects}\label{sec:recast}

In this section we explain the technical aspects of this work.
We begin by describing our recasting toolchain, followed by detailing our methodology for combining bounds across multiple channels.
Next, we describe the data base where we published our recasting results and finally discuss its limitations.

\subsection{Recasting setup}

All VLQ channels described in \cref{sec:model:imp} can be simulated with the publicly available \texttt{eVLQ} \cite{Banerjee:2022xmu} implementation in the \texttt{FeynRules} \cite{Alloul:2013bka} format, producing a \texttt{UFO} library \cite{Degrande:2011ua}.
For simplicity, we only use left-handed couplings\footnote{We verified on several processes that choosing different chiralities did not impact the results.} for all decays except for a pseudoscalar coupling for $S^0 \to t\bar t$. 
We choose small couplings such that we can work within the narrow width approximation.

Current bounds on VLQ masses are of the order $1.0-1.5$~TeV \cite{Banerjee:2024zvg}.
We therefore scan over the VLQ mass $m_Q$ in the range $800~\mathrm{GeV} \leq m_Q \leq 2000$~GeV in steps of 100~GeV.
The mass of the electroweak scalars can in principle be constrained via Drell-Yan processes.
Pair production of the set of scalars considered in this work was studied in \cite{Cacciapaglia:2022bax}, where however no bounds could be obtained for individual states.
We therefore allow the scalar mass $m_S$ to go as low as possible while keeping the decay on-shell, scanning $400 \leq m_S \leq 1800$~GeV in steps of 200~GeV with the constraint $m_Q > m_S + m_t$.
For $S^+/S^{++}$ we also simulate $m_S = 200/300$~GeV to also cover the region where the scalar decay products have little phase space.

We use \texttt{MadGraph5\_aMC@NLO} \cite{Alwall:2014hca} v3.6.6 to generate $10^5$ BSM signal events for each parameter point.
The events are generated at leading order in QCD at $\sqrt s = 13$~TeV with the \texttt{NNPDF} 2.3 PDF set \cite{Ball:2012cx} provided by \texttt{LHAPDF6} \cite{Buckley:2014ana}.
We fix the renormalisation and factorisation scales to the VLQ mass, $\mu_R = \mu_F = m_Q$.
We decay the VLQs and when applicable the exotic scalars to SM particles within \texttt{MadGraph}.
For this we use \texttt{MadSpin} \cite{Artoisenet:2012st} for processes containing only two-body decays to preserve spin correlations.
Since \texttt{MadSpin} does not support three-body decays, we use the inline decay syntax for channels involving $S^{++} \to W^+ t \bar b$.
The hard scattering events are then passed to \texttt{Pythia8} \cite{Bierlich:2022pfr} for hadronisation and showering, producing a \texttt{HepMC} \cite{Dobbs:2001ck} file.
While the events are generated at leading order, for determining mass bounds we rescale the cross section to NNLO+NNLL accuracy calculated with \texttt{Top++2.0} \cite{Czakon:2011xx} with the \texttt{MSTW2008nnlo68cl} PDF set and the same scale choices as above.

The hadronised events are passed to \texttt{MadAnalysis5} \cite{Conte:2012fm,Conte:2014zja,Dumont:2014tja,Conte:2018vmg,Araz:2023mda} and \texttt{CheckMATE} \cite{Drees:2013wra,Dercks:2016npn} to apply the recast analyses.
Detector simulation is performed with \texttt{Delphes 3} \cite{deFavereau:2013fsa} and jets are clustered with the anti-$k_T$ algorithm available in the \texttt{FastJet} \cite{Cacciari:2011ma} library.
The events are passed through the recast analyses to set bounds with the CL$_s$ method \cite{Read:2002hq}.
For each process we ran several parameter points against all implemented searches to identify the most sensitive searches for various kinematic configurations.
The full scans were then only performed using these sensitive searches.
In \cref{tab:analyses} we summarise the searches that gave relevant contributions to at least some of the processes.

We run \texttt{MadAnalysis5} on the substructure branch  with commit \texttt{75fbc8f}. The public implemented searches are collected in the public analysis database\footnote{\url{https://madanalysis.irmp.ucl.ac.be/wiki/PublicAnalysisDatabase}} along with detailed notes validating the implementation. 
Besides the \texttt{Delphes 3}-based implementations, several are also using a simplified fast simulation (SFS) framework \cite{Araz:2020lnp}.
For \texttt{CheckMATE} we use commit \texttt{f407403}. 
Of the many implemented searches, not all have a public validation note. 
In the main text we restrict ourselves to results based on searches with a validation note. 
This is not a judgment on the quality of the recast, but of our ability to assess the extent of the validation.
In \cref{fig:comparison_allsearches} in \cref{app:allplots} we discuss the impact of including all searches for a sample process.
In the constraint data base, we include the recasting results of all searches.

\subsection{Combining channels}\label{sec:recast:method}

We obtain recasting bounds on all VLQ channels.
This allows us to set bounds on the VLQ masses by comparing the obtained upper limits on the cross section times branching ratio with the prediction in a given model.
This works very well if a single decay channel strongly dominates, leading to a branching ratio $\approx 1$. 
However, if the decays are split over multiple channels with appreciable branching ratios --- as expected in a realistic model --- then the bounds obtained from any individual channel are strongly reduced by the branching ratio.
This is often an artificial reduction however, since in many cases the different channels have similar detector signatures and will therefore contribute to the same signal regions.
We should therefore combine the contributions of all channels to all signal regions.

We label a VLQ pair production channel as
\begin{align}
    Q \bar Q \to A\bar B, \quad \text{with } Q \to A, \, \bar Q \to \bar B,
\end{align}
where $A$ and $B$ are short for the decay chain into SM particles, e.g. $A = th$ or $B = tS^0 \to t t\bar t$.
The total branching ratio of a channel is given by
\begin{align}
    \mathrm{Br}(Q\bar Q \to A\bar B) = (2 - \delta_{AB}) \times \mathrm{Br}(Q \to A) \times \mathrm{Br}(Q \to B).
\end{align}
For $A\ne B$ we only simulate one combination, so we have to include a factor of two in the branching ratio for these channels. 
For two-stage decays involving scalars, $\mathrm{Br}(Q\to A)$ factorises into the branching ratio of the VLQ times the one of the scalar.
We collect the efficiencies $\epsilon_{AB}$ of each signal region for each channel.
Given a model with one or more VLQs $Q$ and known branching ratios, we calculate the number of events contributing to a given signal region as
\begin{align}
    N = \sum_Q \sum_{A\bar B\in \text{channels}} \epsilon_{AB} \times \sigma(pp\to Q\bar Q) \times \text{Br}(Q\bar Q \to A\bar B) \times \mathcal L_\text{int}.
\end{align}
with $\mathcal L_\mathrm{int}$ the integrated luminosity utilised by the respective search.
This can be compared to the expected ($N_{95}^\text{exp}$) and observed ($N_{95}^\text{obs}$) upper limits on the number of events of the searches. 
These values are provided by \texttt{CheckMATE} for the searches implemented therein, and can be readily obtained for the searches in \texttt{MadAnalysis5} with a routine from \texttt{HackAnalysis} \cite{Goodsell:2024aig}.
From $N$ and $N_{95}^i$ we construct an exclusion value
\begin{align}
    r_i = \frac{N}{N_{95}^i}
\end{align}
such that the parameter point is excluded if $r>1$.
We repeat this for all searches and signal regions.
We first calculate the expected bound $r_\text{exp}$ to select the most sensitive signal region within each search, continue with the corresponding observed bound and take the strongest $r_\mathrm{obs}$ over all searches as the final result.

A few of the implemented searches allow for statistical combination of signal regions. 
As seen in \cref{fig:bounds_standard}, this can lead to a significantly stronger constraint.
For our purposes however, these combinations cannot be used:
The above method only allows to combine contributions from multiple channels to one specific signal region with a specified efficiency.
The statistical combination would need to be recalculated for each parameter point, which is feasible in principle but significantly increases the computational effort.

\subsection{Constraint data base}\label{sec:constraintdatabase}

\begin{table}[]
\footnotesize
\setlength{\tabcolsep}{4pt}
\centering
\begin{tabular}{@{}llllllll@{}}
\toprule
point & mvlq & tool & search & sr & efficiency & sig95exp(pb) & sig95obs(pb) \\ \midrule
0000 & 800.0 & MadAnalysis5/SFS & atlas\_susy\_2018\_31 & SRB & 0.0 & -1.0 & -1.0 \\
0000 & 800.0 & MadAnalysis5/SFS & atlas\_susy\_2018\_31 & SRA\_H & 0.0004 & 0.0858633 & 0.0774374 \\
0000 & 800.0 & MadAnalysis5/SFS & atlas\_susy\_2018\_31 & SRA & 0.00086 & 0.0895867 & 0.0891231 \\
0000 & 800.0 & MadAnalysis5/SFS & atlas\_susy\_2018\_31 & SRA\_M & 0.00034 & 0.1230578 & 0.0858176 \\ 
\vdots & \vdots & \vdots & \vdots & \vdots & \vdots & \vdots & \vdots \\ \bottomrule
\end{tabular}
\caption{Excerpt of the constraint data base for $B\bar B\to bhbh$. \texttt{point} is an internal numbering, \texttt{mvlq} is the VLQ mass $m_Q$. \texttt{tool}, \texttt{search} and \texttt{sr} indicate the search and signal region, as well as the tool in which is has been recast. A list of all searches is given in \cref{tab:analyses}. The last three columns show the results of the recast: the number of events passing all cuts divided by the number of initial events, and the expected and observed 95\% CL upper limits on the cross section in pico barn.
For processes featuring an exotic scalar, another column \texttt{ms} for the scalar mass is added.}
\label{tab:database}
\end{table}

The advantage of the method for combining bounds on several channels described above is that bounds can be set on any VLQ model without the need for rerunning recasting tools.
To facilitate this, we have published the results of our recasting runs as a constraint data base: it is available at \cite{zenodo}.
Specifically, the data base contains the efficiencies and expected and observed upper limits on the cross section for all signal regions of the searches in \cref{tab:analyses}, which turned out to be the most sensitive searches after an initial scan.
The format of the data base is indicated for a sample process in \cref{tab:database}.

Besides the recasting results, we also include a table of the upper limits on the number of events for each signal region and the pair production cross sections used in this paper.
Furthermore, we publish two Python files: 
\texttt{exclusion.py} contains all routines necessary for extracting the upper limits from the data base and for performing the combination of bounds across several VLQ decay modes.
In \texttt{example.py} we show how to use these routines in practice, giving the code needed for reproducing \cref{fig:tptp_application}.
Further documentation can be found in the repository.

\subsection{Shortcomings of the method}\label{sec:recast:flaws}

The constraint data base presented with this article represents a simple and useful tool to determine constraints on models which contain top-philic VLQs. As efficiencies and upper limits on the (expected and observed) number of events for each channel and search region are accessible, constraints can be determined by specifying the VLQ particle content, mass spectrum and branching ratios without the need of event simulation or recasting. Nevertheless, the data base and method is not without limitations and shortcomings which we list here for the awareness of the reader.

\paragraph{Limited search inclusion.}  The data base includes only searches by ATLAS and CMS for which recasts are publicly available (at this time) in \texttt{MadAnalysis5} or \texttt{CheckMATE}.
Only a limited number of search recasts are publicly available. 
In particular, none of the direct searches for VLQ pair production using the full Run-2 data set are currently available as a publicly available recast. 
Data of recasts which become available in the future can be added to the data base. 
A more complete set of BSM search recast would be highly desirable, but recasting, validating and publishing those is beyond the scope of this work. 

\paragraph{Simple statistics.} Some of the recast searches support routines of the recasting tools for combining bounds from several signal regions.
While this can increase the bound, it is not compatible with our method of combining contributions of different processes to a single signal region.
The efficiencies of the signal regions that are contained in our data base can in principle be used to construct the combinations of signal regions.
However, this would require rerunning the statistical combination for each combination for masses and branching ratios, and we do not provide any routines to this end.

\begin{table}[]
\centering
\small
\renewcommand{\arraystretch}{0.85}
\begin{tabular}{lll}
    \toprule
    Analysis & Description & Recast \\ \midrule
    \begin{tabular}[c]{@{}l@{}}ATLAS JHEP \cite{ATLAS:2019gdh}\\ $139~\mathrm{fb}^{-1}$\end{tabular} & \begin{tabular}[c]{@{}l@{}}Bottom-squark pair production; \\ no leptons, $\geq 3$ $b$-jets and MET\end{tabular} & \begin{tabular}[c]{@{}l@{}}\texttt{MadAnalysis5} \cite{Araz:2020stn}\\ \texttt{atlas\_susy\_2018\_31}\end{tabular} \\[3ex]
    \begin{tabular}[c]{@{}l@{}}CMS-SUS-16-033 \cite{CMS:2017abv}\\ $35.9~\mathrm{fb}^{-1}$\end{tabular} & \begin{tabular}[c]{@{}l@{}}Gluinos and stops;\\ no leptons, multiple jets and MET\end{tabular} & \begin{tabular}[c]{@{}l@{}}\texttt{MadAnalysis5} \cite{DVN/GBDC91_2021}\\ \texttt{cms\_sus\_16\_033}\end{tabular} \\[3ex] 
    \begin{tabular}[c]{@{}l@{}}CMS PAS SUS-19-006 \cite{CMS:2019xjf}\\ $137~\mathrm{fb}^{-1}$\end{tabular} & \begin{tabular}[c]{@{}l@{}}Gluinos and squarks;\\ no leptons, multiple jets and MET\end{tabular} & \begin{tabular}[c]{@{}l@{}}\texttt{MadAnalysis5} \cite{Mrowietz:2020ztq}\\ \texttt{cms\_sus\_19\_006}\end{tabular} \\[3ex]
    \begin{tabular}[c]{@{}l@{}}ATLAS-CONF-2019-040 \cite{ATLAS:2019vcq}\\ $139~\mathrm{fb}^{-1}$\end{tabular} & \begin{tabular}[c]{@{}l@{}}$R$-parity conserving SUSY;\\ jets, MET, no leptons\end{tabular} & \begin{tabular}[c]{@{}l@{}}\texttt{MadAnalysis5} \cite{DVN/NW3NPG_2021}\\ \texttt{atlas\_conf\_2019\_040}\end{tabular} \\[3ex]
    \begin{tabular}[c]{@{}l@{}}CMS PAS EXO-19-002 \cite{CMS:2019xud}\\ $137~\mathrm{fb}^{-1}$\end{tabular} & \begin{tabular}[c]{@{}l@{}}Type-III seesaw and light scalars;\\ at least 3 charged leptons\end{tabular} & \begin{tabular}[c]{@{}l@{}}\texttt{MadAnalysis5} \cite{Conte:2021xtt}\\ \texttt{cms\_exo\_19\_002}\end{tabular} \\[3ex]
    \begin{tabular}[c]{@{}l@{}}CMS PAS TOP-18-003 \cite{CMS:4top}\\ $137~\mathrm{fb}^{-1}$\end{tabular} & \begin{tabular}[c]{@{}l@{}}Search for SM $4t$ production \\ 2SSL or 3 leptons and jets\end{tabular} & \begin{tabular}[c]{@{}l@{}}\texttt{MadAnalysis5} \cite{Darme:2020hxc}\\ \texttt{cms\_top\_18\_003}\end{tabular} \\[3ex]
    \begin{tabular}[c]{@{}l@{}}ATLAS JHEP \cite{ATLAS:2019fag}\\ $139~\mathrm{fb}^{-1}$\end{tabular} & \begin{tabular}[c]{@{}l@{}}Gluino and squark production \\ 2SSL or 3 leptons and jets\end{tabular} & \begin{tabular}[c]{@{}l@{}}\texttt{CheckMATE}\\ \texttt{atlas\_1909\_08457}\end{tabular} \\[3ex]
    \begin{tabular}[c]{@{}l@{}}ATLAS PRD \cite{ATLAS:2017mjy}\\ $36.1~\mathrm{fb}^{-1}$\end{tabular} & \begin{tabular}[c]{@{}l@{}}Gluino and squark production \\ fully hadronic final state\end{tabular} & \begin{tabular}[c]{@{}l@{}}\texttt{CheckMATE}\\ \texttt{atlas\_1712\_02332}\end{tabular} \\[3ex]
    \begin{tabular}[c]{@{}l@{}}ATLAS JHEP \cite{ATLAS:2017tmw}\\ $36.1~\mathrm{fb}^{-1}$\end{tabular} & \begin{tabular}[c]{@{}l@{}}Gluino and squark production \\ 2SSL or 3 leptons, $b$-jets and MET \end{tabular} & \begin{tabular}[c]{@{}l@{}}\texttt{CheckMATE}\\ \texttt{atlas\_1706\_03731}\end{tabular} \\
    \midrule
    \begin{tabular}[c]{@{}l@{}}ATLAS EPJ C \cite{ATLAS:2021twp}\\ $139~\mathrm{fb}^{-1}$\end{tabular} & \begin{tabular}[c]{@{}l@{}}Squarks and gluinos;\\ 1 lepton, jets and MET\end{tabular} & \begin{tabular}[c]{@{}l@{}}\texttt{CheckMATE}\\ \texttt{atlas\_2101\_01629}\end{tabular} \\[3ex]
    \begin{tabular}[c]{@{}l@{}}ATLAS EPJ C \cite{ATLAS:2022ihe}\\$139~\mathrm{fb}^{-1}$\end{tabular} & \begin{tabular}[c]{@{}l@{}}Gluino pair production; \\ $\leq 1$ leptons, $\geq 3$ b-jets and MET\end{tabular} & \begin{tabular}[c]{@{}l@{}}\texttt{CheckMATE}\\ \texttt{atlas\_2211\_08028}\end{tabular} \\[3ex]
    \begin{tabular}[c]{@{}l@{}}ATLAS EPJ C \cite{ATLAS:2021fbt}\\ $139~\mathrm{fb}^{-1}$\end{tabular} & \begin{tabular}[c]{@{}l@{}}RPV SUSY; many jets,\\ $\geq 1$ leptons and 0 or $\geq 3$ $b$-jets\end{tabular} & \begin{tabular}[c]{@{}l@{}}\texttt{CheckMATE}\\ \texttt{atlas\_2106\_09609}\end{tabular} \\ \bottomrule
\end{tabular}
\caption{Summary of the analyses used in the constraint data base. Ref.~\cite{ATLAS:2019gdh} is also implemented in \texttt{CheckMATE} as \texttt{atlas\_1908\_03122}. The searches in the lower panel do not have a public validation note and are not included in the results in the main text.}
\label{tab:analyses}
\end{table}

\paragraph{Simple-minded signal event generation.} 
Signal simulation of the QCD pair production was only performed at leading order. 
We rescaled the cross section to NNLO+NNLL calculations afterwards.
We did not consider signal-BSM nor signal-signal interference (in case of multiple VLQs). 
We set all couplings small enough to use the narrow width approximation, neglecting all finite width effects.
For the VLQ and BSM scalar decays we only used left-handed BSM couplings. 
We checked on several processes that for the currently most dominant search recasts, the constraints obtained from the recasts with opposite chirality couplings do not differ beyond statistical fluctuations from those presented.  

\paragraph{Limited channels.}
We only considered decays of VLQs into third generation quarks and colour-neutral SM or BSM scalars (not including, for example, coloured scalars which also are present in composite Higgs models \cite{Flacke:2025xwl}).
Furthermore, we assumed the BSM scalars to decay as $S^{0}\rightarrow t\bar{t}$, $S^{+}\rightarrow t\bar{b}$, and $S^{++}\rightarrow t\bar{b}W^+$, neglecting other potential scalar decay modes such as into dibosons \cite{Cacciapaglia:2022bax}. We also do not consider ``VLQ-chain-decays'' such as $Q\rightarrow Q' S $ or $Q\to Q' V$ with a scalar $S$ or vector $V$. 

\paragraph{Mass grid.} 
We simulated a regular grid of $(m_Q,m_S)$ combinations. Our results are not directly applicable to a more general spectrum. However, the database can be used to get a good estimate also for a generic masses, e.g.\ by interpolating the efficiencies between simulated mass points.

\section{Bounds on VLQ pair production}\label{sec:bounds}

In this section we first highlight some bounds on specific channels and compare them to experimental searches where available. 
Then we give some examples for combining limits across decay channels using the constraint data base.

\subsection{Individual channels}\label{sec:individual}

\begin{figure}
    \centering
    \begin{subfigure}{0.32\linewidth}
        \includegraphics[width=\linewidth]{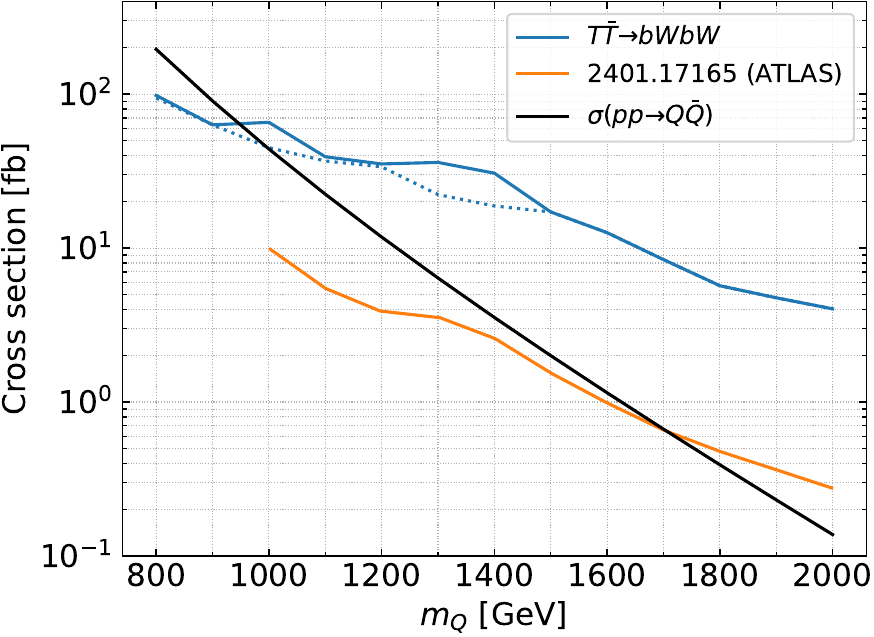}
        \caption{$T\bar T/Y\bar Y \to bWbW$}
        \label{fig:bwbw}
    \end{subfigure}
    \begin{subfigure}{0.32\linewidth}
        \includegraphics[width=\linewidth]{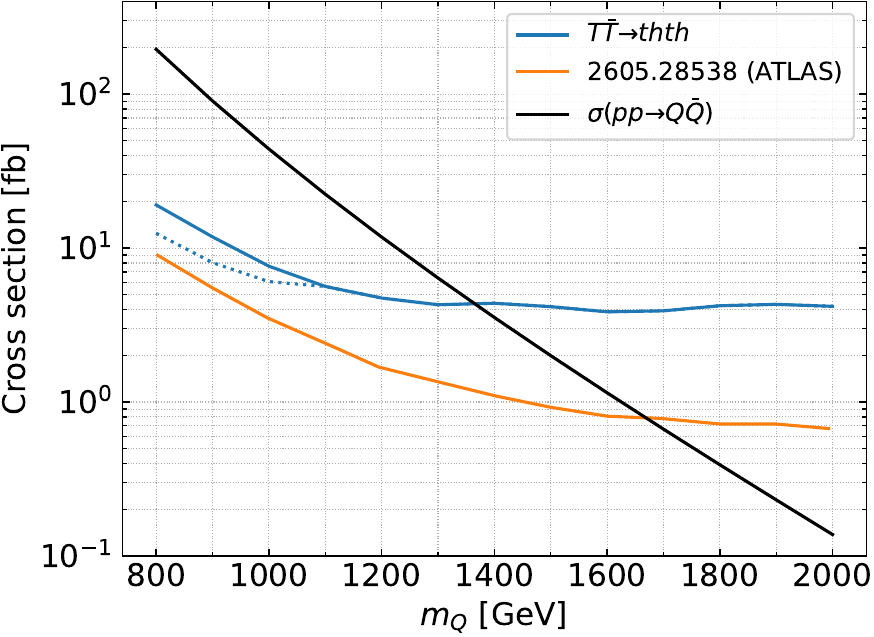}
        \caption{$T\bar T \to thth$}
        \label{fig:thth}
    \end{subfigure}
    \begin{subfigure}{0.32\linewidth}
        \includegraphics[width=\linewidth]{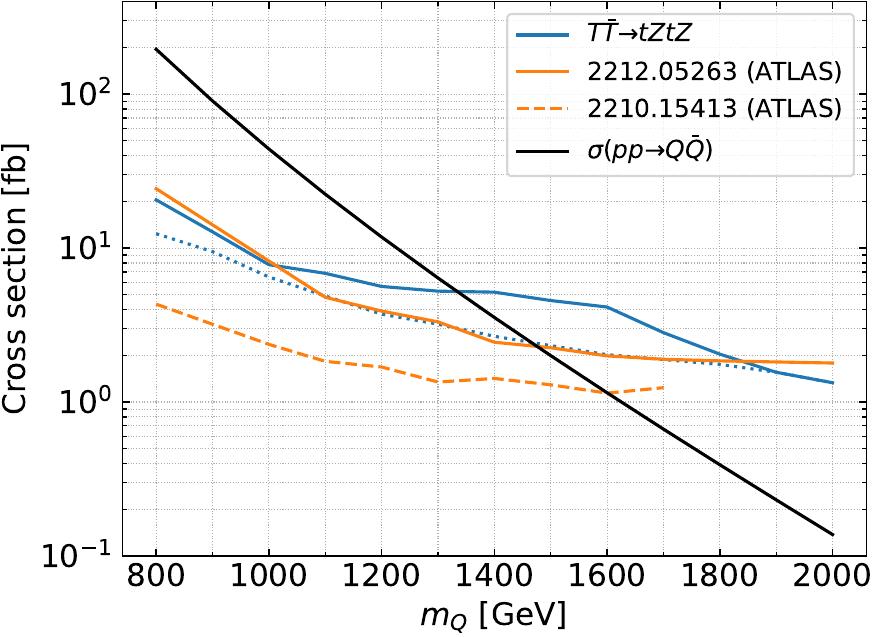}
        \caption{$T\bar T \to tZtZ$}
        \label{fig:tztz}
    \end{subfigure}
    \vspace{1ex}
    
    \begin{subfigure}{0.32\linewidth}
        \includegraphics[width=\linewidth]{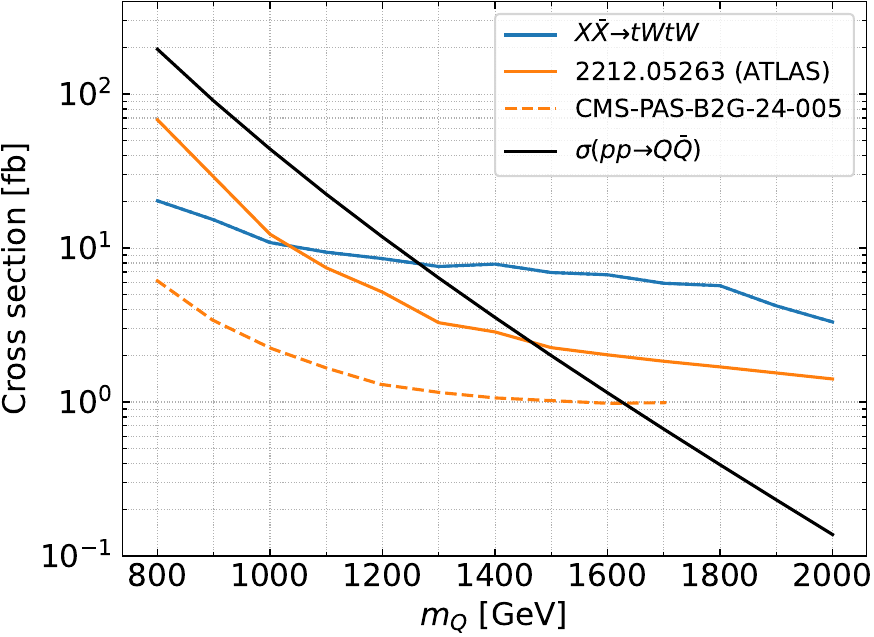}
        \caption{$B\bar B/X\bar X \to tWtW$}
        \label{fig:twtw}
    \end{subfigure}
    \begin{subfigure}{0.32\linewidth} 
        \includegraphics[width=\linewidth]{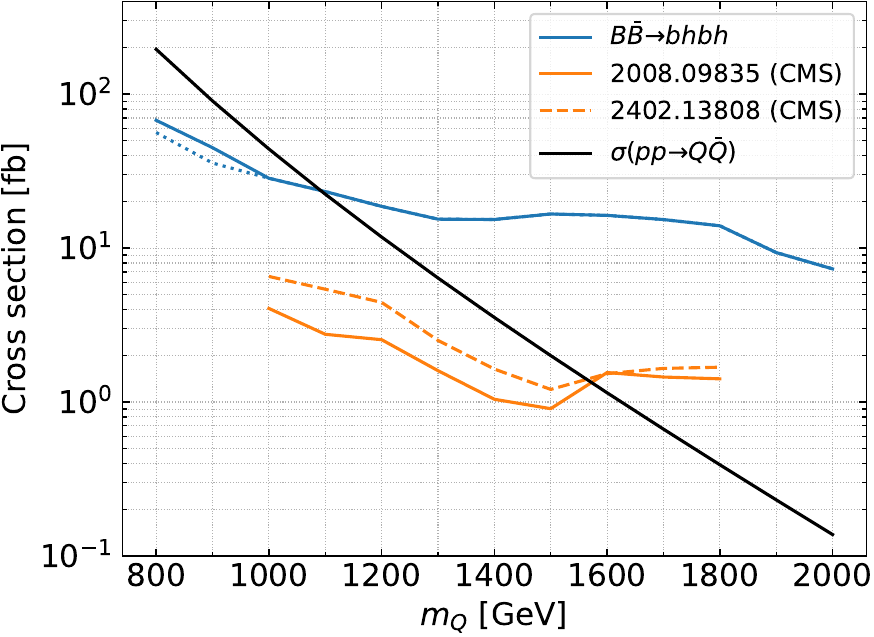}
        \caption{$B\bar B\to bhbh$}
        \label{fig:bhbh}
    \end{subfigure}
    \begin{subfigure}{0.32\linewidth}
        \includegraphics[width=\linewidth]{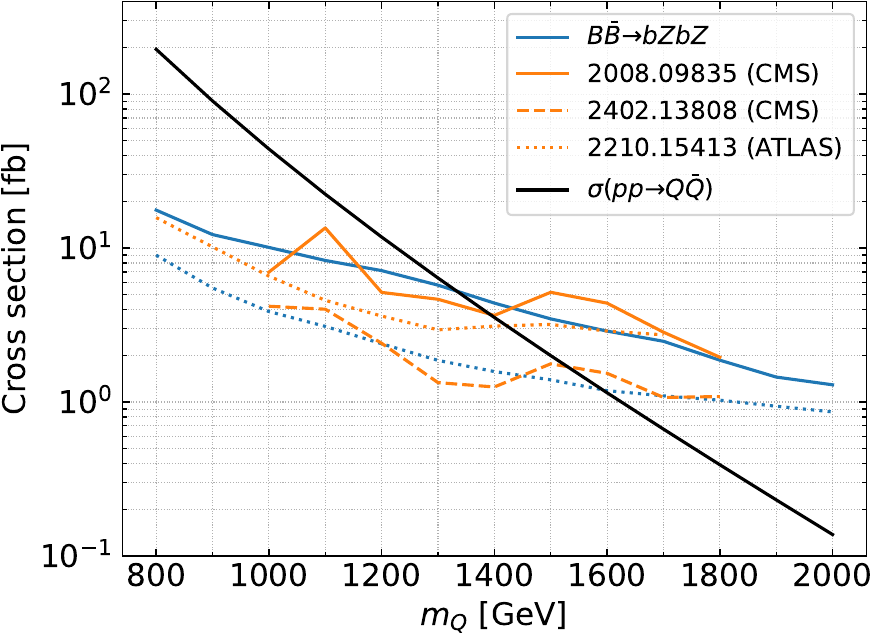}
        \caption{$B\bar B \to bZbZ$}
        \label{fig:bzbz}
    \end{subfigure}
    \caption{Bounds on standard decay channels of VLQs. The solid blue lines show the recasting bounds used in the following analysis, the dotted blue lines include statistical combination of signal regions where available (see text). For comparison, we show limits from experimental searches in orange \cite{ATLAS:2022tla,ATLAS:2024gyc,ATLAS:2026ojf,ATLAS:2022hnn,CMS:2024xbc,CMS:2020ttz,CMS:2026ggz}. All bounds in (a) and (d) apply to both production channels, except for the green line which targets only $X\bar X \to tWtW$. The theory cross section in black is calculated at NNLO+NNLL accuracy.}
    \label{fig:bounds_standard}
\end{figure}

In \cref{fig:bounds_standard} we study all standard channels where both VLQs decay via the same decay channel.
We show the upper limits obtained from the recasts in blue, with solid lines using only single signal region bounds and dotted lines including combinations of signal regions.
We can read off mass bounds by comparing with the QCD cross section, shown in black for NNLO+NNLL accuracy.
We also show all limits on these channels from dedicated VLQ searches by ATLAS and CMS that use the full Run-2 dataset \cite{ATLAS:2022tla,ATLAS:2024gyc,ATLAS:2026ojf,ATLAS:2022hnn,CMS:2024xbc,CMS:2020ttz,CMS:2026ggz}. 
In \cref{fig:bwbw,fig:twtw}, the bounds are not charge sensitive and therefore apply to both production channels.

We differentiate between bounds including (dotted blue lines) and excluding (solid blue lines) combinations of signal regions because the method for combining channels described in \cref{sec:recast:method} is only applicable to single signal regions. 
The results obtainable with the constraint data base are therefore based on the solid lines in \cref{fig:bounds_standard}.
The combinations make the most notable differences in \cref{fig:tztz,fig:bzbz}.
The dominant bounds for these channels arise from Ref.~\cite{CMS:2019xjf}, a search for supersymmetry including $\tilde t \tilde t^* \to t\bar t+\met$ and $\tilde b\tilde b^* \to b\bar b + \met$.
These are very sensitive to the VLQ decays into quarks and $Z(\to \nu\bar \nu)$, leading to recasting bounds on $B\bar B \to bZbZ$ in \cref{fig:bzbz} that are competitive with the direct searches.
Otherwise the experimental bounds are stronger than the recast bounds, as expected from dedicated searches.
The difference in mass bounds is especially stark for $T\bar T\to bWbW$ in \cref{fig:bwbw}. 
This is because this channel suffers from the large SM $t\bar t$ background, which requires a dedicated search strategy to be overcome.
The searches available in the recasting tools do not specifically target the $t\bar t$ signature, so they yield only weak constraints for this channel.

\begin{figure}
    \centering
    \begin{subfigure}{0.4\linewidth}
        \includegraphics[width=\linewidth]{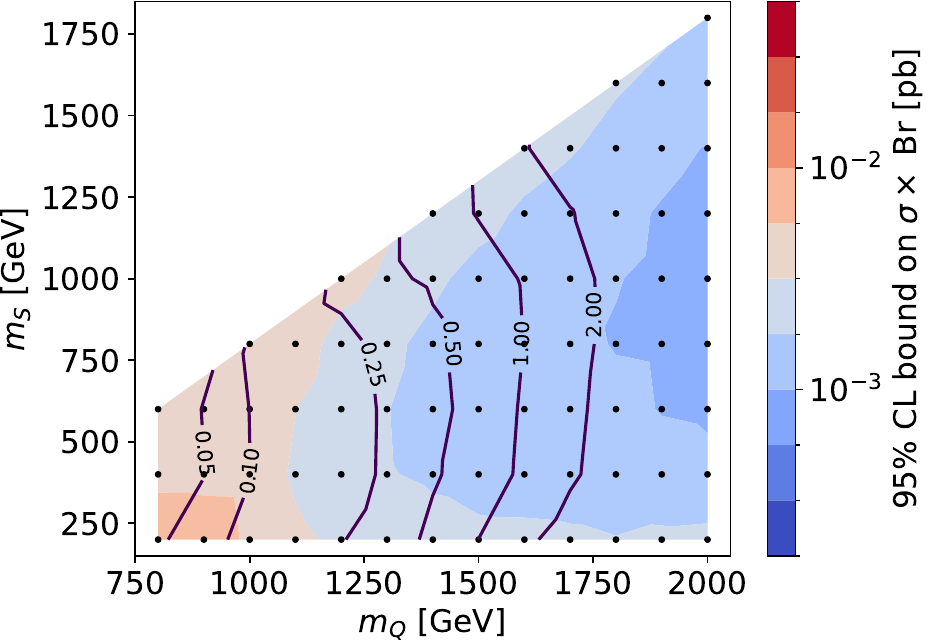}
    \end{subfigure}\quad 
    \begin{subfigure}{0.4\linewidth}
        \includegraphics[width=\linewidth]{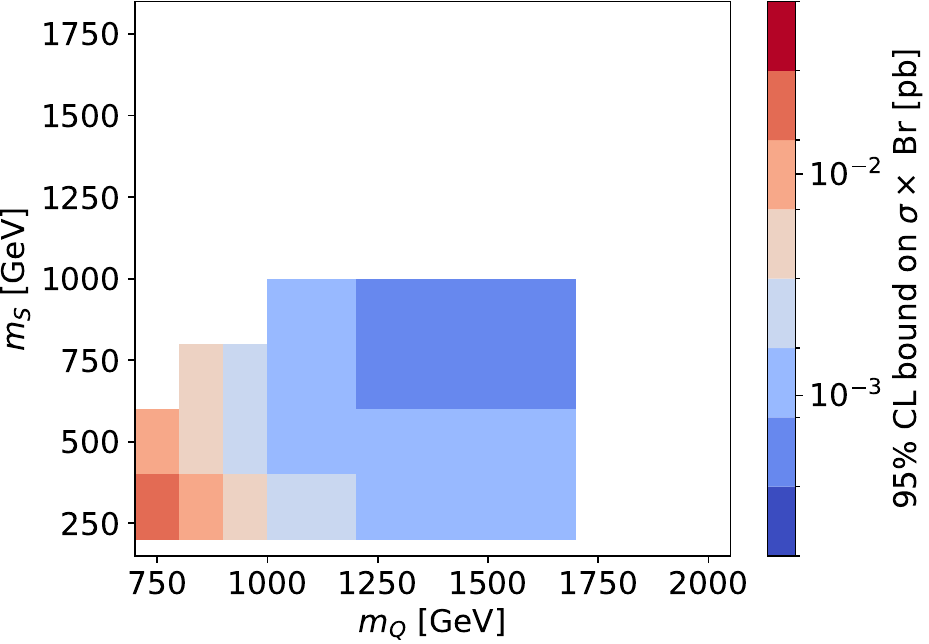}
    \end{subfigure}
    \caption{Bounds on $X \bar X \to t S^+ \bar tS^- \to ttbttb$. The left panel shows recasting bounds with mass bounds for fixed values of the total branching ratio to SM states, the right panel are the results from a direct search for this channel \cite{CMS:2026ggz}. In colour we show the upper limit on cross section times branching ratio, referring to the effective branching ratio of the full decay chain into SM particles.}
    \label{fig:xxts11ts11}
\end{figure}

While the standard decay channels have been searched for extensively by ATLAS and CMS, to our knowledge $X\bar X \to tS^+ \bar t S^- \to ttbttb$ is the only exotic channel with top-philic decays that has been targeted so far \cite{CMS:2026ggz}.
In \cref{fig:xxts11ts11} we compare our recasting bounds (left) to the experimentally observed bound (right).
We present the results as upper limits on cross section times total branching ratio to SM states with a heatmap in the mass-mass-plane.
Assuming a fixed branching ratio, we draw mass bounds as solid lines.
We also include a line for an effective branching ratio of 200\%, which is to be interpreted as two copies of $X$ with the same mass decaying the same way.
As expected, the direct search provides stronger bounds in the high $m_Q$ region, leading to a mass bound of up to $1.71$~TeV compared to $1.59$~TeV from the recast.

\begin{figure}
    \centering
    \begin{subfigure}{0.3\linewidth}
        \includegraphics[height=4.0cm]{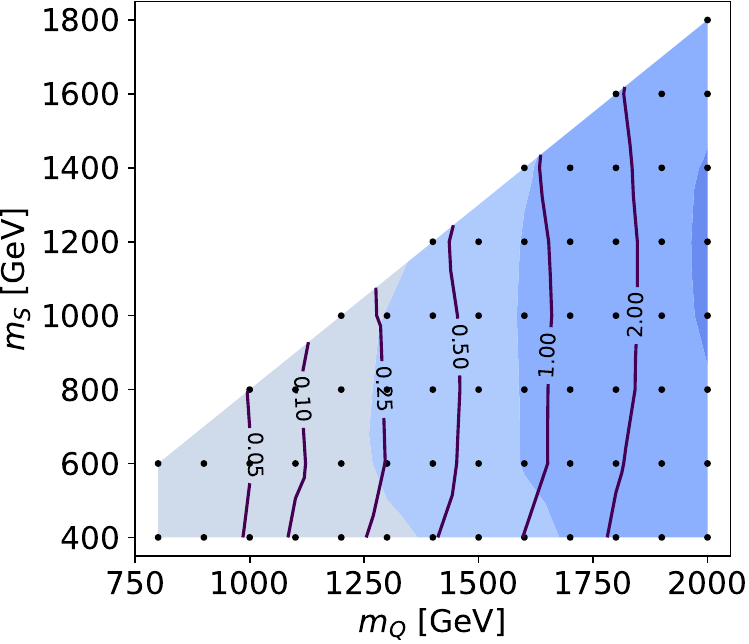}
        \caption{$T\bar T \to tS^0 \bar tS^0 \to 6t$}
        \label{fig:bounds_main_6t}
    \end{subfigure}
    \begin{subfigure}{0.3\linewidth}
        \includegraphics[height=4.0cm]{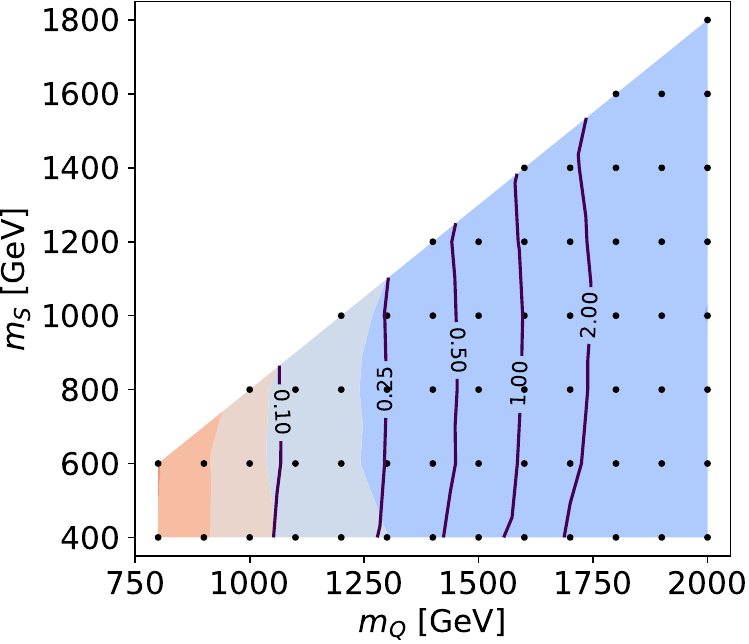}
        \caption{$T\bar T \to tS^0 \bar t Z \to ttt tZ$}
        \label{fig:bounds_main_4tz}
    \end{subfigure}
    \begin{subfigure}{0.36\linewidth}
        \includegraphics[height=4.0cm]{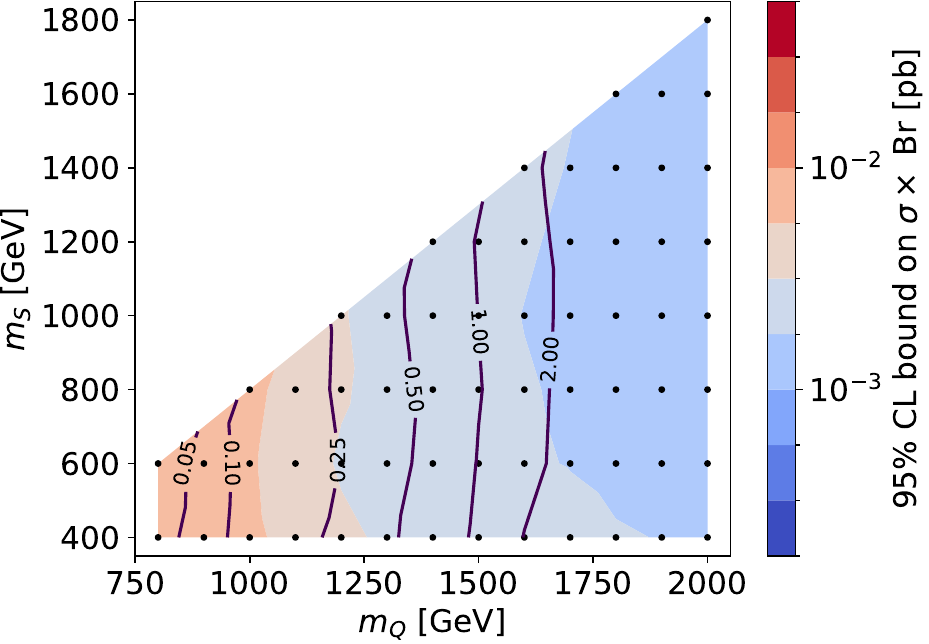}
        \caption{$T\bar T \to th \bar tS^0 \to th ttt$}
        \label{fig:bounds_main_3tbw}
    \end{subfigure}
    \caption{Bounds on $T\bar T$ production in purely exotic and mixed standard-exotic decay channels. In colour we show the upper limit on cross section times branching ratio, referring to the effective branching ratio of the full decay chain into SM particles. The solid lines show mass bounds assuming the indicated effective branching ratio.}
    \label{fig:bounds_exotic_sample}
\end{figure}

In \cref{fig:bounds_exotic_sample} we show recasting bounds on some selected processes for which no direct search is available yet. 
We highlight the process $T\bar T \to t S^0 \bar tS^0 \to 6t$ in \cref{fig:bounds_main_6t}, which is particularly interesting for composite Higgs models as it is a generic prediction of the model class in \cite{Belyaev:2016ftv}. 
If we take one single VLQ with a total branching ratio of 100\%, this processes is excluded up to a VLQ mass of 1.65~TeV, even exceeding the typical experimental bounds on standard channels. 
The reason is the busy nature of the final state with many $b$-quarks and leptons, which has very little SM background. 
Our strong recasting bounds underline the importance of experimental searches for this channel, as a dedicated search will be even more sensitive and can therefore test VLQ masses well above 1.5~TeV. 

In a full model, the $T$ will however not exclusively decay to BSM scalars but also to SM bosons, thus reducing the branching ratio into the $6t$ channel. 
For example, for a top partner with branching ratios
\begin{align}
    \mathrm{Br}(T\to tS^0) = 50\%, \quad \mathrm{Br}(T\to tZ) = 25\%, \quad \mathrm{Br}(T\to th) = 25\%, 
\end{align}
we have to read off the mass bound from the 25\% contour in \cref{fig:bounds_main_6t}, thus reducing the bound on $m_Q$ to 1.30~TeV.
We can see from \cref{fig:thth,fig:tztz} that the bounds $tZtZ$ and $thth$ are negligible for a total branching ratio of $(25\%)^2$.
However, in scenarios with multiple decay channels, also mixed decay channels as in \cref{fig:bounds_main_4tz,fig:bounds_main_3tbw} have to be considered.
For $T\bar T \to tS^0 tZ$ we have to take into account both combinations of the decays of $T$ and $\bar T$ (which have identical bounds).
Thus, this mixed channel has an effective branching ratio of 25\%, which also excludes $m_Q$ up to 1.30~TeV.
All other channels are subdominant and thus do not contribute to the bound despite making up half of the events.
In the next section we will show how this bound can be improved by taking into account all channels with the constraint data base.
We show the upper limits on all 43 channels in \cref{app:allplots}.

\subsection{Combination of channels}\label{sec:app}

In a realistic model, VLQs have several decay channels that can all contribute to the signal regions of the same search.
With the efficiencies for the individual channels that we have published, see \cref{sec:constraintdatabase}, bounds on arbitrary combinations of channels can be calculated using the method in \cref{sec:recast:method}.
In this section we show several applications of the constraint data base by setting bounds on both simplified and state-of-the-art VLQ models.

\begin{figure}
    \centering
    \includegraphics[width=0.47\linewidth]{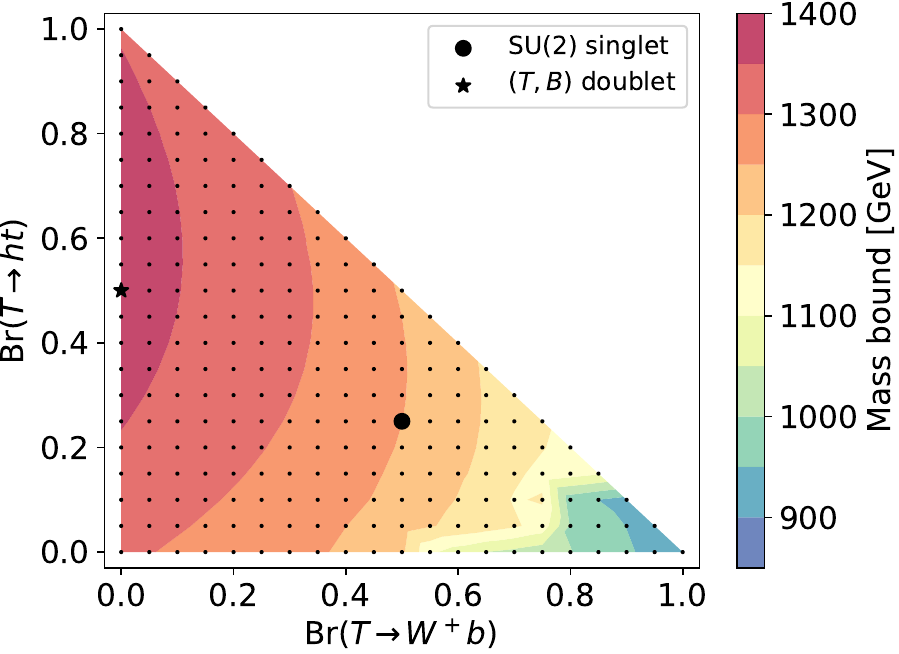} \quad 
    \includegraphics[width=0.47\linewidth]{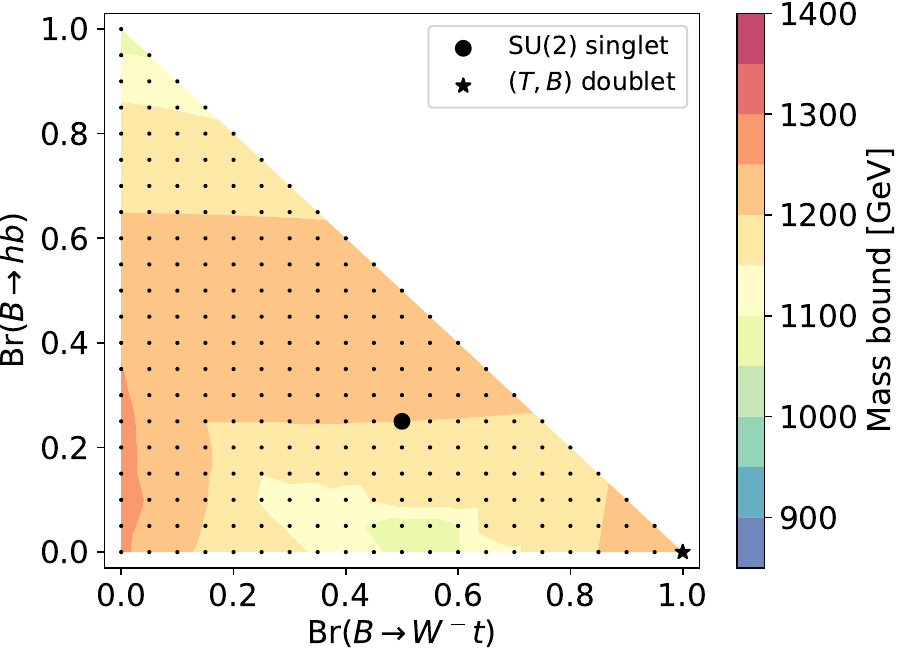}
    \caption{Mass bounds on VLQ pair production with standard decay channels in the plane of branching ratios.}
    \label{fig:triangle}
\end{figure}

As a first application, we only consider $T$ and $B$ standard decay channels.
Each VLQ can decay into three combinations of quark and SM boson. 
The parameter space of decays can therefore be fully specified using two branching ratios, the third being fixed by the condition the sum add up to unity.
This allows for a convenient visualisation of mass bounds in terms of a heat map in the Br-Br-plane, which are commonly employed by ATLAS and CMS.
In \cref{fig:triangle} we show these ``triangle plots'' derived from the recasting bounds\footnote{ATLAS has published triangle plots using 139~fb$^{-1}$ in the multi-lepton \cite{ATLAS:2022hnn} and large-MET \cite{ATLAS:2022tla} final states, and a combination of both final states has been performed for $36.1$~fb$^{-1}$ \cite{ATLAS:2018ziw}.
Our recast searches mostly use 139~fb$^{-1}$ of data and include both leptonic and hadronic analyses.
Therefore, none of the above papers can be used for a direct comparison. To our knowledge, also CMS has not published a corresponding combination of final states. We therefore don't show a comparison for the triangle plots.}.
We again see that the $T\to Wb$ region is weakly constrained, whereas $T\to th, tZ$ are excluded for $m_T$ below about $1.35$~TeV.
The bounds on $B$ are generally weaker but less dependent on the decay channel.
In the large mass limit, the branching ratios are fixed by the Goldstone Equivalence theorem and depend only on the $\SU(2)$ representations \cite{Aguilar-Saavedra:2009xmz,DeSimone:2012fs}.
Singlets decay into the channels featuring a $W:Z:h$ with relative branching ratios $2:1:1$.
Top partners from a $(T,B)$ doublet on the other hand decay as $T\to tZ, th$ with equal branching ratio and exclusively $B \to tW^-$.
In \cref{fig:triangle} we mark the corresponding points in the plane of branching ratios.

\begin{figure}[h]
    \centering
    \begin{subfigure}{0.47\linewidth}
        \includegraphics[width=\linewidth]{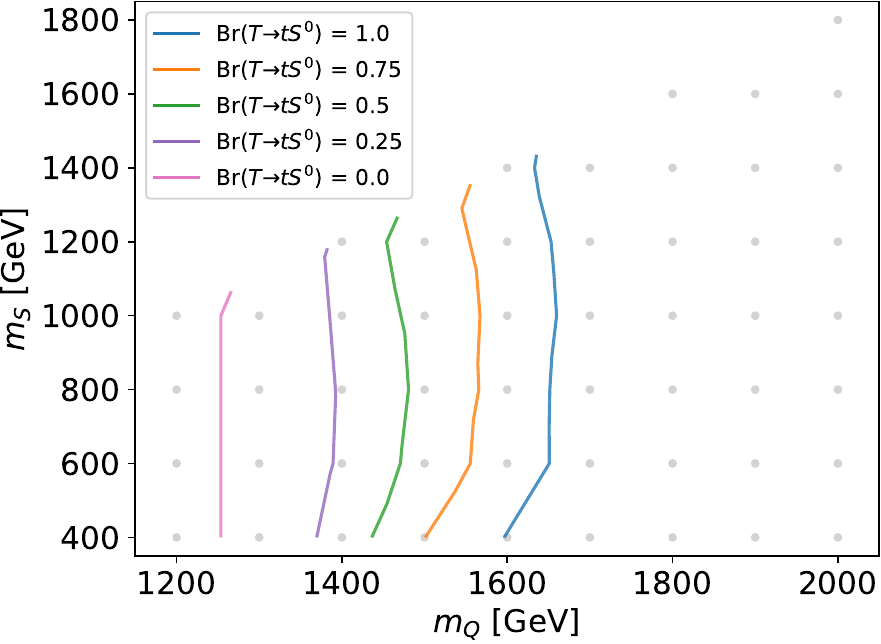}
        \caption{Singlet top partner}
        \label{fig:tptp_singlet}
    \end{subfigure}\quad 
    \begin{subfigure}{0.47\linewidth}
        \includegraphics[width=\linewidth]{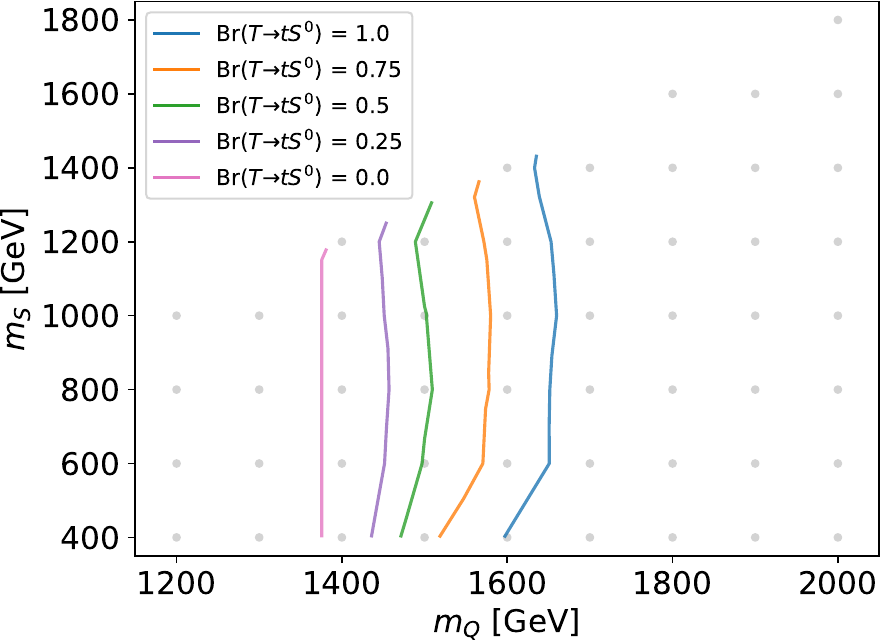}
        \caption{Doublet top partner}
        \label{fig:tptp_doublet}
    \end{subfigure}
    \caption{Bounds on pair production of a charge 2/3 top partner $T$ for different branching ratios into $T\to tS^0$. In (a), $T$ is an $\SU(2)_L$ singlet with standard decays $bW:tZ:th=2:1:1$. In (b), $T$ emerges from an $\SU(2)_L$ doublet and the standard decays are $bW:tZ:th=0:1:1$.}
    \label{fig:tptp_application}
\end{figure}

As a next step, we also allow for the exotic decay channel $T \to tS^0$, which occurs in a $\SU(4)/\Sp(4)$ composite Higgs model.
We take $T$ to emerge from a singlet or doublet, so the decays are fixed by specifying $\mathrm{Br}(T\to tS^0)$.
In \cref{fig:tptp_application} we show the exclusion limits in the $m_Q$-$m_S$-plane for different branching ratios.
The blue lines mark the case of purely exotic decays and match with the 100\% contour in \cref{fig:bounds_main_6t}.
At the end of \cref{sec:individual} we found that for $\mathrm{Br}(T\to tS^0) = 50\%$, the strongest limit from an individual channel is 1.30~TeV.
From the green line in \cref{fig:tptp_doublet} we find a limit of up to 1.51~TeV from the combination of channels, marking a significant improvement.

\begin{figure}
    \centering
    \includegraphics[width=0.5\linewidth]{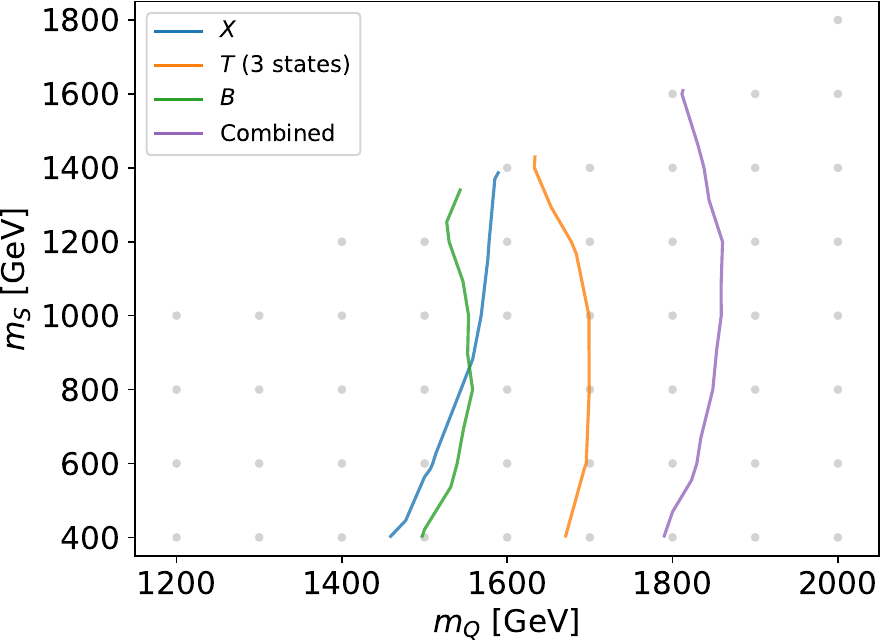}
    \caption{Bounds on pair production of VLQs in the model M5 including both standard and exotic decay channels.}
    \label{fig:m5_application}
\end{figure}

As a more complicated example we take the VLQ content of a full composite Higgs model with $\SU(5)/\SO(5)$ symmetry breaking and featuring top partners in the $\mathbf 5$ of $\SO(5)$ \cite{Cacciapaglia:2021uqh}.
The VLQs consist of one species of $X$, three $T$, and one $B$. 
Their couplings to scalars are calculated in appendix B of \cite{Cacciapaglia:2021uqh}. 
From the couplings to the Higgs and the $\SU(2)_L$ representations, we can reconstruct all branching ratios of the VLQs.
While we assumed constant branching ratios in the examples above, for this model we calculate the branching ratio for each point of the mass grid.
To this end we assume all pNGBs besides the Higgs to have the same mass $m_S$.
We also take all VLQs to have the same mass $m_Q$, which is necessary since the constraint data base assumes a regular mass grid. 
In \cref{fig:m5_application} we show the mass bounds of each kind of VLQ separately, as well as the combined limit from all five states.
The bound on $X$ increases with $m_S$, reaching from $1.46$~TeV to $1.59$~TeV.
The bottom partner is excluded up to $1.56$~TeV, while the multiplicity of the three top partners pushes their mass bound to $1.70$~TeV.
When taking all five VLQs into account, the bound increases to $m_Q \geq 1.79$-$1.86$~TeV.
Realistically, the VLQs that participate in the top mixing receive additional contributions to their mass, breaking the mass degeneracy.
We can interpret the bounds with all states at $m_Q$ as conservative if we take $m_Q$ as the mass of the heaviest VLQ, since other states will be lighter and therefore have a larger cross section.
Conversely, if we take $m_Q$ as the mass of the lightest VLQ, our bounds are too aggressive.

\section{Conclusions and outlook}\label{sec:conclusion}
We have presented a study of pair production of vector-like quarks, including partners $T$ and $B$ of the top and bottom quarks as well as exotic states $X_{8/3}$, $X$ and $Y$ with charges $8/3$, $5/3$, and $-4/3$, respectively. 
Besides the standard decay channels into SM bosons we have also allowed for decays into BSM scalars $S^{++}$, $S^+$, and $S^0$.
Motivated by composite Higgs models, we assumed the scalars to decay to third generation quarks.
We identified 43 independent channels of pair production and decays and obtained recasting bounds for each of them on a large mass grid (cf.\ \cref{tab:channels}).
For many exotic channels, the recasting bounds on the VLQ mass reach up to $1.6$~TeV, which is competitive with direct searches for standard decay channels.
When the branching ratios of a VLQ are spread out over several decay channels, setting limits on each of them individually artificially limits the total bound. 
On the other hand, re-deriving the recasting bounds for each set of branching ratio implies a significant workload. 
We therefore described a method for obtaining a combined bound without the need for additional recasting, based on calculating the contributions of each channel to a signal region using the efficiencies. 
We show on a sample model that this significantly improves the bound.
The reader can apply our recasting results to their VLQ model using the public constraint data base\footnote{\url{https://zenodo.org/records/22142863}}.

There are multiple avenues for expanding on the presented work.
First and foremost, the approach taken in this work is not limited to vector-like quarks but can be applied to any BSM process as long as the production cross section is specified. 
Pooling the significant technical and time effort of producing recasting bounds into systematic studies such as this one, which allow for reuse without rerunning tools, is a more effective use of resources in the field and therefore deserves further attention. 

We only considered VLQ pair production, but at high masses we expect single production to also give a sizeable contribution. 
The data base can be readily extended to include single production, which we leave for a future work.
For the VLQ channels studied here, we limited ourselves to decays of the scalars into third-generation quarks. 
Additional decay channels, such as VLQ decays through coloured scalars or scalar decays to di-bosons, would be an interesting extension of the data base to cover an even broader set of options in underlying models with VLQs.
Finally, the bounds can be improved by including further searches.
This is currently bottlenecked by the considerable but not yet large number of public recast implementations, which however grows steadily with time.
Recasting bounds from novel recasts can easily be added to the data base.

\section*{Acknowledgement} We thank Giacomo Cacciapaglia, Mark Goodsell and Werner Porod for useful discussions and Rosy Caliri and Yu Chen for comments on the draft. 
This work has been supported by the Center for Advanced Computation at Korea Institute for Advanced Study.
TF is supported by a KIAS Individual Grant (QP083701) via the Quantum Universe Center at the Korea Institute for Advanced Study.
MK is supported by DFG research training group GRK 2994.

\appendix

\clearpage

\section{Upper limits for all channels}\label{app:allplots}

\begin{figure}[b]
    \centering
    \includegraphics[width=0.75\linewidth]{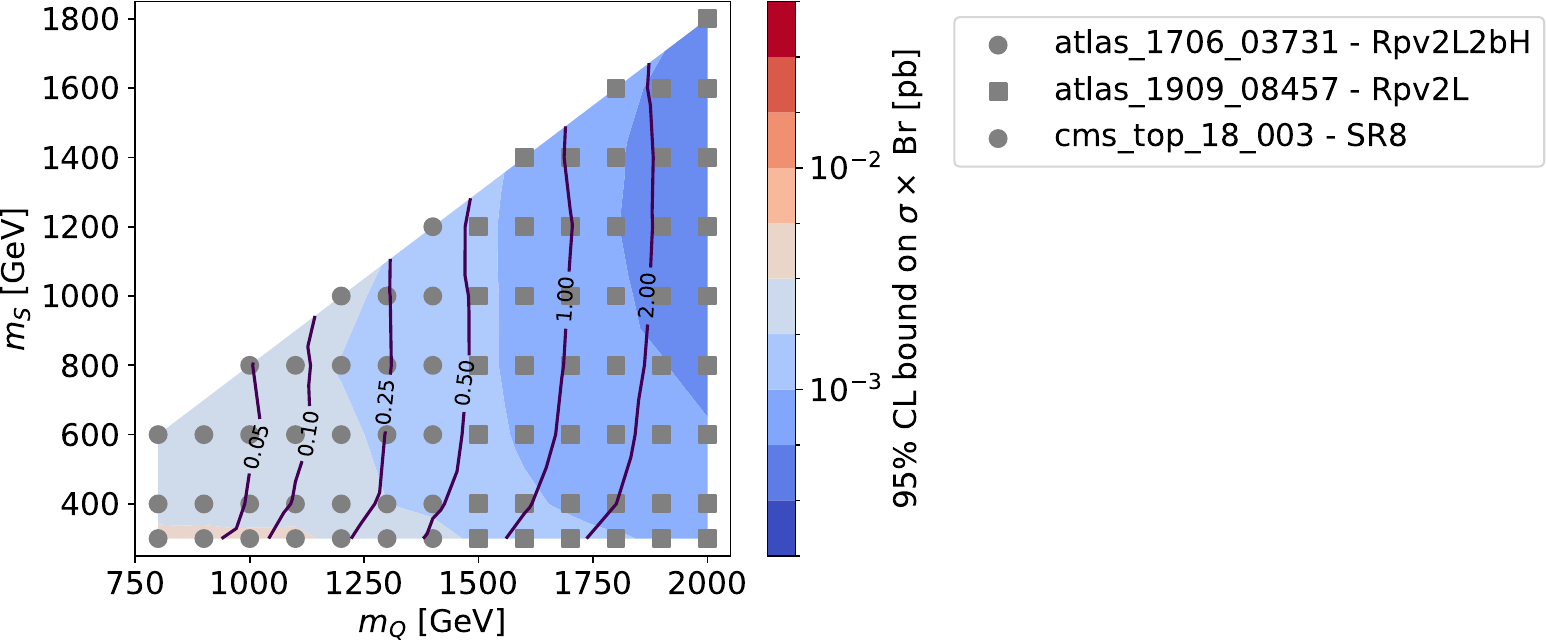}
    \caption{Bounds on $X_{8/3}$ pair production with decay $X_{8/3} \to t S^{++}$. The bounds match the ones for $Y\bar Y \to t S^{--} \bar t S^{++}$ as the relevant searches are not charge sensitive.}
    \label{fig:x83x83}
\end{figure}

\begin{figure}
    \centering

    \begin{subfigure}[t]{0.34\linewidth}
        \vspace{0pt}
        \includegraphics[width=\linewidth]{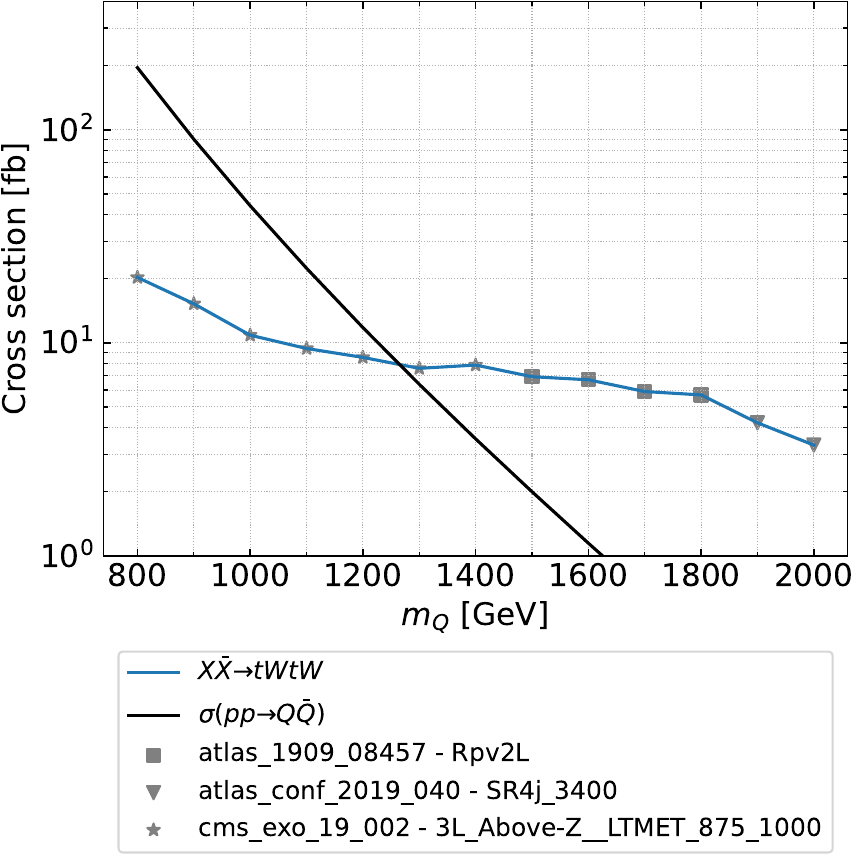}
        \caption{$X\bar X \to tW^+ \bar t W^-$}
        \label{sfig:xxtwtw}
    \end{subfigure}
    \begin{subfigure}[t]{0.29\linewidth}
        \vspace{0pt}
        \includegraphics[width=\linewidth]{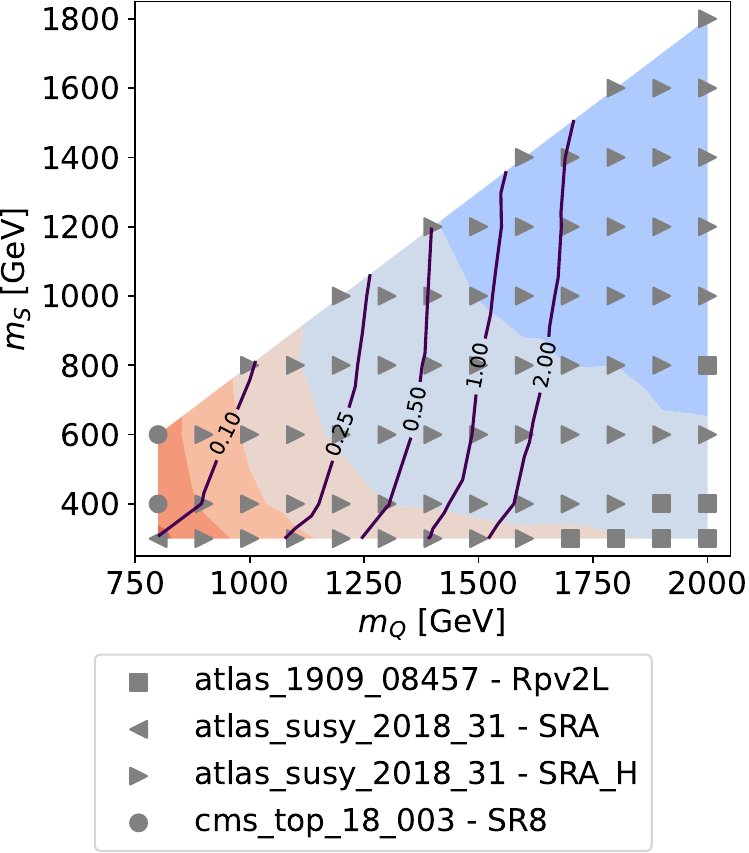}
        \caption{$X\bar X \to bS^{++} \bar t W^-$}
        \label{sfig:xxbstw}
    \end{subfigure}
    \begin{subfigure}[t]{0.352\linewidth}
        \vspace{0pt}
        \includegraphics[width=\linewidth]{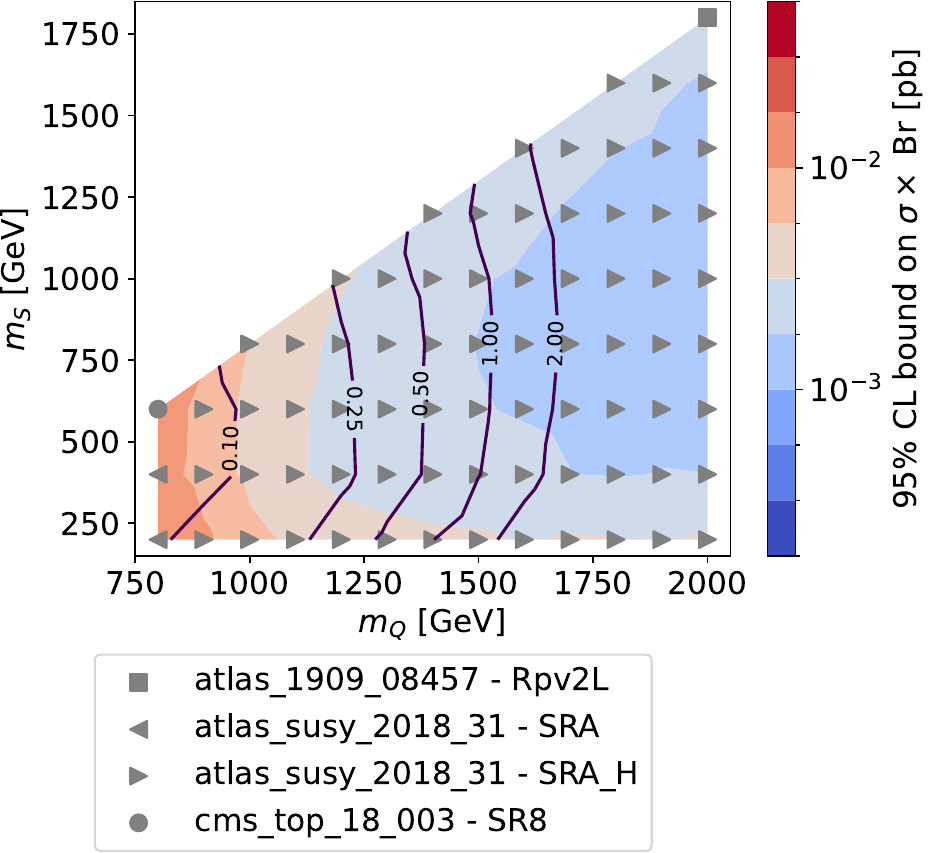}
        \caption{$X\bar X \to tS^{+} \bar t W^-$}
        \label{sfig:xxtstw}
    \end{subfigure}
    \vspace{1ex}
    
    \begin{subfigure}[t]{0.3\linewidth}
        \vspace{0pt}
        \includegraphics[width=\linewidth]{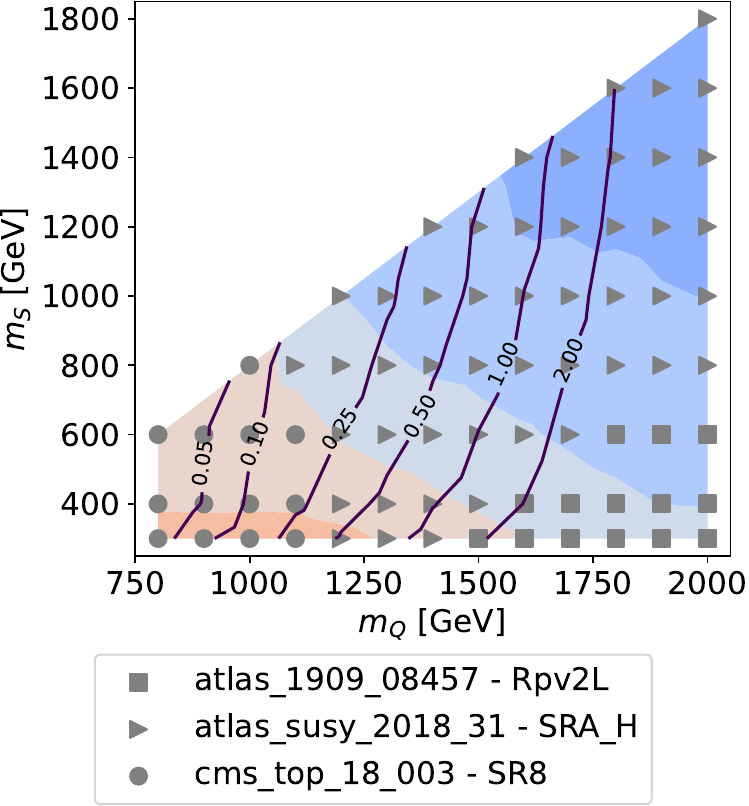}
        \caption{$X\bar X \to bS^{++} \bar b S^{--}$}
        \label{sfig:xxbsbs}
    \end{subfigure}
    \begin{subfigure}[t]{0.3\linewidth}
        \vspace{0pt}
        \includegraphics[width=\linewidth]{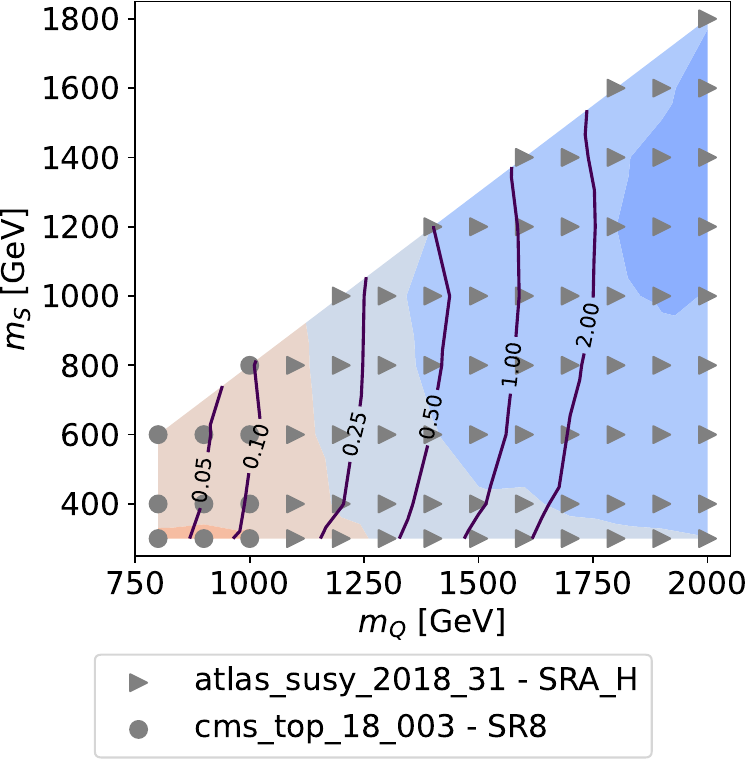}
        \caption{$X\bar X \to bS^{++} \bar t S^{-}$}
        \label{sfig:xxbsts}
    \end{subfigure}
    \begin{subfigure}[t]{0.367\linewidth}
        \vspace{0pt}
        \includegraphics[width=\linewidth]{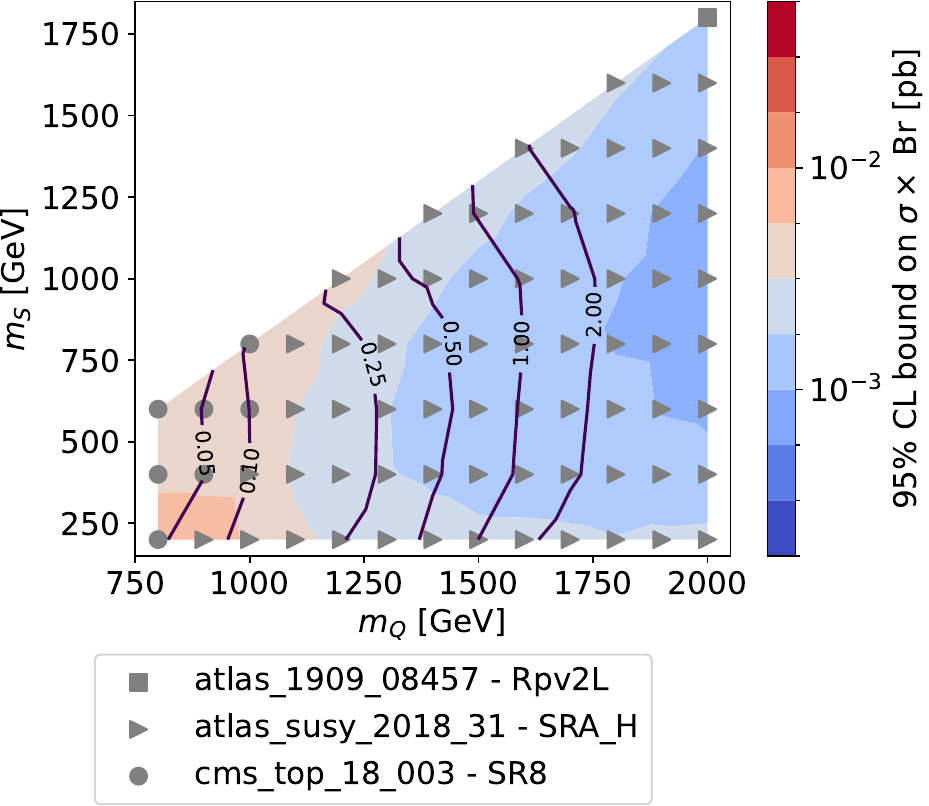}
        \caption{$X\bar X \to tS^{+} \bar t S^{-}$}
        \label{sfig:xxtsts}
    \end{subfigure}

    \caption{Bounds on pair production of $X$}
    \label{fig:xx}
\end{figure}

\begin{figure}
    \centering

    \begin{subfigure}[t]{0.34\linewidth}
        \vspace{0pt}
        \includegraphics[width=\linewidth]{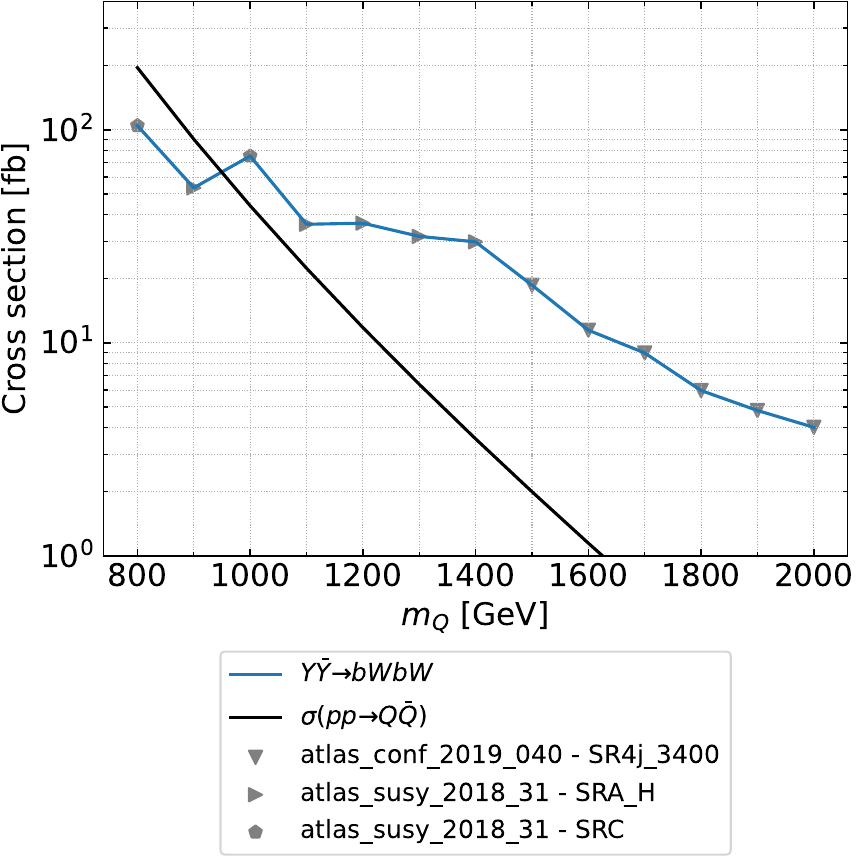}
        \caption{$Y\bar Y \to bW^- \bar b W^+$}
        \label{sfig:yybwbw}
    \end{subfigure}
    \begin{subfigure}[t]{0.29\linewidth}
        \vspace{0pt}
        \includegraphics[width=\linewidth]{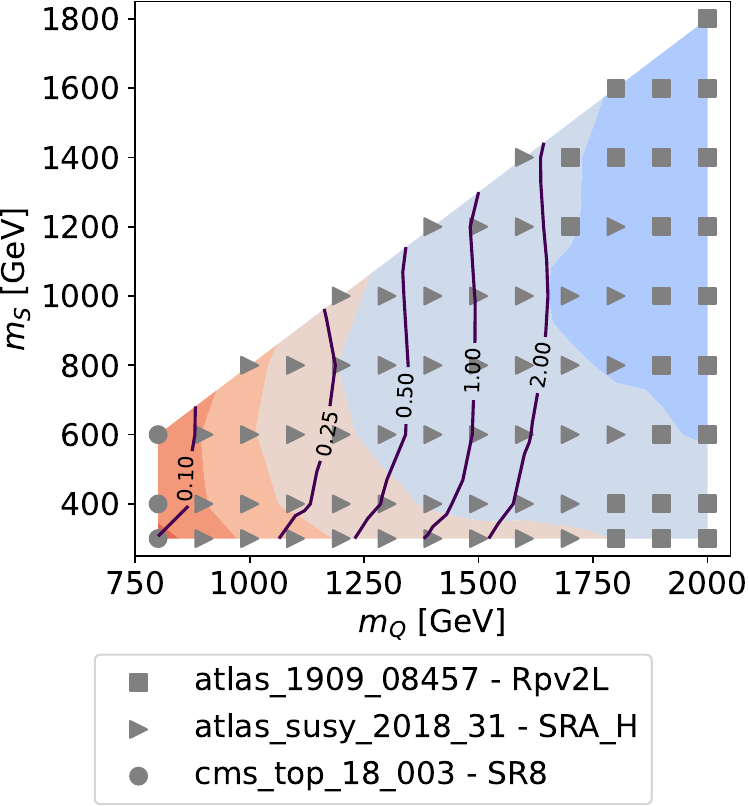}
        \caption{$Y\bar Y \to bW^- \bar tS^{--}$}
        \label{sfig:yybwts}
    \end{subfigure}
    \begin{subfigure}[t]{0.352\linewidth}
        \vspace{0pt}
        \includegraphics[width=\linewidth]{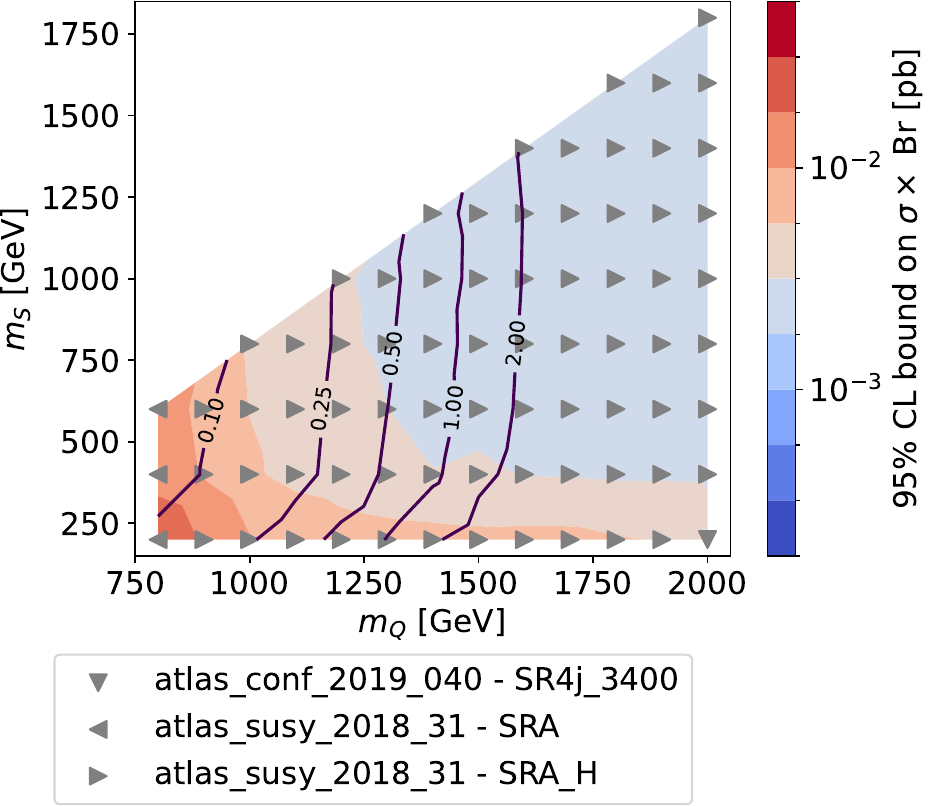}
        \caption{$Y \bar Y \to bW^- \bar b S^+ $}
        \label{sfig:yybwbs}
    \end{subfigure}
    
    \begin{subfigure}[t]{0.3\linewidth}
        \vspace{0pt}
        \includegraphics[width=\linewidth]{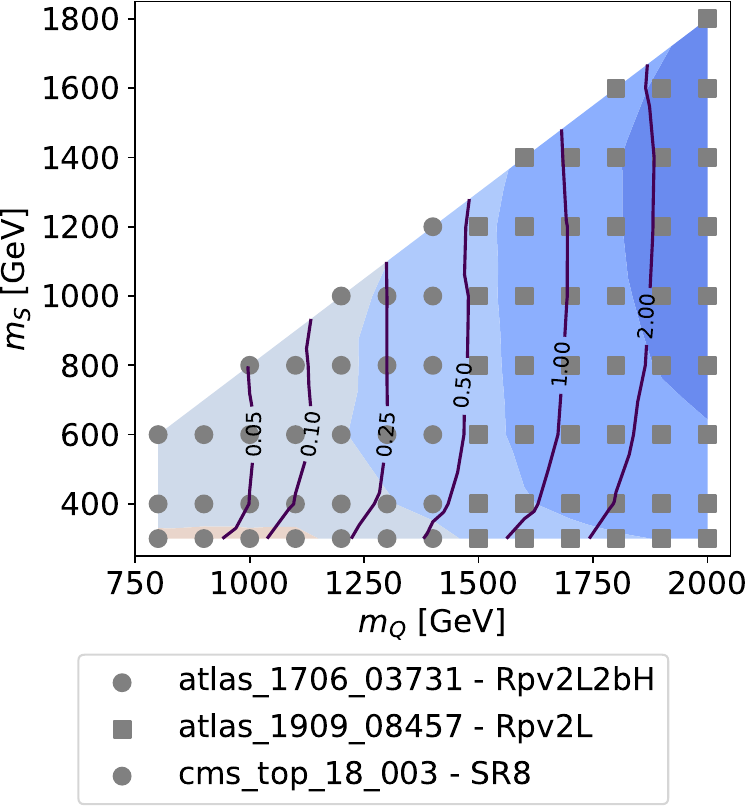}
        \caption{$Y\bar Y \to tS^{--} \bar t S^{++}$}
        \label{sfig:yytsts}
    \end{subfigure}
    \begin{subfigure}[t]{0.3\linewidth}
        \vspace{0pt}
        \includegraphics[width=\linewidth]{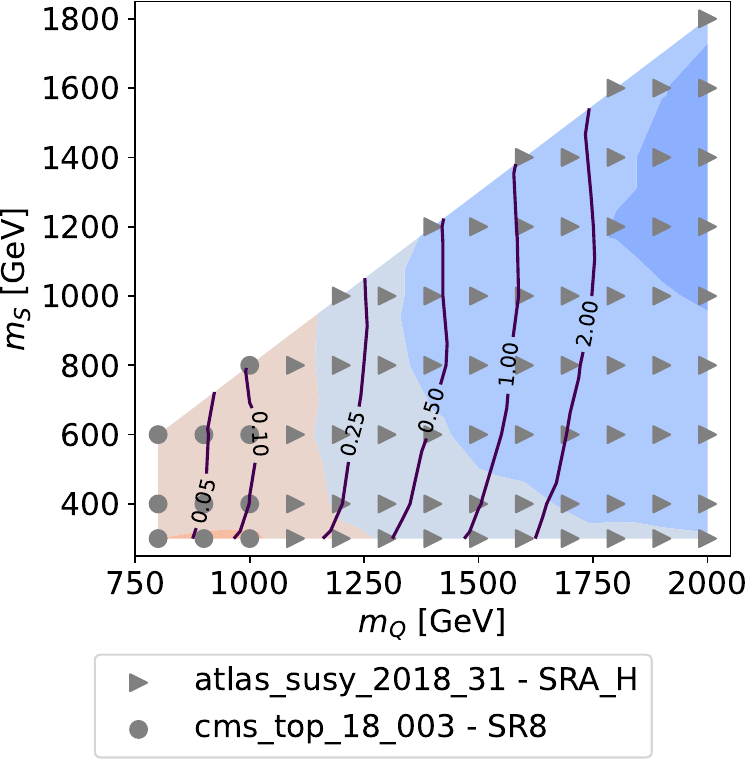}
        \caption{$Y\bar Y \to tS^{--} \bar b S^{+}$}
        \label{sfig:yytsbs}
    \end{subfigure}
    \begin{subfigure}[t]{0.367\linewidth}
        \vspace{0pt}
        \includegraphics[width=\linewidth]{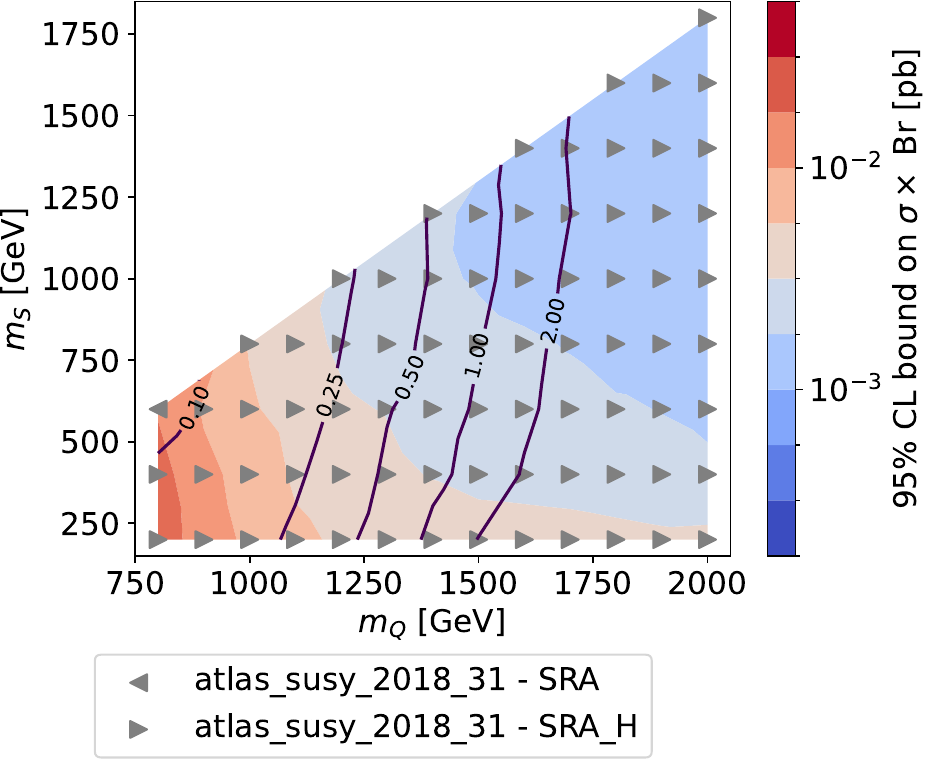}
        \caption{$Y\bar Y\to bS^{-} \bar b S^{+}$}
        \label{sfig:yybsbs}
    \end{subfigure}

    \caption{Bounds on pair production of $Y$}
    \label{fig:yy}
\end{figure}

\begin{figure}[p]
    \centering

    \begin{subfigure}[t]{0.6\linewidth}
        \vspace{0pt}
        \includegraphics[width=\linewidth]{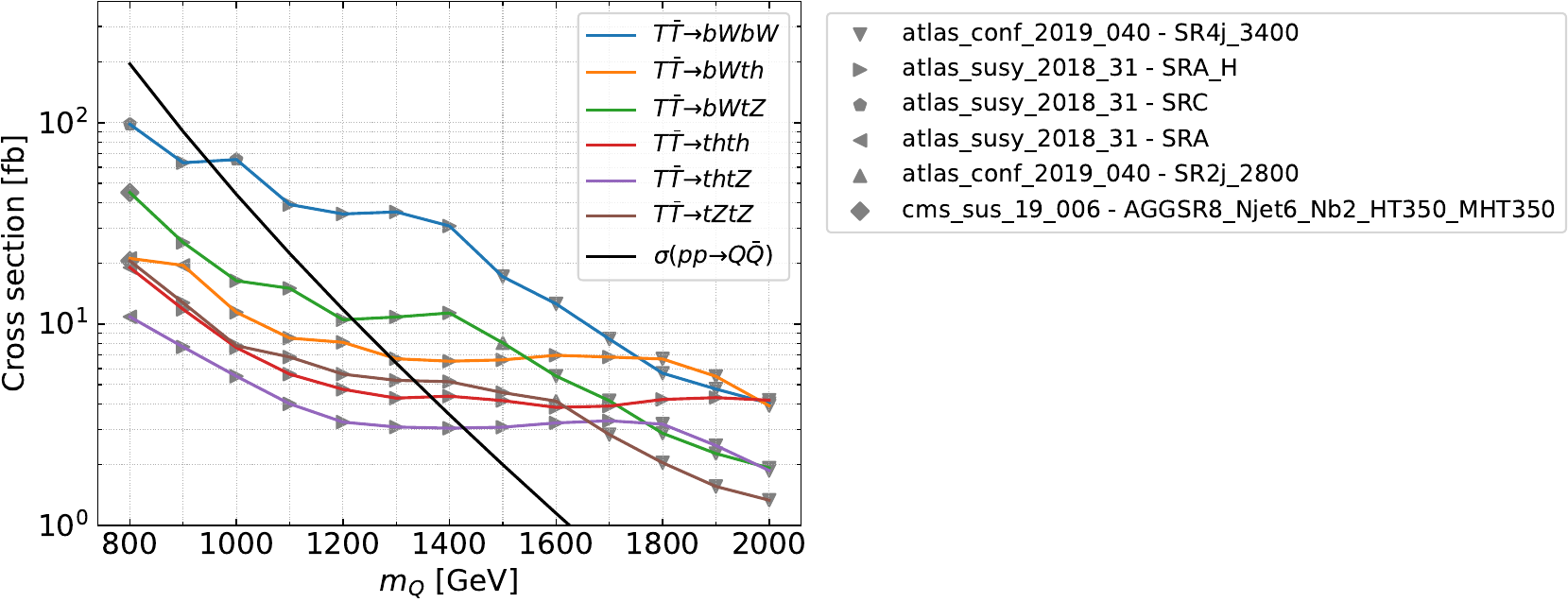}
        \caption{Standard channels}
        \label{sfig:ttstd}
    \end{subfigure}\quad
    \begin{subfigure}[t]{0.36\linewidth}
        \vspace{0pt}
        \includegraphics[width=\linewidth]{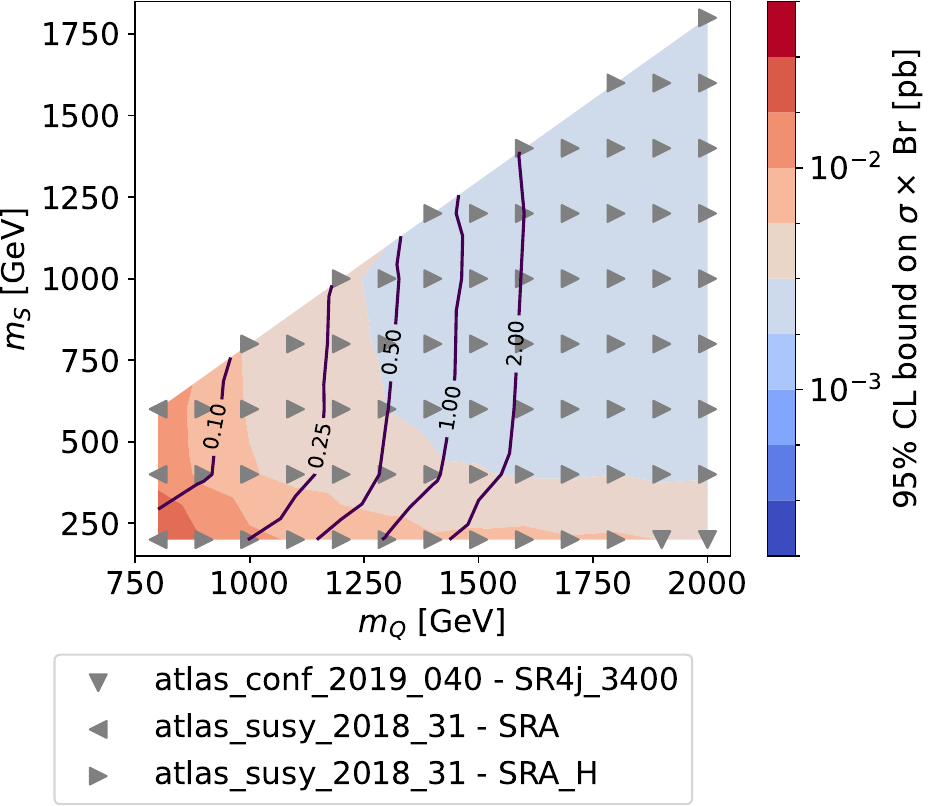}
        \caption{$T\bar{T} \to b S^+ \bar{b} W^-$}
        \label{sfig:ttbsbw}
    \end{subfigure}
    
    \vspace{1ex}
    
    \begin{subfigure}[t]{0.3\linewidth}
        \vspace{0pt}
        \includegraphics[width=\linewidth]{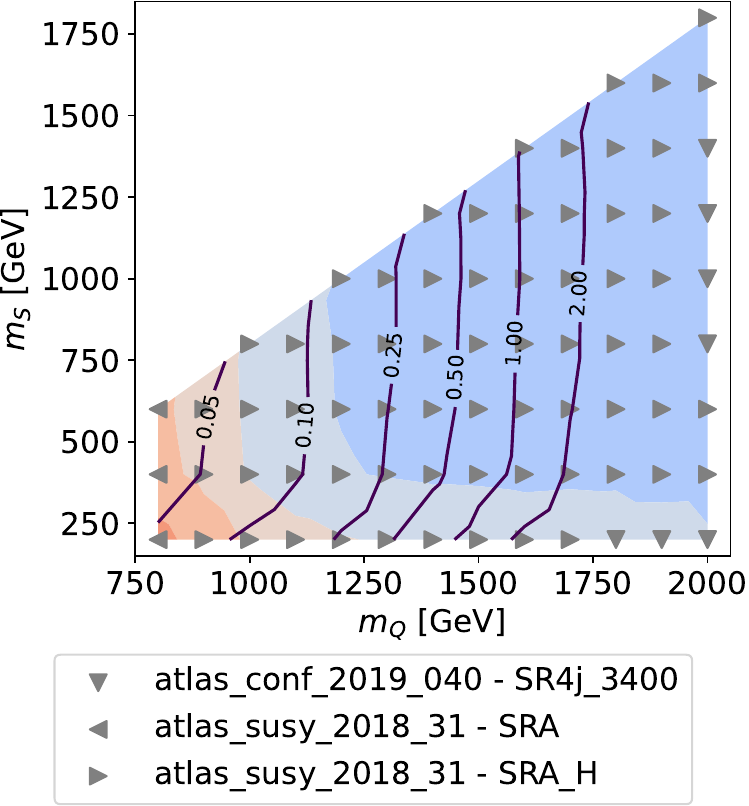}
        \caption{$T\bar{T} \to b S^+ \bar{t} Z$}
        \label{sfig:ttbstz}
    \end{subfigure}
    \begin{subfigure}[t]{0.3\linewidth}
        \vspace{0pt}
        \includegraphics[width=\linewidth]{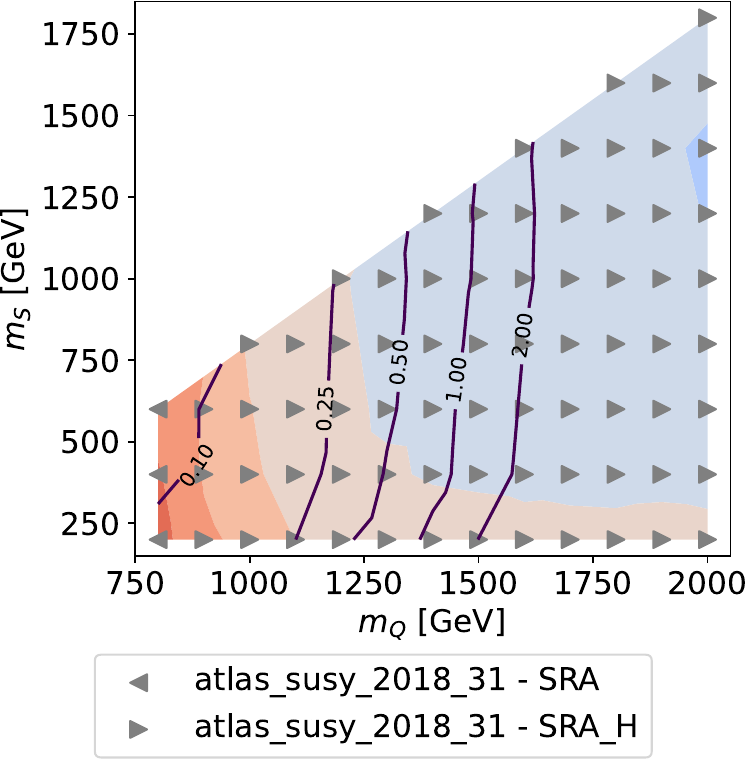}
        \caption{$T\bar{T} \to b S^+ \bar{t} h$}
        \label{sfig:ttbsth}
    \end{subfigure}
    \begin{subfigure}[t]{0.367\linewidth}
        \vspace{0pt}
        \includegraphics[width=\linewidth]{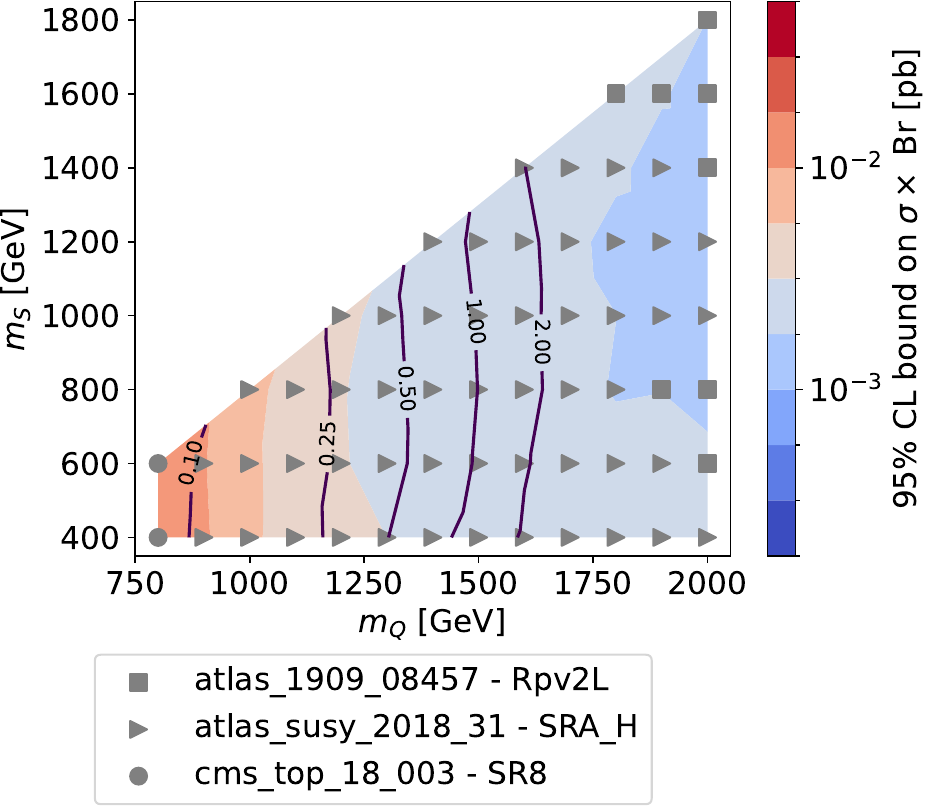}
        \caption{$T\bar{T} \to b W^+ \bar{t} S^0$}
        \label{sfig:ttbwts}
    \end{subfigure}
    \vspace{1ex}
    
    \begin{subfigure}[t]{0.3\linewidth}
        \vspace{0pt}
        \includegraphics[width=\linewidth]{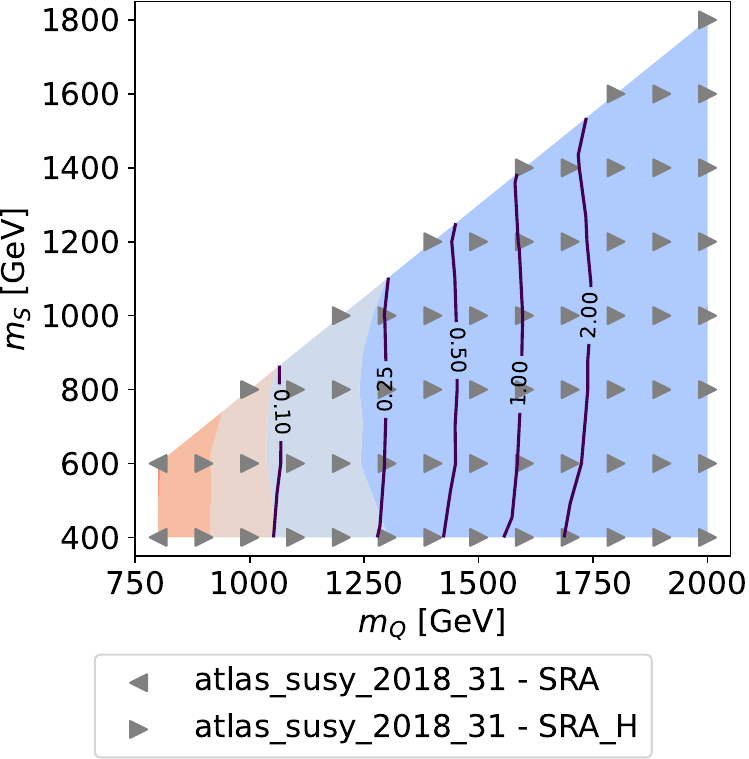}
        \caption{$T\bar{T} \to t S^0 \bar{t} Z$}
        \label{sfig:tttstz}
    \end{subfigure}
    \begin{subfigure}[t]{0.3\linewidth}
        \vspace{0pt}
        \includegraphics[width=\linewidth]{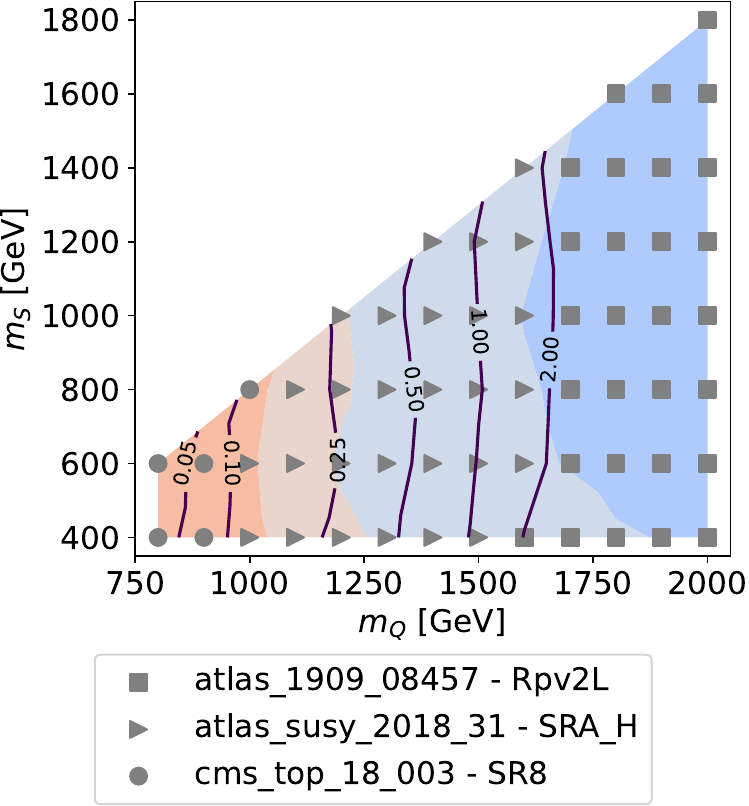}
        \caption{$T\bar{T} \to t h \bar{t} S^0$}
        \label{sfig:ttthts}
    \end{subfigure}
    \begin{subfigure}[t]{0.367\linewidth}
        \vspace{0pt}
        \includegraphics[width=\linewidth]{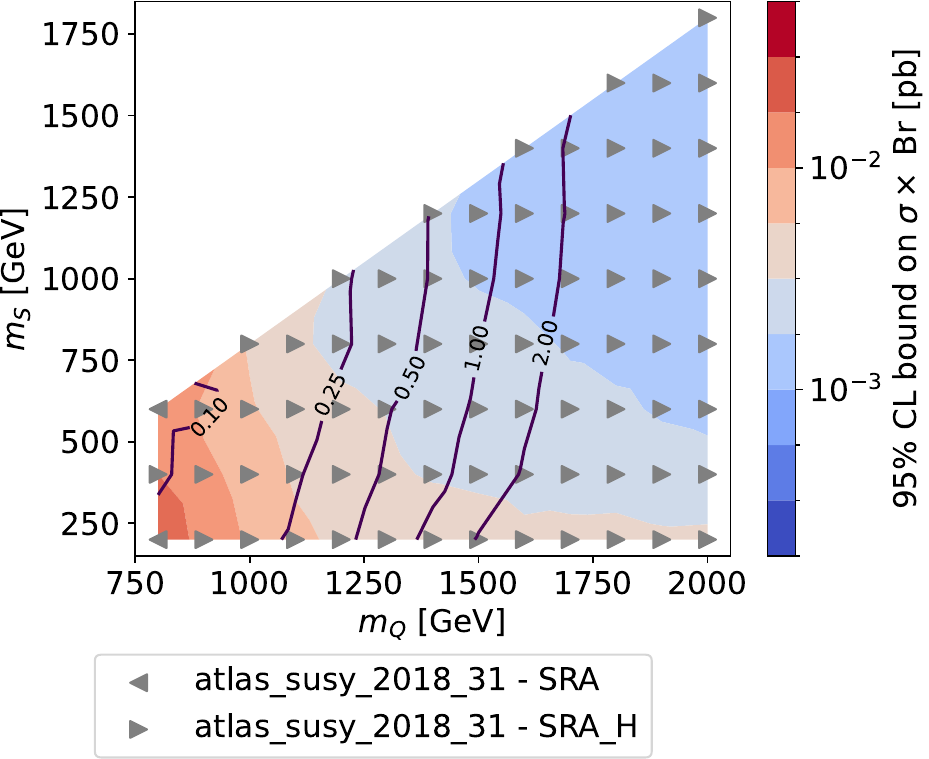}
        \caption{$T\bar{T} \to b S^+ \bar{b} S^-$}
        \label{sfig:ttbsbs}
    \end{subfigure}
    \vspace{1ex}
    
    \begin{subfigure}[t]{0.3\linewidth}
        \vspace{0pt}
        \includegraphics[width=\linewidth]{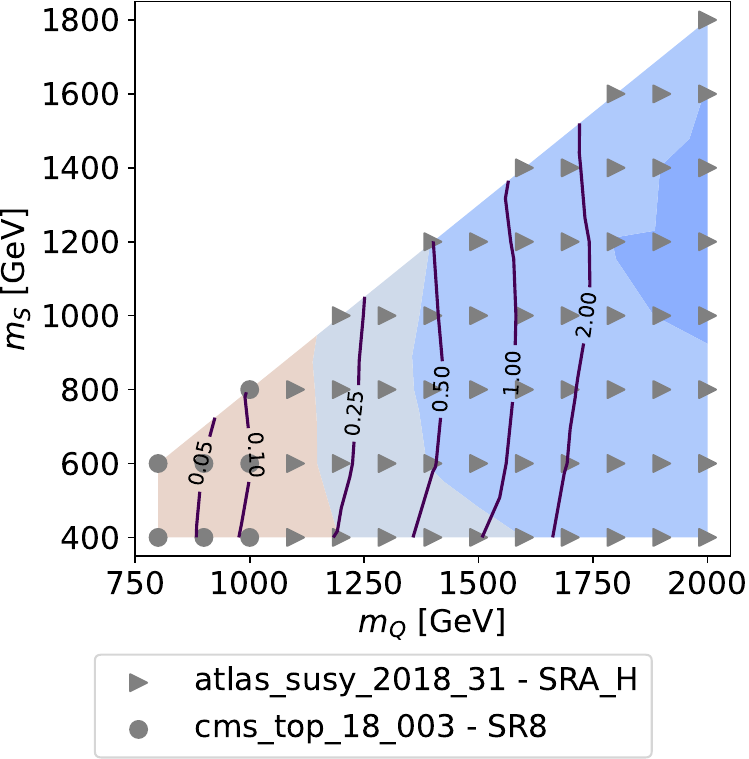}
        \caption{$T\bar{T} \to b S^+ \bar{t} S^0$}
        \label{sfig:ttbsts}
    \end{subfigure}
    \begin{subfigure}[t]{0.367\linewidth}
        \vspace{0pt}
        \includegraphics[width=\linewidth]{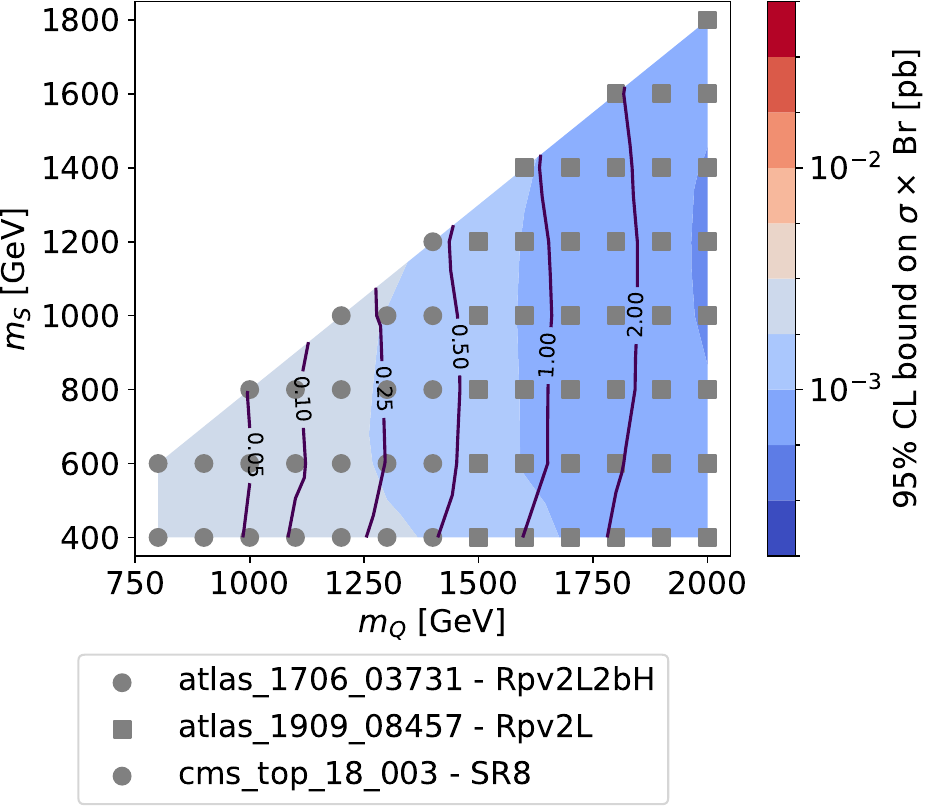}
        \caption{$T\bar{T} \to t S^0 \bar{t} S^0$}
        \label{sfig:tttsts}
    \end{subfigure}

    \caption{Bounds on pair production of $T$}
    \label{fig:tptp}
\end{figure}

\begin{figure}[p]
    \centering

    \begin{subfigure}[t]{0.6\linewidth}
        \vspace{0pt}
        \includegraphics[width=\linewidth]{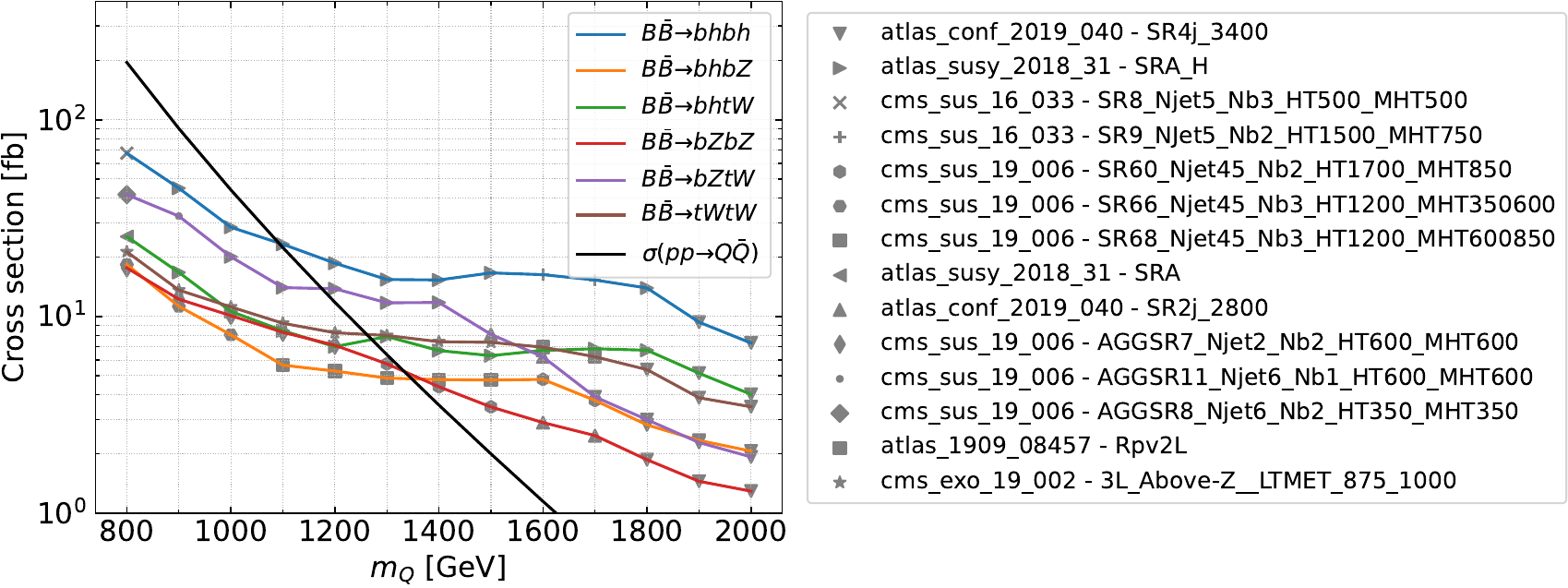}
        \caption{Standard channels}
        \label{sfig:bbstd}
    \end{subfigure}\quad
    \begin{subfigure}[t]{0.36\linewidth}
        \vspace{0pt}
        \includegraphics[width=\linewidth]{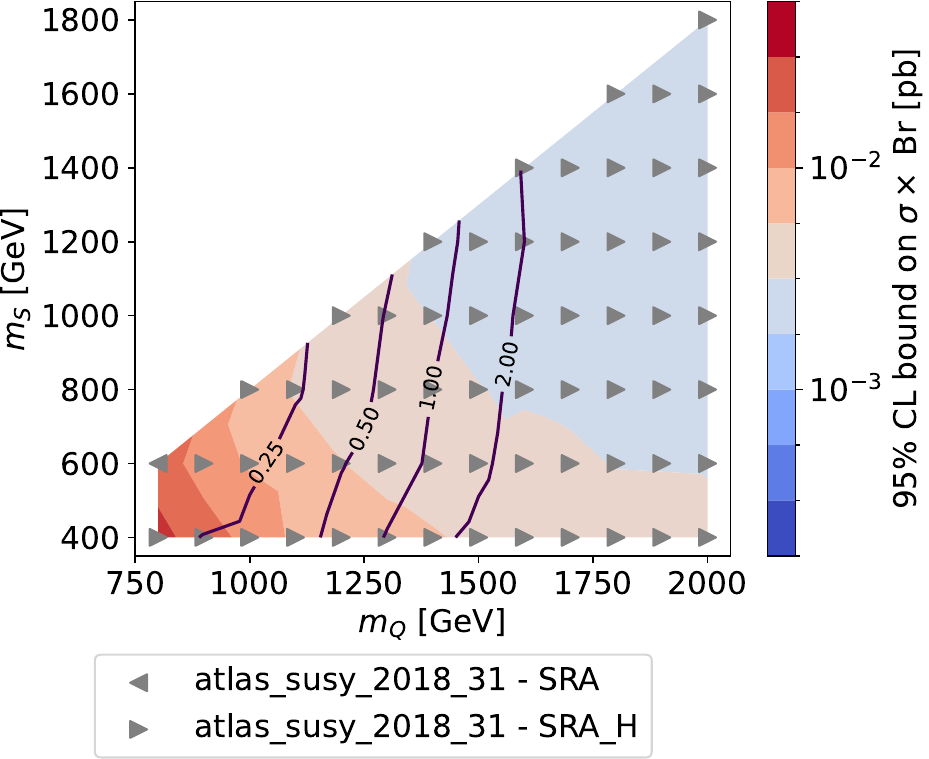}
        \caption{$B\bar{B} \to b h \bar{b} S^0$}
        \label{sfig:bbbhbs}
    \end{subfigure}
    \vspace{1ex}
    
    \begin{subfigure}[t]{0.3\linewidth}
        \vspace{0pt}
        \includegraphics[width=\linewidth]{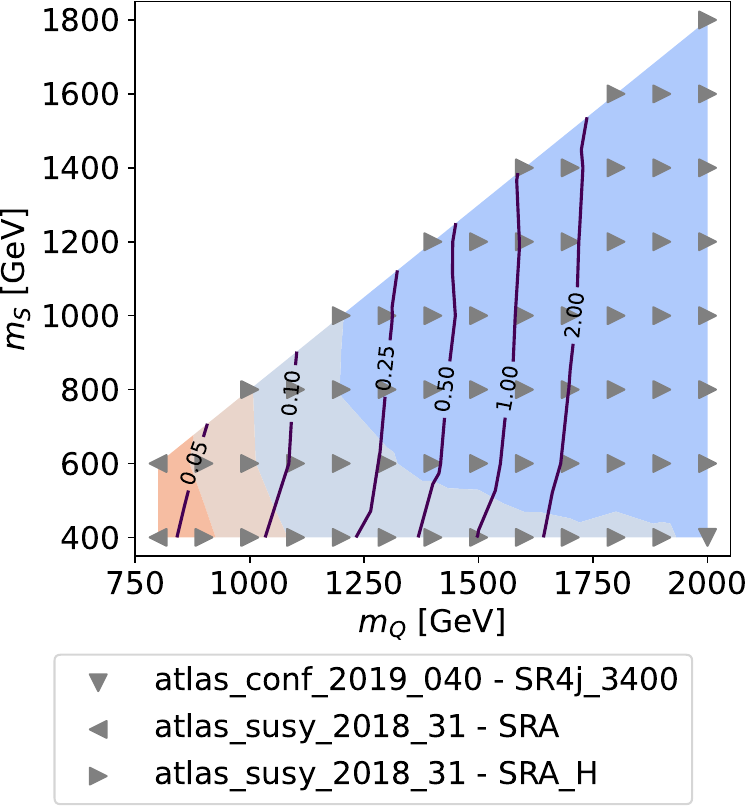}
        \caption{$B\bar{B} \to b S^0 \bar{b} Z$}
        \label{sfig:bbbsbz}
    \end{subfigure}
    \begin{subfigure}[t]{0.3\linewidth}
        \vspace{0pt}
        \includegraphics[width=\linewidth]{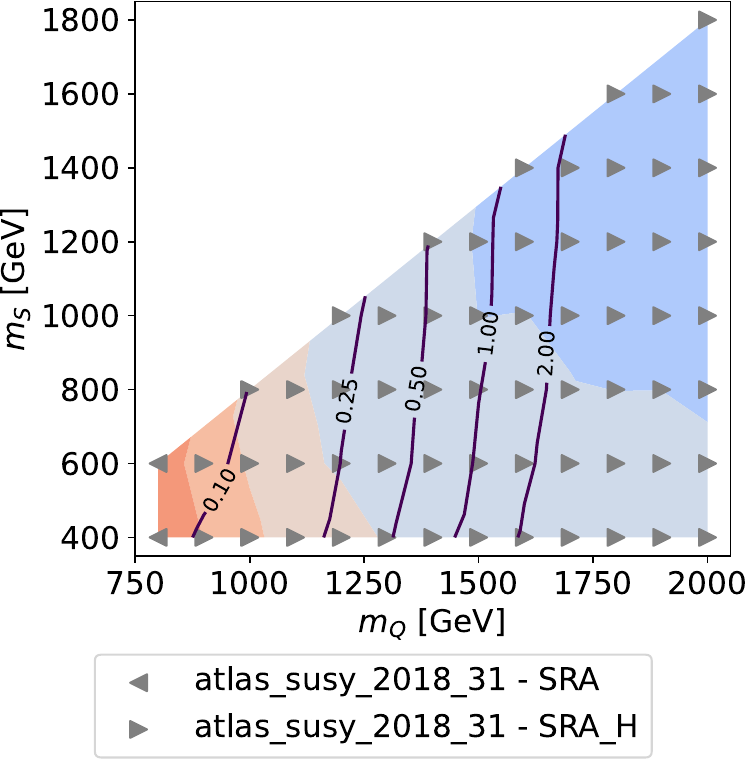}
        \caption{$B\bar{B} \to b S^0 \bar{t} W^+$}
        \label{sfig:bbbstw}
    \end{subfigure}
    \begin{subfigure}[t]{0.367\linewidth}
        \vspace{0pt}
        \includegraphics[width=\linewidth]{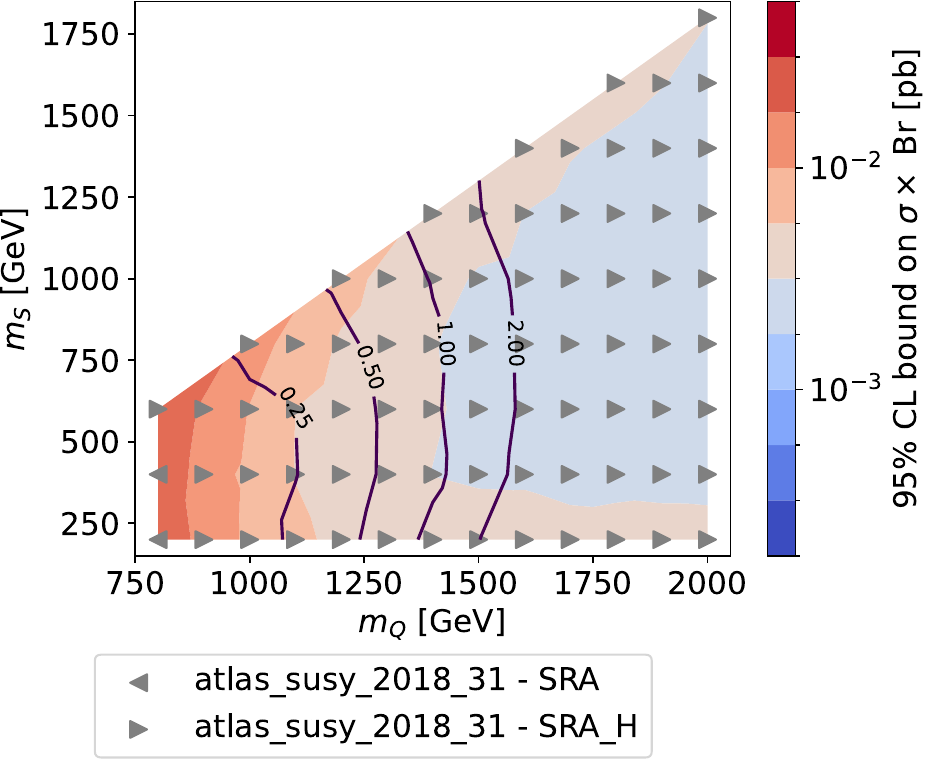}
        \caption{$B\bar{B} \to b h \bar{t} S^+$}
        \label{sfig:bbbhts}
    \end{subfigure}
    \vspace{1ex}
    
    \begin{subfigure}[t]{0.3\linewidth}
        \vspace{0pt}
        \includegraphics[width=\linewidth]{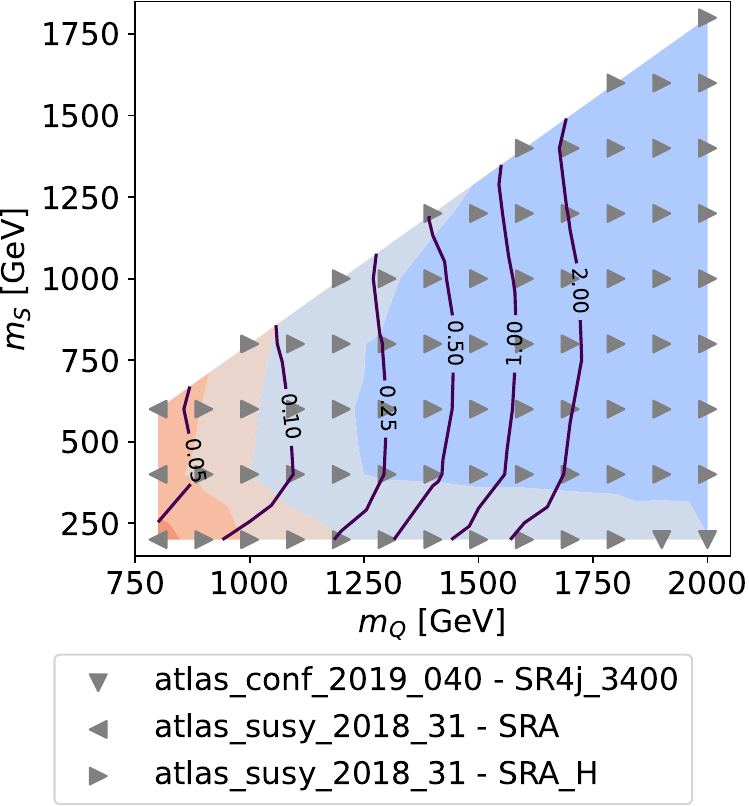}
        \caption{$B\bar{B} \to b Z \bar{t} S^+$}
        \label{sfig:bbbzts}
    \end{subfigure}
    \begin{subfigure}[t]{0.3\linewidth}
        \vspace{0pt}
        \includegraphics[width=\linewidth]{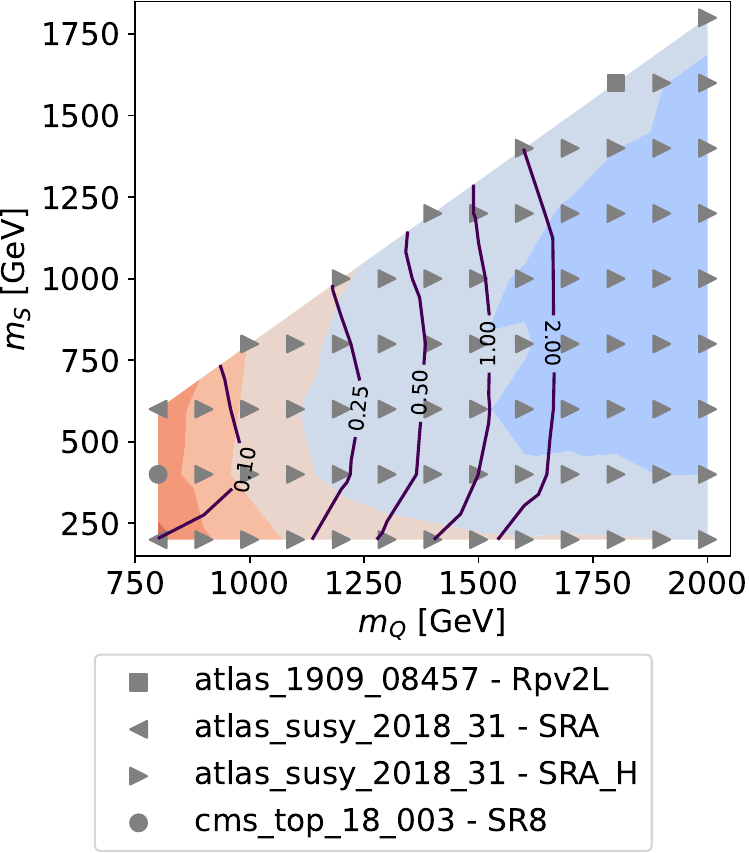}
        \caption{$B\bar{B} \to t S^- \bar{t} W^+$}
        \label{sfig:bbtstw}
    \end{subfigure}
   \begin{subfigure}[t]{0.367\linewidth}
        \vspace{0pt}
        \includegraphics[width=\linewidth]{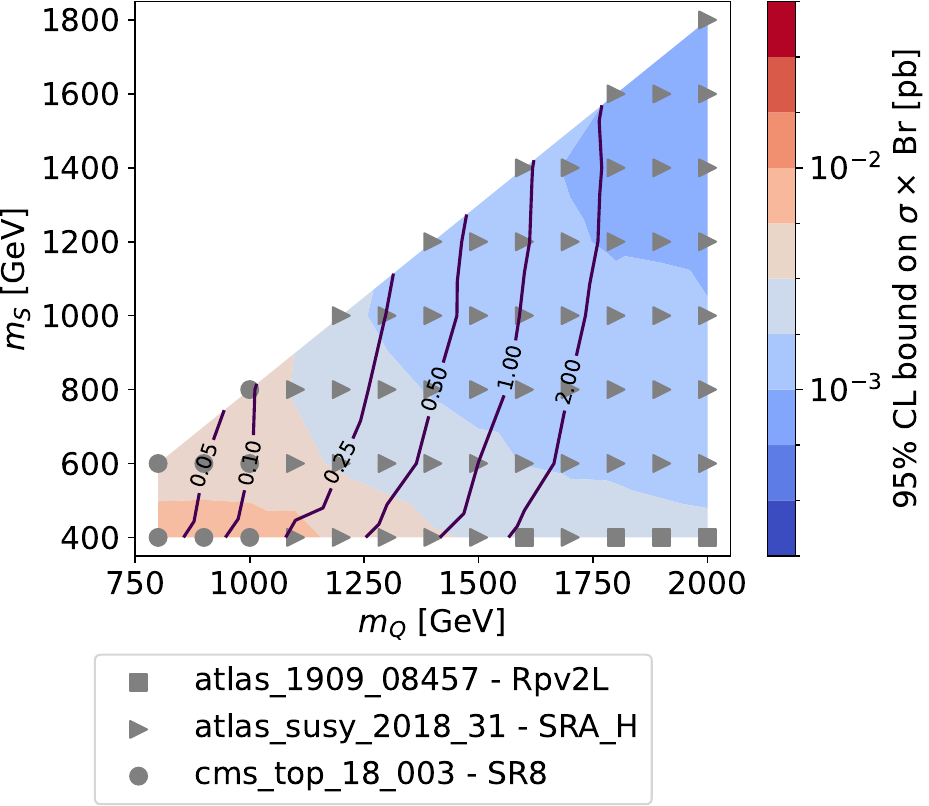}
        \caption{$B\bar{B} \to b S^0 \bar{b} S^0$}
        \label{sfig:bbbsbs}
    \end{subfigure}
    \vspace{1ex}

    \begin{subfigure}[t]{0.3\linewidth}
        \vspace{0pt}
        \includegraphics[width=\linewidth]{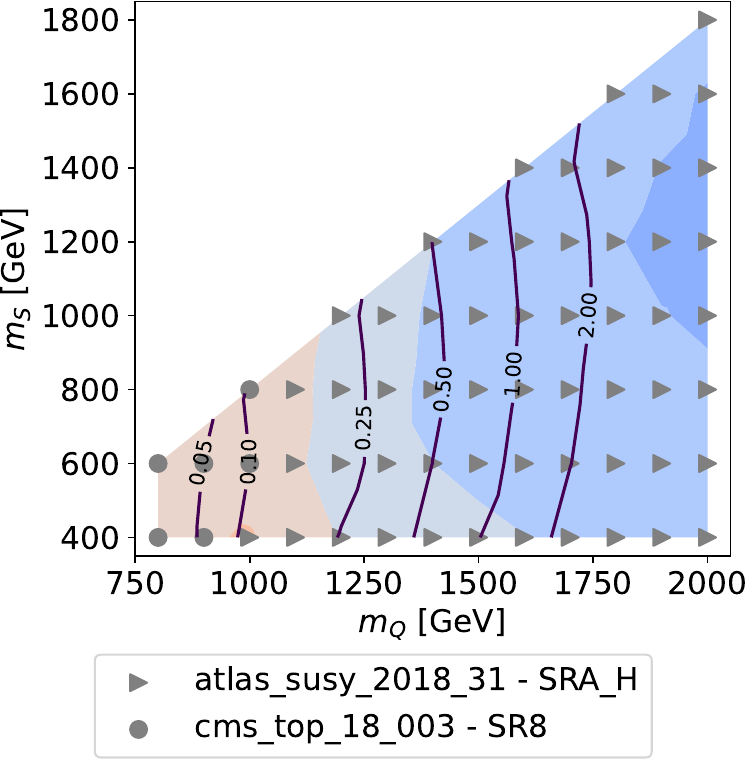}
        \caption{$B\bar{B} \to b S^0 \bar{t} S^+$}
        \label{sfig:bbbsts}
    \end{subfigure}
    \begin{subfigure}[t]{0.367\linewidth}
        \vspace{0pt}
        \includegraphics[width=\linewidth]{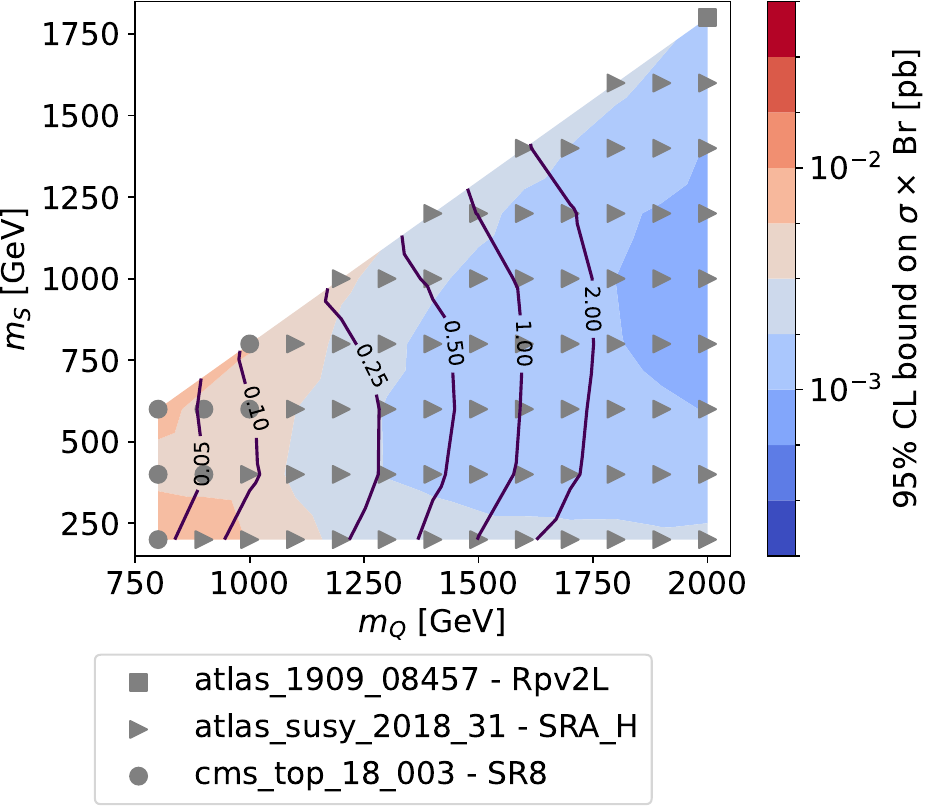}
        \caption{$B\bar{B} \to t S^- \bar{t} S^+$}
        \label{sfig:bbtsts}
    \end{subfigure}

    \caption{Bounds on pair production of $B$}
    \label{fig:bpbp}
\end{figure}

In this appendix we show the upper limits on the cross section times branching ratio for all VLQ processes that were simulated.
Processes involving an exotic scalar are shown in the $m_Q$-$m_S$ plane in the format of \cref{fig:x83x83}. 
The markers show which points were simulated: we chose an even grid scanning $m_Q$ in steps of 100~GeV and $m_S$ in steps of 200~GeV.
At low masses, a different value for $m_S$ was sometimes chosen to probe the region where the scalar decay products have little boost.
The marker symbols indicate the search and signal region that yielded the strongest constraint at that parameter point.
The heat map shows the upper limits on the cross section into the chosen channel.
Finally, the solid lines show mass bounds assuming a specific branching ratio of $Q\bar Q$ into the chosen channel.
This includes branching ratios that are too large to be physical for a single VLQ. 
These can be interpreted as multiple copies of a VLQ with the same mass and branching ratios populating the same signal region.
For processes without exotic scalars, we show the upper limit as a function of $m_Q$, see e.g.\ \cref{sfig:xxtwtw}.
Again, the marker symbols show the dominant search and signal region.
We split the results into several figures by the VLQ that is produced:
\begin{itemize}
    \item Figure \ref{fig:x83x83}: $X_{8/3}$
    \item Figure \ref{fig:xx}: $X$
    \item Figure \ref{fig:yy}: $Y$
    \item Figure \ref{fig:tptp}: $T$
    \item Figure \ref{fig:bpbp}: $B$
\end{itemize}

\begin{figure}
    \centering
    \includegraphics[width=0.47\linewidth]{figures/channels-tptp/tptp-ts10ts10-tttttt-markers.pdf} \quad 
    \includegraphics[width=0.47\linewidth]{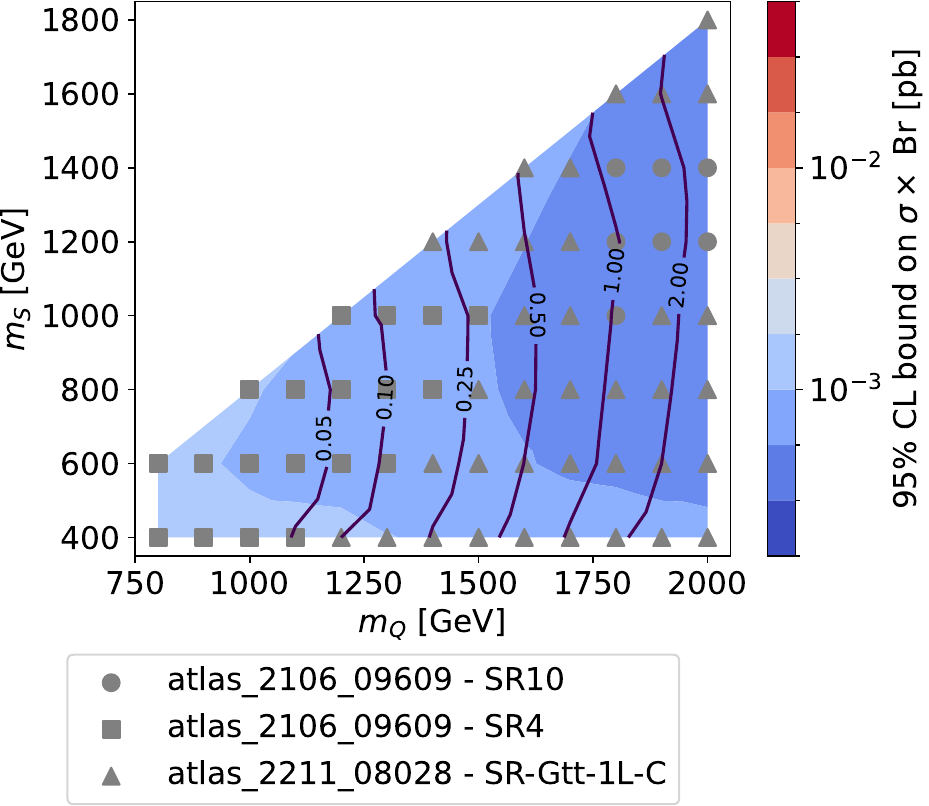}
    \caption{Comparison of the impact of the pool of searches for the process $T\bar T \to tS^0 \bar t S^0 \to 6t$. The left panel uses only searches where the recast has a public validation note, the right panel uses all available searches.}
    \label{fig:comparison_allsearches}
\end{figure}

In the main text and in the figures referenced above we are only considering searches that have a public validation note. 
The constraint data base also contains results from three further searches without such notes that are also sensitive to the VLQ processes.
When using the data base, the user can decide whether to include these or not by setting the flag \texttt{only\_validated} to \texttt{False} or \texttt{True}, respectively.
Including all searches typically leads to stronger bounds, with the precise impact depending on the channel.
As an example, we compare the bounds with only validated and with all searches in \cref{fig:comparison_allsearches} for the channel $T\bar T \to tS^0 \bar t S^0 \to 6t$.

\clearpage

\bibliography{bibliography}

@article{Contino:2003ve,
    author = "Contino, Roberto and Nomura, Yasunori and Pomarol, Alex",
    title = "{Higgs as a Holographic Pseudo Goldstone Boson}",
    eprint = "hep-ph/0306259",
    archivePrefix = "arXiv",
    reportNumber = "FT-UAM-03-11, FERMILAB-PUB-03-195-T, UAB-FT-549",
    doi = "10.1016/j.nuclphysb.2003.08.027",
    journal = "Nucl. Phys. B",
    volume = "671",
    pages = "148--174",
    year = "2003"
}

@article{Contino:2004vy,
    author = "Contino, Roberto and Pomarol, Alex",
    title = "{Holography for fermions}",
    eprint = "hep-th/0406257",
    archivePrefix = "arXiv",
    reportNumber = "UAB-FT-566",
    doi = "10.1088/1126-6708/2004/11/058",
    journal = "JHEP",
    volume = "11",
    pages = "058",
    year = "2004"
}

@article{Aguilar-Saavedra:2017giu,
    author = "Aguilar-Saavedra, J. A. and L{\'o}pez-Fogliani, D. E. and Mu{\~n}oz, C.",
    title = "{Novel signatures for vector-like quarks}",
    eprint = "1705.02526",
    archivePrefix = "arXiv",
    primaryClass = "hep-ph",
    doi = "10.1007/JHEP06(2017)095",
    journal = "JHEP",
    volume = "06",
    pages = "095",
    year = "2017"
}

@article{Barnard:2013zea,
    author = "Barnard, James and Gherghetta, Tony and Ray, Tirtha Sankar",
    title = "{UV descriptions of composite Higgs models without elementary scalars}",
    eprint = "1311.6562",
    archivePrefix = "arXiv",
    primaryClass = "hep-ph",
    doi = "10.1007/JHEP02(2014)002",
    journal = "JHEP",
    volume = "02",
    pages = "002",
    year = "2014"
}

@article{Buchkremer:2013bha,
    author = "Buchkremer, Mathieu and Cacciapaglia, Giacomo and Deandrea, Aldo and Panizzi, Luca",
    title = "{Model Independent Framework for Searches of Top Partners}",
    eprint = "1305.4172",
    archivePrefix = "arXiv",
    primaryClass = "hep-ph",
    reportNumber = "LYCEN-2013-03, SHEP-13-10, CP3-13-22",
    doi = "10.1016/j.nuclphysb.2013.08.010",
    journal = "Nucl. Phys. B",
    volume = "876",
    pages = "376--417",
    year = "2013"
}

@article{ATLAS:2024zlo,
    author = "Aad, Georges and others",
    collaboration = "ATLAS",
    title = "{Search for pair-produced vectorlike quarks coupling to light quarks in the lepton plus jets final state using 13~TeV pp collisions with the ATLAS detector}",
    eprint = "2405.19862",
    archivePrefix = "arXiv",
    primaryClass = "hep-ex",
    reportNumber = "CERN-EP-2024-130",
    doi = "10.1103/PhysRevD.110.052009",
    journal = "Phys. Rev. D",
    volume = "110",
    number = "5",
    pages = "052009",
    year = "2024"
}

@article{ATLAS:2026ojf,
    author = "Aad, Georges and others",
    collaboration = "ATLAS",
    title = "{Search for pair-produced vector-like $T$-quarks decaying into $Ht$ final states in the lepton-plus-jets channel in $pp$ collisions at $\sqrt{s}$=13 TeV with the ATLAS detector}",
    eprint = "2605.28538",
    archivePrefix = "arXiv",
    primaryClass = "hep-ex",
    reportNumber = "CERN-EP-2026-131",
    month = "5",
    year = "2026"
}

@article{ATLAS:2022hnn,
    author = "Aad, Georges and others",
    collaboration = "ATLAS",
    title = "{Search for pair-production of vector-like quarks in pp collision events at $\sqrt s=13$ TeV with at least one leptonically decaying Z boson and a third-generation quark with the ATLAS detector}",
    eprint = "2210.15413",
    archivePrefix = "arXiv",
    primaryClass = "hep-ex",
    reportNumber = "CERN-EP-2021-207",
    doi = "10.1016/j.physletb.2023.138019",
    journal = "Phys. Lett. B",
    volume = "843",
    pages = "138019",
    year = "2023"
}

@article{CMS:2026ggz,
    collaboration = "CMS",
    title = "{Search for heavy top quark partners with charge 5/3 in the single-lepton final state in proton-proton collisions at $\sqrt{s}=13$~TeV}",
    reportNumber = "CMS-PAS-B2G-24-005",
    year = "2026",
    note = {\href{https://cds.cern.ch/record/2961473}{CMS-PAS-B2G-24-005}}
}

@article{CMS:2026fky,
    collaboration = "CMS",
    title = "{Search for pair-produced vector-like top quarks decaying to Lorentz-boosted top quarks and scalars in proton-proton collisions at $\sqrt{s}=$ 13 TeV}",
    reportNumber = "CMS-PAS-B2G-24-016",
    year = "2026",
    note = {\href{https://cds.cern.ch/record/2961613}{CMS-PAS-B2G-24-016}}
}

@article{CMS:2024xbc,
    author = "Hayrapetyan, Aram and others",
    collaboration = "CMS",
    title = "{Search for bottom-type vectorlike quark pair production in dileptonic and fully hadronic final states in proton-proton collisions at $\sqrt s=13$ TeV}",
    eprint = "2402.13808",
    archivePrefix = "arXiv",
    primaryClass = "hep-ex",
    reportNumber = "CMS-B2G-20-014, CERN-EP-2024-016",
    doi = "10.1103/PhysRevD.110.052004",
    journal = "Phys. Rev. D",
    volume = "110",
    number = "5",
    pages = "052004",
    year = "2024"
}

@article{CMS:2020ttz,
    author = "Sirunyan, Albert M and others",
    collaboration = "CMS",
    title = "{A search for bottom-type, vector-like quark pair production in a fully hadronic final state in proton-proton collisions at $\sqrt{s} =$ 13 TeV}",
    eprint = "2008.09835",
    archivePrefix = "arXiv",
    primaryClass = "hep-ex",
    reportNumber = "CMS-B2G-19-005, CERN-EP-2020-154",
    doi = "10.1103/PhysRevD.102.112004",
    journal = "Phys. Rev. D",
    volume = "102",
    pages = "112004",
    year = "2020"
}

@article{Cacciapaglia:2021uqh,
    author = "Cacciapaglia, Giacomo and Flacke, Thomas and Kunkel, Manuel and Porod, Werner",
    title = "{Phenomenology of unusual top partners in composite Higgs models}",
    eprint = "2112.00019",
    archivePrefix = "arXiv",
    primaryClass = "hep-ph",
    reportNumber = "KIAS-A21002",
    doi = "10.1007/JHEP02(2022)208",
    journal = "JHEP",
    volume = "02",
    pages = "208",
    year = "2022"
}

@article{Banerjee:2022xmu,
    author = "Banerjee, Avik and others",
    title = "{Phenomenological aspects of composite Higgs scenarios: exotic scalars and vector-like quarks}",
    eprint = "2203.07270",
    archivePrefix = "arXiv",
    primaryClass = "hep-ph",
    reportNumber = "KIAS-A22001",
    month = "3",
    year = "2022",
    note = "{contribution to Snowmass 2021}"
}

@article{Cacciapaglia:2022bax,
    author = "Cacciapaglia, G. and Flacke, T. and Kunkel, M. and Porod, W. and Schwarze, L.",
    title = "{Exploring extended Higgs sectors via pair production at the LHC}",
    eprint = "2210.01826",
    archivePrefix = "arXiv",
    primaryClass = "hep-ph",
    reportNumber = "KIAS-A22008",
    doi = "10.1007/JHEP12(2022)087",
    journal = "JHEP",
    volume = "12",
    pages = "087",
    year = "2022"
}

@article{Cacciapaglia:2024wdn,
    author = "Cacciapaglia, Giacomo and Deandrea, Aldo and Kunkel, Manuel and Porod, Werner",
    title = "{Coloured spin-1 states in composite Higgs models}",
    eprint = "2404.02198",
    archivePrefix = "arXiv",
    primaryClass = "hep-ph",
    doi = "10.1007/JHEP06(2024)092",
    journal = "JHEP",
    volume = "06",
    pages = "092",
    year = "2024"
}

@article{Caliri:2024jdk,
    author = "Caliri, Rosy and Hadlik, Jan and Kunkel, Manuel and Porod, Werner and Verollet, Ch. and Verollet, Christian",
    title = "{Electroweak spin-1 resonances in Composite Higgs models}",
    eprint = "2412.08720",
    archivePrefix = "arXiv",
    primaryClass = "hep-ph",
    doi = "10.1007/JHEP04(2025)160",
    journal = "JHEP",
    volume = "04",
    pages = "160",
    year = "2025"
}

@article{Kaplan:1983fs,
    author = "Kaplan, David B. and Georgi, Howard",
    title = "{SU(2) x U(1) Breaking by Vacuum Misalignment}",
    reportNumber = "HUTP-83/A069",
    doi = "10.1016/0370-2693(84)91177-8",
    journal = "Phys. Lett. B",
    volume = "136",
    pages = "183--186",
    year = "1984"
}

@article{Kaplan:1983sm,
    author = "Kaplan, David B. and Georgi, Howard and Dimopoulos, Savas",
    title = "{Composite Higgs Scalars}",
    reportNumber = "HUTP-83/A079",
    doi = "10.1016/0370-2693(84)91178-X",
    journal = "Phys. Lett. B",
    volume = "136",
    pages = "187--190",
    year = "1984"
}

@article{Dugan:1984hq,
    author = "Dugan, Michael J. and Georgi, Howard and Kaplan, David B.",
    title = "{Anatomy of a Composite Higgs Model}",
    reportNumber = "HUTP-84/A073",
    doi = "10.1016/0550-3213(85)90221-4",
    journal = "Nucl. Phys. B",
    volume = "254",
    pages = "299--326",
    year = "1985"
}

@article{Kaplan:1991dc,
    author = "Kaplan, David B.",
    title = "{Flavor at SSC energies: A New mechanism for dynamically generated fermion masses}",
    reportNumber = "UCSD-PTH-91-04",
    doi = "10.1016/S0550-3213(05)80021-5",
    journal = "Nucl. Phys. B",
    volume = "365",
    pages = "259--278",
    year = "1991"
}

@article{Wess:1971yu,
    author = "Wess, J. and Zumino, B.",
    title = "{Consequences of anomalous Ward identities}",
    doi = "10.1016/0370-2693(71)90582-X",
    journal = "Phys. Lett. B",
    volume = "37",
    pages = "95--97",
    year = "1971"
}

@article{Witten:1983tw,
    author = "Witten, Edward",
    title = "{Global Aspects of Current Algebra}",
    reportNumber = "PRINT-83-0262 (PRINCETON)",
    doi = "10.1016/0550-3213(83)90063-9",
    journal = "Nucl. Phys. B",
    volume = "223",
    pages = "422--432",
    year = "1983"
}

@article{Witten:1983tx,
    author = "Witten, Edward",
    title = "{Current Algebra, Baryons, and Quark Confinement}",
    reportNumber = "Print-83-0261 (PRINCETON)",
    doi = "10.1016/0550-3213(83)90064-0",
    journal = "Nucl. Phys. B",
    volume = "223",
    pages = "433--444",
    year = "1983"
}

@article{BuarqueFranzosi:2016ooy,
    author = "Buarque Franzosi, Diogo and Cacciapaglia, Giacomo and Cai, Haiying and Deandrea, Aldo and Frandsen, Mads",
    title = "{Vector and Axial-vector resonances in composite models of the Higgs boson}",
    eprint = "1605.01363",
    archivePrefix = "arXiv",
    primaryClass = "hep-ph",
    reportNumber = "CP3-ORIGINS-2016-022, LYCEN-2016-05",
    doi = "10.1007/JHEP11(2016)076",
    journal = "JHEP",
    volume = "11",
    pages = "076",
    year = "2016"
}

@article{Ferretti:2013kya,
    author = "Ferretti, Gabriele and Karateev, Denis",
    title = "{Fermionic UV completions of Composite Higgs models}",
    eprint = "1312.5330",
    archivePrefix = "arXiv",
    primaryClass = "hep-ph",
    doi = "10.1007/JHEP03(2014)077",
    journal = "JHEP",
    volume = "03",
    pages = "077",
    year = "2014"
}

@article{Ferretti:2016upr,
    author = "Ferretti, Gabriele",
    title = "{Gauge theories of Partial Compositeness: Scenarios for Run-II of the LHC}",
    eprint = "1604.06467",
    archivePrefix = "arXiv",
    primaryClass = "hep-ph",
    doi = "10.1007/JHEP06(2016)107",
    journal = "JHEP",
    volume = "06",
    pages = "107",
    year = "2016"
}

@article{Belyaev:2016ftv,
    author = "Belyaev, Alexander and Cacciapaglia, Giacomo and Cai, Haiying and Ferretti, Gabriele and Flacke, Thomas and Parolini, Alberto and Serodio, Hugo",
    title = "{Di-boson signatures as Standard Candles for Partial Compositeness}",
    eprint = "1610.06591",
    archivePrefix = "arXiv",
    primaryClass = "hep-ph",
    reportNumber = "CTPU-16-24, LU-TP-16-56, LYCEN-2016-08",
    doi = "10.1007/JHEP01(2017)094",
    journal = "JHEP",
    volume = "01",
    pages = "094",
    year = "2017",
    note = "[Erratum: JHEP 12, 088 (2017)]"
}

@article{Banerjee:2022izw,
    author = "Banerjee, Avik and Franzosi, Diogo Buarque and Ferretti, Gabriele",
    title = "{Modelling vector-like quarks in partial compositeness framework}",
    eprint = "2202.00037",
    archivePrefix = "arXiv",
    primaryClass = "hep-ph",
    doi = "10.1007/JHEP03(2022)200",
    journal = "JHEP",
    volume = "03",
    pages = "200",
    year = "2022"
}

@article{Cacciapaglia:2015eqa,
    author = "Cacciapaglia, Giacomo and Cai, Haiying and Deandrea, Aldo and Flacke, Thomas and Lee, Seung J. and Parolini, Alberto",
    title = "{Composite scalars at the LHC: the Higgs, the Sextet and the Octet}",
    eprint = "1507.02283",
    archivePrefix = "arXiv",
    primaryClass = "hep-ph",
    reportNumber = "LYCEN-2015-05",
    doi = "10.1007/JHEP11(2015)201",
    journal = "JHEP",
    volume = "11",
    pages = "201",
    year = "2015"
}

@article{Banerjee:2024zvg,
    author = "Banerjee, Avik and Bergeaas Kuutmann, Elin and Ellajosyula, Venugopal and Enberg, Rikard and Ferretti, Gabriele and Panizzi, Luca",
    title = "{Vector-like quarks: Status and new directions at the LHC}",
    eprint = "2406.09193",
    archivePrefix = "arXiv",
    primaryClass = "hep-ph",
    doi = "10.21468/SciPostPhysCore.7.4.079",
    journal = "SciPost Phys. Core",
    volume = "7",
    pages = "079",
    year = "2024"
}

@article{Cacciapaglia:2019zmj,
    author = "Cacciapaglia, Giacomo and Flacke, Thomas and Park, Myeonghun and Zhang, Mengchao",
    title = "{Exotic decays of top partners: mind the search gap}",
    eprint = "1908.07524",
    archivePrefix = "arXiv",
    primaryClass = "hep-ph",
    doi = "10.1016/j.physletb.2019.135015",
    journal = "Phys. Lett. B",
    volume = "798",
    pages = "135015",
    year = "2019"
}

@article{Xie:2019gya,
    author = "Xie, Ke-Pan and Cacciapaglia, Giacomo and Flacke, Thomas",
    title = "{Exotic decays of top partners with charge 5/3: bounds and opportunities}",
    eprint = "1907.05894",
    archivePrefix = "arXiv",
    primaryClass = "hep-ph",
    reportNumber = "CTPU-PTC-19-20",
    doi = "10.1007/JHEP10(2019)134",
    journal = "JHEP",
    volume = "10",
    pages = "134",
    year = "2019"
}

@article{Bizot:2018tds,
    author = "Bizot, Nicolas and Cacciapaglia, Giacomo and Flacke, Thomas",
    title = "{Common exotic decays of top partners}",
    eprint = "1803.00021",
    archivePrefix = "arXiv",
    primaryClass = "hep-ph",
    reportNumber = "CTPU-PTC-18-05, LYCEN 2018-03, LYCEN-2018-03",
    doi = "10.1007/JHEP06(2018)065",
    journal = "JHEP",
    volume = "06",
    pages = "065",
    year = "2018"
}

@article{Benbrik:2019zdp,
    author = "Benbrik, Rachid and others",
    title = "{Signatures of vector-like top partners decaying into new neutral scalar or pseudoscalar bosons}",
    eprint = "1907.05929",
    archivePrefix = "arXiv",
    primaryClass = "hep-ph",
    doi = "10.1007/JHEP05(2020)028",
    journal = "JHEP",
    volume = "05",
    pages = "028",
    year = "2020"
}

@article{Banerjee:2023upj,
    author = "Banerjee, Avik and Ellajosyula, Venugopal and Panizzi, Luca",
    title = "{Heavy vector-like quarks decaying to exotic scalars: a case study with triplets}",
    eprint = "2311.17877",
    archivePrefix = "arXiv",
    primaryClass = "hep-ph",
    doi = "10.1007/JHEP01(2024)187",
    journal = "JHEP",
    volume = "01",
    pages = "187",
    year = "2024"
}

@article{Aguilar-Saavedra:2009xmz,
    author = "Aguilar-Saavedra, J. A.",
    title = "{Identifying top partners at LHC}",
    eprint = "0907.3155",
    archivePrefix = "arXiv",
    primaryClass = "hep-ph",
    doi = "10.1088/1126-6708/2009/11/030",
    journal = "JHEP",
    volume = "11",
    pages = "030",
    year = "2009"
}

@article{DeSimone:2012fs,
    author = "De Simone, Andrea and Matsedonskyi, Oleksii and Rattazzi, Riccardo and Wulzer, Andrea",
    title = "{A First Top Partner Hunter's Guide}",
    eprint = "1211.5663",
    archivePrefix = "arXiv",
    primaryClass = "hep-ph",
    reportNumber = "CERN-PH-TH-2012-323, SISSA-31-2012-EP",
    doi = "10.1007/JHEP04(2013)004",
    journal = "JHEP",
    volume = "04",
    pages = "004",
    year = "2013"
}

@article{ATLAS:2019gdh,
    author = "Aad, Georges and others",
    collaboration = "ATLAS",
    title = "{Search for bottom-squark pair production with the ATLAS detector in final states containing Higgs bosons, $b$-jets and missing transverse momentum}",
    eprint = "1908.03122",
    archivePrefix = "arXiv",
    primaryClass = "hep-ex",
    reportNumber = "CERN-EP-2019-142",
    doi = "10.1007/JHEP12(2019)060",
    journal = "JHEP",
    volume = "12",
    pages = "060",
    year = "2019"
}

@article{ATLAS:2022ihe,
    author = "Aad, Georges and others",
    collaboration = "ATLAS",
    title = "{Search for supersymmetry in final states with missing transverse momentum and three or more b-jets in 139 fb$^{-1}$ of proton\textendash{}proton collisions at $\sqrt{s} = 13$~TeV with the ATLAS detector}",
    eprint = "2211.08028",
    archivePrefix = "arXiv",
    primaryClass = "hep-ex",
    reportNumber = "CERN-EP-2022-213",
    doi = "10.1140/epjc/s10052-023-11543-6",
    journal = "Eur. Phys. J. C",
    volume = "83",
    number = "7",
    pages = "561",
    year = "2023"
}

@article{ATLAS:2021fbt,
    author = "Aad, Georges and others",
    collaboration = "ATLAS",
    title = "{Search for R-parity-violating supersymmetry in a final state containing leptons and many jets with the ATLAS experiment using $\sqrt{s} = 13$~TeV proton\textendash{}proton collision data}",
    eprint = "2106.09609",
    archivePrefix = "arXiv",
    primaryClass = "hep-ex",
    reportNumber = "CERN-EP-2021-066",
    doi = "10.1140/epjc/s10052-021-09761-x",
    journal = "Eur. Phys. J. C",
    volume = "81",
    number = "11",
    pages = "1023",
    year = "2021"
}

@article{CMS:2019xud,
    collaboration = "CMS",
    title = "{Search for new physics in multilepton final states in pp collisions at $\sqrt{s}=13~\mathrm{TeV}$}",
    reportNumber = "CMS-PAS-EXO-19-002",
    year = "2019",
    note = {\href{https://cds.cern.ch/record/2668721}{CMS PAS EXO-19-002}}
}

@article{ATLAS:2021twp,
    author = "Aad, Georges and others",
    collaboration = "ATLAS",
    title = "{Search for squarks and gluinos in final states with one isolated lepton, jets, and missing transverse momentum at $\sqrt{s}=13$~TeV with the ATLAS detector}",
    eprint = "2101.01629",
    archivePrefix = "arXiv",
    primaryClass = "hep-ex",
    reportNumber = "CERN-EP-2020-228",
    doi = "10.1140/epjc/s10052-021-09748-8",
    journal = "Eur. Phys. J. C",
    volume = "81",
    number = "7",
    pages = "600",
    year = "2021",
    note = "[Erratum: Eur.Phys.J.C 81, 956 (2021)]"
}

@article{CMS:2019xjf,
    collaboration = "CMS",
    title = "{Search for supersymmetry in proton-proton collisions at 13 TeV in final states with jets and missing transverse momentum}",
    reportNumber = "CMS-PAS-SUS-19-006",
    year = "2019",
    note = {\href{https://cds.cern.ch/record/2682103}{CMS PAS SUS-19-006}}
}

@article{CMS:2017abv,
    author = "Sirunyan, Albert M and others",
    collaboration = "CMS",
    title = "{Search for supersymmetry in multijet events with missing transverse momentum in proton-proton collisions at 13 TeV}",
    eprint = "1704.07781",
    archivePrefix = "arXiv",
    primaryClass = "hep-ex",
    reportNumber = "CMS-SUS-16-033, CERN-EP-2017-072",
    doi = "10.1103/PhysRevD.96.032003",
    journal = "Phys. Rev. D",
    volume = "96",
    number = "3",
    pages = "032003",
    year = "2017"
}

@article{ATLAS:2019fag,
    author = "Aad, Georges and others",
    collaboration = "ATLAS",
    title = "{Search for squarks and gluinos in final states with same-sign leptons and jets using 139 fb$^{-1}$ of data collected with the ATLAS detector}",
    eprint = "1909.08457",
    archivePrefix = "arXiv",
    primaryClass = "hep-ex",
    reportNumber = "CERN-EP-2019-161",
    doi = "10.1007/JHEP06(2020)046",
    journal = "JHEP",
    volume = "06",
    pages = "046",
    year = "2020"
}

@article{ATLAS:2024gyc,
    author = "Aad, Georges and others",
    collaboration = "ATLAS",
    title = "{Search for pair-production of vector-like quarks in lepton+jets final states containing at least one b-tagged jet using the Run 2 data from the ATLAS experiment}",
    eprint = "2401.17165",
    archivePrefix = "arXiv",
    primaryClass = "hep-ex",
    reportNumber = "CERN-EP-2023-254",
    doi = "10.1016/j.physletb.2024.138743",
    journal = "Phys. Lett. B",
    volume = "854",
    pages = "138743",
    year = "2024"
}

@article{CMS:2022fck,
    author = "Tumasyan, Armen and others",
    collaboration = "CMS",
    title = "{Search for pair production of vector-like quarks in leptonic final states in proton-proton collisions at $ \sqrt{s} $ = 13 TeV}",
    eprint = "2209.07327",
    archivePrefix = "arXiv",
    primaryClass = "hep-ex",
    reportNumber = "CMS-B2G-20-011, CERN-EP-2022-175",
    doi = "10.1007/JHEP07(2023)020",
    journal = "JHEP",
    volume = "07",
    pages = "020",
    year = "2023"
}

@article{ATLAS:2022tla,
    author = "Aad, Georges and others",
    collaboration = "ATLAS",
    title = "{Search for pair-produced vector-like top and bottom partners in events with large missing transverse momentum in pp collisions with the ATLAS detector}",
    eprint = "2212.05263",
    archivePrefix = "arXiv",
    primaryClass = "hep-ex",
    reportNumber = "CERN-EP-2022-201",
    doi = "10.1140/epjc/s10052-023-11790-7",
    journal = "Eur. Phys. J. C",
    volume = "83",
    number = "8",
    pages = "719",
    year = "2023"
}

@article{ATLAS:2017tmw,
    author = "Aaboud, Morad and others",
    collaboration = "ATLAS",
    title = "{Search for supersymmetry in final states with two same-sign or three leptons and jets using 36 fb$^{-1}$ of $\sqrt{s}=13$ TeV $pp$ collision data with the ATLAS detector}",
    eprint = "1706.03731",
    archivePrefix = "arXiv",
    primaryClass = "hep-ex",
    reportNumber = "CERN-EP-2017-108",
    doi = "10.1007/JHEP09(2017)084",
    journal = "JHEP",
    volume = "09",
    pages = "084",
    year = "2017",
    note = "[Erratum: JHEP 08, 121 (2019)]"
}

@article{Goodsell:2024aig,
    author = "Goodsell, Mark D.",
    title = "{HackAnalysis 2: A powerful and hackable recasting tool}",
    eprint = "2406.10042",
    archivePrefix = "arXiv",
    primaryClass = "hep-ph",
    month = "6",
    year = "2024"
}

@article{Artoisenet:2012st,
    author = "Artoisenet, Pierre and Frederix, Rikkert and Mattelaer, Olivier and Rietkerk, Robbert",
    title = "{Automatic spin-entangled decays of heavy resonances in Monte Carlo simulations}",
    eprint = "1212.3460",
    archivePrefix = "arXiv",
    primaryClass = "hep-ph",
    reportNumber = "NIKHEF-2012-021, CERN-PH-TH-2012-329",
    doi = "10.1007/JHEP03(2013)015",
    journal = "JHEP",
    volume = "03",
    pages = "015",
    year = "2013"
}

@article{Alloul:2013bka,
    author = "Alloul, Adam and Christensen, Neil D. and Degrande, C\'eline and Duhr, Claude and Fuks, Benjamin",
    title = "{FeynRules  2.0 - A complete toolbox for tree-level phenomenology}",
    eprint = "1310.1921",
    archivePrefix = "arXiv",
    primaryClass = "hep-ph",
    reportNumber = "CERN-PH-TH-2013-239, MCNET-13-14, IPPP-13-71, DCPT-13-142, PITT-PACC-1308",
    doi = "10.1016/j.cpc.2014.04.012",
    journal = "Comput. Phys. Commun.",
    volume = "185",
    pages = "2250--2300",
    year = "2014"
}

@article{Degrande:2011ua,
    author = "Degrande, Celine and Duhr, Claude and Fuks, Benjamin and Grellscheid, David and Mattelaer, Olivier and Reiter, Thomas",
    title = "{UFO - The Universal FeynRules Output}",
    eprint = "1108.2040",
    archivePrefix = "arXiv",
    primaryClass = "hep-ph",
    reportNumber = "CP3-11-25, IPHC-PHENO-11-04, IPPP-11-39, DCPT-11-78, MPP-2011-68",
    doi = "10.1016/j.cpc.2012.01.022",
    journal = "Comput. Phys. Commun.",
    volume = "183",
    pages = "1201--1214",
    year = "2012"
}

@article{Alwall:2014hca,
    author = "Alwall, J. and Frederix, R. and Frixione, S. and Hirschi, V. and Maltoni, F. and Mattelaer, O. and Shao, H. -S. and Stelzer, T. and Torrielli, P. and Zaro, M.",
    title = "{The automated computation of tree-level and next-to-leading order differential cross sections, and their matching to parton shower simulations}",
    eprint = "1405.0301",
    archivePrefix = "arXiv",
    primaryClass = "hep-ph",
    reportNumber = "CERN-PH-TH-2014-064, CP3-14-18, LPN14-066, MCNET-14-09, ZU-TH-14-14",
    doi = "10.1007/JHEP07(2014)079",
    journal = "JHEP",
    volume = "07",
    pages = "079",
    year = "2014"
}

@article{Dobbs:2001ck,
    author = "Dobbs, Matt and Hansen, Jorgen Beck",
    title = "{The HepMC C++ Monte Carlo event record for High Energy Physics}",
    reportNumber = "ATL-SOFT-2000-001",
    doi = "10.1016/S0010-4655(00)00189-2",
    journal = "Comput. Phys. Commun.",
    volume = "134",
    pages = "41--46",
    year = "2001"
}

@article{Ball:2012cx,
    author = "Ball, Richard D. and others",
    title = "{Parton distributions with LHC data}",
    eprint = "1207.1303",
    archivePrefix = "arXiv",
    primaryClass = "hep-ph",
    reportNumber = "EDINBURGH-2012-08, IFUM-FT-997, FR-PHENO-2012-014, RWTH-TTK-12-25, CERN-PH-TH-2012-037, SFB-CPP-12-47",
    doi = "10.1016/j.nuclphysb.2012.10.003",
    journal = "Nucl. Phys. B",
    volume = "867",
    pages = "244--289",
    year = "2013"
}

@article{Buckley:2014ana,
    author = {Buckley, Andy and Ferrando, James and Lloyd, Stephen and Nordstr\"om, Karl and Page, Ben and R\"ufenacht, Martin and Sch\"onherr, Marek and Watt, Graeme},
    title = "{LHAPDF6: parton density access in the LHC precision era}",
    eprint = "1412.7420",
    archivePrefix = "arXiv",
    primaryClass = "hep-ph",
    reportNumber = "GLAS-PPE-2014-05, MCNET-14-29, IPPP-14-111, DCPT-14-222",
    doi = "10.1140/epjc/s10052-015-3318-8",
    journal = "Eur. Phys. J. C",
    volume = "75",
    pages = "132",
    year = "2015"
}

@article{Conte:2012fm,
    author = "Conte, Eric and Fuks, Benjamin and Serret, Guillaume",
    title = "{MadAnalysis 5, A User-Friendly Framework for Collider Phenomenology}",
    eprint = "1206.1599",
    archivePrefix = "arXiv",
    primaryClass = "hep-ph",
    reportNumber = "IPHC-PHENO-06",
    doi = "10.1016/j.cpc.2012.09.009",
    journal = "Comput. Phys. Commun.",
    volume = "184",
    pages = "222--256",
    year = "2013"
}

@article{Conte:2014zja,
    author = "Conte, Eric and Dumont, B\'eranger and Fuks, Benjamin and Wymant, Chris",
    title = "{Designing and recasting LHC analyses with MadAnalysis 5}",
    eprint = "1405.3982",
    archivePrefix = "arXiv",
    primaryClass = "hep-ph",
    reportNumber = "CERN-PH-TH-2014-088, LPSC-14-079, LAPTH-033-14, MCNET-14-11",
    doi = "10.1140/epjc/s10052-014-3103-0",
    journal = "Eur. Phys. J. C",
    volume = "74",
    number = "10",
    pages = "3103",
    year = "2014"
}

@article{Dumont:2014tja,
    author = "Dumont, B. and Fuks, B. and Kraml, S. and Bein, S. and Chalons, G. and Conte, E. and Kulkarni, S. and Sengupta, D. and Wymant, C.",
    title = "{Toward a public analysis database for LHC new physics searches using MADANALYSIS 5}",
    eprint = "1407.3278",
    archivePrefix = "arXiv",
    primaryClass = "hep-ph",
    reportNumber = "CERN-PH-TH-2014-109, LAPTH-048-14, LPSC14143",
    doi = "10.1140/epjc/s10052-014-3242-3",
    journal = "Eur. Phys. J. C",
    volume = "75",
    number = "2",
    pages = "56",
    year = "2015"
}

@article{Conte:2018vmg,
    author = "Conte, Eric and Fuks, Benjamin",
    title = "{Confronting new physics theories to LHC data with MADANALYSIS 5}",
    eprint = "1808.00480",
    archivePrefix = "arXiv",
    primaryClass = "hep-ph",
    doi = "10.1142/S0217751X18300272",
    journal = "Int. J. Mod. Phys. A",
    volume = "33",
    number = "28",
    pages = "1830027",
    year = "2018"
}

@article{Araz:2020lnp,
    author = "Araz, Jack Y. and Fuks, Benjamin and Polykratis, Georgios",
    title = "{Simplified fast detector simulation in MADANALYSIS 5}",
    eprint = "2006.09387",
    archivePrefix = "arXiv",
    primaryClass = "hep-ph",
    doi = "10.1140/epjc/s10052-021-09052-5",
    journal = "Eur. Phys. J. C",
    volume = "81",
    number = "4",
    pages = "329",
    year = "2021"
}

@article{Drees:2013wra,
    author = "Drees, Manuel and Dreiner, Herbi and Schmeier, Daniel and Tattersall, Jamie and Kim, Jong Soo",
    title = "{CheckMATE: Confronting your Favourite New Physics Model with LHC Data}",
    eprint = "1312.2591",
    archivePrefix = "arXiv",
    primaryClass = "hep-ph",
    reportNumber = "ADP-13-29-T849, IFT-UAM-CSIC-13-133, ADP-13-29-T849-IFT-UAM-CSIC-13-133",
    doi = "10.1016/j.cpc.2014.10.018",
    journal = "Comput. Phys. Commun.",
    volume = "187",
    pages = "227--265",
    year = "2015"
}

@article{Dercks:2016npn,
    author = "Dercks, Daniel and Desai, Nishita and Kim, Jong Soo and Rolbiecki, Krzysztof and Tattersall, Jamie and Weber, Torsten",
    title = "{CheckMATE 2: From the model to the limit}",
    eprint = "1611.09856",
    archivePrefix = "arXiv",
    primaryClass = "hep-ph",
    reportNumber = "CTPU-16-36, CSIC-16-116, TTK-16-47",
    doi = "10.1016/j.cpc.2017.08.021",
    journal = "Comput. Phys. Commun.",
    volume = "221",
    pages = "383--418",
    year = "2017"
}

@article{deFavereau:2013fsa,
    author = "de Favereau, J. and Delaere, C. and Demin, P. and Giammanco, A. and Lema\^\i{}tre, V. and Mertens, A. and Selvaggi, M.",
    collaboration = "DELPHES 3",
    title = "{DELPHES 3, A modular framework for fast simulation of a generic collider experiment}",
    eprint = "1307.6346",
    archivePrefix = "arXiv",
    primaryClass = "hep-ex",
    doi = "10.1007/JHEP02(2014)057",
    journal = "JHEP",
    volume = "02",
    pages = "057",
    year = "2014"
}

@article{Cacciari:2011ma,
    author = "Cacciari, Matteo and Salam, Gavin P. and Soyez, Gregory",
    title = "{FastJet User Manual}",
    eprint = "1111.6097",
    archivePrefix = "arXiv",
    primaryClass = "hep-ph",
    reportNumber = "CERN-PH-TH-2011-297",
    doi = "10.1140/epjc/s10052-012-1896-2",
    journal = "Eur. Phys. J. C",
    volume = "72",
    pages = "1896",
    year = "2012"
}

@article{Read:2002hq,
    author = "Read, Alexander L.",
    editor = "Whalley, M. R. and Lyons, L.",
    title = "{Presentation of search results: The $CL_s$ technique}",
    doi = "10.1088/0954-3899/28/10/313",
    journal = "J. Phys. G",
    volume = "28",
    pages = "2693--2704",
    year = "2002"
}

@article{Araz:2023mda,
    author = "Araz, Jack Y. and others",
    title = "{Les Houches guide to reusable ML models in LHC analyses}",
    eprint = "2312.14575",
    archivePrefix = "arXiv",
    primaryClass = "hep-ph",
    doi = "10.21468/SciPostPhysCommRep.3",
    journal = "SciPost Phys. Comm. Rep. ",
    pages = "3",
    year = "2024"
}

@article{Matsedonskyi:2014lla,
    author = "Matsedonskyi, Oleksii and Riva, Francesco and Vantalon, Thibaud",
    title = "{Composite Charge 8/3 Resonances at the LHC}",
    eprint = "1401.3740",
    archivePrefix = "arXiv",
    primaryClass = "hep-ph",
    doi = "10.1007/JHEP04(2014)059",
    journal = "JHEP",
    volume = "04",
    pages = "059",
    year = "2014"
}

@article{ATLAS:2019vcq,
    collaboration = "ATLAS",
    title = "{Search for squarks and gluinos in final states with jets and missing transverse momentum using 139 fb$^{-1}$ of $\sqrt{s}$ =13 TeV $pp$ collision data with the ATLAS detector}",
    reportNumber = "ATLAS-CONF-2019-040",
    month = "8",
    year = "2019",
    note = {\href{https://cds.cern.ch/record/2686254}{ATLAS-CONF-2019-040}}
}

@article{CMS:4top,
    collaboration = "CMS",
    title = "{Search for standard model production of four top quarks in final states with same-sign and multiple leptons in proton-proton collisions at $\sqrt s = 13$ TeV}",
    reportNumber = " CMS-PAS-TOP-18-003",
    note = "\href{https://cds.cern.ch/record/2668710}{CMS-PAS-TOP-18-003}"
}

@article{ATLAS:2017mjy,
    author = "Aaboud, Morad and others",
    collaboration = "ATLAS",
    title = "{Search for squarks and gluinos in final states with jets and missing transverse momentum using 36  fb$^{-1}$ of $\sqrt{s}=13$  TeV pp collision data with the ATLAS detector}",
    eprint = "1712.02332",
    archivePrefix = "arXiv",
    primaryClass = "hep-ex",
    reportNumber = "CERN-EP-2017-136",
    doi = "10.1103/PhysRevD.97.112001",
    journal = "Phys. Rev. D",
    volume = "97",
    number = "11",
    pages = "112001",
    year = "2018"
}

@article{ATLAS:2018ziw,
    author = "Aaboud, Morad and others",
    collaboration = "ATLAS",
    title = "{Combination of the searches for pair-produced vector-like partners of the third-generation quarks at $\sqrt{s} =$ 13 TeV with the ATLAS detector}",
    eprint = "1808.02343",
    archivePrefix = "arXiv",
    primaryClass = "hep-ex",
    reportNumber = "CERN-EP-2018-205",
    doi = "10.1103/PhysRevLett.121.211801",
    journal = "Phys. Rev. Lett.",
    volume = "121",
    number = "21",
    pages = "211801",
    year = "2018"
}

@article{Flacke:2025xwl,
    author = "Flacke, Thomas and Kim, Jeong Han and Kunkel, Manuel and Pi, Jun Seung and Porod, Werner",
    title = "{Hunting and identifying coloured resonances in four top events with machine learning}",
    eprint = "2506.04318",
    archivePrefix = "arXiv",
    primaryClass = "hep-ph",
    reportNumber = "KIAS - A25019",
    doi = "10.1007/JHEP02(2026)028",
    journal = "JHEP",
    volume = "02",
    pages = "028",
    year = "2026"
}

@article{Czakon:2011xx,
    author = "Czakon, Michal and Mitov, Alexander",
    title = "{Top++: A Program for the Calculation of the Top-Pair Cross-Section at Hadron Colliders}",
    eprint = "1112.5675",
    archivePrefix = "arXiv",
    primaryClass = "hep-ph",
    reportNumber = "CERN-PH-TH-2011-303, TTK-11-58",
    doi = "10.1016/j.cpc.2014.06.021",
    journal = "Comput. Phys. Commun.",
    volume = "185",
    pages = "2930",
    year = "2014"
}

@article{Bierlich:2022pfr,
    author = "Bierlich, Christian and others",
    title = "{A comprehensive guide to the physics and usage of PYTHIA 8.3}",
    eprint = "2203.11601",
    archivePrefix = "arXiv",
    primaryClass = "hep-ph",
    reportNumber = "LU-TP 22-16, MCNET-22-04, FERMILAB-PUB-22-227-SCD",
    doi = "10.21468/SciPostPhysCodeb.8",
    journal = "SciPost Phys. Codeb.",
    volume = "2022",
    pages = "8",
    year = "2022"
}

@article{Buckley:2020wzk,
    author = "Buckley, A. and Butterworth, J. M. and Corpe, L. and Huang, D. and Sun, P.",
    title = "{New sensitivity of current LHC measurements to vector-like quarks}",
    eprint = "2006.07172",
    archivePrefix = "arXiv",
    primaryClass = "hep-ph",
    reportNumber = "MCnet-20",
    doi = "10.21468/SciPostPhys.9.5.069",
    journal = "SciPost Phys.",
    volume = "9",
    number = "5",
    pages = "069",
    year = "2020"
}

@article{Arkani-Hamed:2002ikv,
    author = "Arkani-Hamed, N. and Cohen, A. G. and Katz, E. and Nelson, A. E.",
    title = "{The Littlest Higgs}",
    eprint = "hep-ph/0206021",
    archivePrefix = "arXiv",
    reportNumber = "UW-PT-01-07, HUTP-02-A017, BUHEP-02-23",
    doi = "10.1088/1126-6708/2002/07/034",
    journal = "JHEP",
    volume = "07",
    pages = "034",
    year = "2002"
}

@article{ATLAS:2026rnb,
    author = "Aad, Georges and others",
    collaboration = "ATLAS",
    title = "{Search for pair-production of vector-like $T$ quarks decaying into a top quark and a spin-0 particle in the diphoton final state in proton proton collisions at $\sqrt{s}=13$ TeV with the ATLAS detector}",
    eprint = "2607.28381",
    archivePrefix = "arXiv",
    primaryClass = "hep-ex",
    reportNumber = "CERN-EP-2026-225",
    month = "7",
    year = "2026"
}

@article{ATLAS:2022ozf,
    author = "Aad, Georges and others",
    collaboration = "ATLAS",
    title = "{Search for single production of a vectorlike $T$ quark decaying into a Higgs boson and top quark with fully hadronic final states using the ATLAS detector}",
    eprint = "2201.07045",
    archivePrefix = "arXiv",
    primaryClass = "hep-ex",
    reportNumber = "CERN-EP-2021-249",
    doi = "10.1103/PhysRevD.105.092012",
    journal = "Phys. Rev. D",
    volume = "105",
    number = "9",
    pages = "092012",
    year = "2022"
}

@article{ATLAS:2023pja,
    author = "Aad, Georges and others",
    collaboration = "ATLAS",
    title = "{Search for single production of vector-like T quarks decaying into Ht or Zt in pp collisions at $ \sqrt{s} $ = 13 TeV with the ATLAS detector}",
    eprint = "2305.03401",
    archivePrefix = "arXiv",
    primaryClass = "hep-ex",
    reportNumber = "CERN-EP-2023-058",
    doi = "10.1007/JHEP08(2023)153",
    journal = "JHEP",
    volume = "08",
    pages = "153",
    year = "2023"
}

@article{ATLAS:2023bfh,
    author = "Aad, Georges and others",
    collaboration = "ATLAS",
    title = "{Search for singly produced vectorlike top partners in multilepton final states with 139~fb$^{-1}$ of pp collision data at $\sqrt s=13$~TeV with the ATLAS detector}",
    eprint = "2307.07584",
    archivePrefix = "arXiv",
    primaryClass = "hep-ex",
    reportNumber = "CERN-EP-2023-074",
    doi = "10.1103/PhysRevD.109.112012",
    journal = "Phys. Rev. D",
    volume = "109",
    number = "11",
    pages = "112012",
    year = "2024"
}

@article{ATLAS:2023ixh,
    author = "Aad, Georges and others",
    collaboration = "ATLAS",
    title = "{Search for single vector-like $B$ quark production and decay via $B\rightarrow bH(b\bar{b})$ in $pp$ collisions at $\sqrt{s} = 13$ TeV with the ATLAS detector}",
    eprint = "2308.02595",
    archivePrefix = "arXiv",
    primaryClass = "hep-ex",
    reportNumber = "CERN-EP-2023-091",
    doi = "10.1007/JHEP11(2023)168",
    journal = "JHEP",
    volume = "11",
    pages = "168",
    year = "2023"
}

@article{ATLAS:2024xne,
    author = "Aad, Georges and others",
    collaboration = "ATLAS",
    title = "{Search for new particles in final states with a boosted top quark and missing transverse momentum in proton-proton collisions at $ \sqrt{s} $ = 13 TeV with the ATLAS detector}",
    eprint = "2402.16561",
    archivePrefix = "arXiv",
    primaryClass = "hep-ex",
    reportNumber = "CERN-EP-2024-046",
    doi = "10.1007/JHEP05(2024)263",
    journal = "JHEP",
    volume = "05",
    pages = "263",
    year = "2024"
}

@article{ATLAS:2024xdc,
    author = "Aad, Georges and others",
    collaboration = "ATLAS",
    title = "{Combination of searches for singly produced vectorlike top quarks in pp collisions at $\sqrt s=13$~TeV with the ATLAS detector}",
    eprint = "2408.08789",
    archivePrefix = "arXiv",
    primaryClass = "hep-ex",
    reportNumber = "CERN-EP-2024-203",
    doi = "10.1103/PhysRevD.111.012012",
    journal = "Phys. Rev. D",
    volume = "111",
    number = "1",
    pages = "012012",
    year = "2025"
}

@article{CMS:2022yxp,
    author = "Tumasyan, Armen and others",
    collaboration = "CMS",
    title = "{Search for single production of a vector-like T quark decaying to a top quark and a Z boson in the final state with jets and missing transverse momentum at $ \sqrt{s} $ = 13 TeV}",
    eprint = "2201.02227",
    archivePrefix = "arXiv",
    primaryClass = "hep-ex",
    reportNumber = "CMS-B2G-19-004, CERN-EP-2021-115",
    doi = "10.1007/JHEP05(2022)093",
    journal = "JHEP",
    volume = "05",
    pages = "093",
    year = "2022"
}

@article{CMS:2023agg,
    author = "Tumasyan, Armen and others",
    collaboration = "CMS",
    title = "{Search for a vector-like quark T$'$$\to$ tH via the diphoton decay mode of the Higgs boson in proton-proton collisions at $\sqrt{s}$ = 13 TeV}",
    eprint = "2302.12802",
    archivePrefix = "arXiv",
    primaryClass = "hep-ex",
    reportNumber = "CMS-B2G-21-007, CERN-EP-2022-253",
    doi = "10.1007/JHEP09(2023)057",
    journal = "JHEP",
    volume = "09",
    pages = "057",
    year = "2023"
}

@article{CMS:2024qdd,
    author = "Hayrapetyan, Aram and others",
    collaboration = "CMS",
    title = "{Search for production of a single vectorlike quark decaying to tH or tZ in the all-hadronic final state in pp collisions at $\sqrt s=13$~TeV}",
    eprint = "2405.05071",
    archivePrefix = "arXiv",
    primaryClass = "hep-ex",
    reportNumber = "CMS-B2G-19-001, CERN-EP-2024-067",
    doi = "10.1103/PhysRevD.110.072012",
    journal = "Phys. Rev. D",
    volume = "110",
    number = "7",
    pages = "072012",
    year = "2024"
}

@article{CMS:2026vwc,
    author = "Hayrapetyan, Aram and others",
    collaboration = "CMS",
    title = "{Search for the single production of vector-like quarks decaying into a W boson and a b quark using single-lepton final states in proton-proton collisions at $\sqrt{s}$ = 13 TeV}",
    eprint = "2604.17564",
    archivePrefix = "arXiv",
    primaryClass = "hep-ex",
    reportNumber = "CMS-B2G-22-004, CERN-EP-2026-021",
    note = "\href{https://cms-results.web.cern.ch/cms-results/public-results/publications/B2G-22-004/index.html}{CMS-B2G-22-004}",
    month = "4",
    year = "2026"
}

@article{CMS:2026iyo,
    author = "Tauqeer, Komal and others",
    collaboration = "CMS",
    title = "{Search for single production of a vector-like B' quark decaying to a top quark and a W boson in the single-lepton final state in proton-proton collisions at $\sqrt{s}$ = 13 TeV}",
    eprint = "2606.01423",
    archivePrefix = "arXiv",
    primaryClass = "hep-ex",
    reportNumber = "CMS-B2G-24-013, CERN-EP-2026-137",
    month = "5",
    year = "2026"
}

@article{Chala:2018qdf,
    author = {Chala, Mikael and Gr{\"o}ber, Ramona and Spannowsky, Michael},
    title = "{Searches for vector-like quarks at future colliders and implications for composite Higgs models with dark matter}",
    eprint = "1801.06537",
    archivePrefix = "arXiv",
    primaryClass = "hep-ph",
    reportNumber = "IPPP-18-6",
    doi = "10.1007/JHEP03(2018)040",
    journal = "JHEP",
    volume = "03",
    pages = "040",
    year = "2018"
}

@article{Ramos:2019qqa,
    author = "Ramos, Maria",
    title = "{Composite dark matter phenomenology in the presence of lighter degrees of freedom}",
    eprint = "1912.11061",
    archivePrefix = "arXiv",
    primaryClass = "hep-ph",
    doi = "10.1007/JHEP07(2020)128",
    journal = "JHEP",
    volume = "07",
    pages = "128",
    year = "2020"
}

@article{Cacciapaglia:2026jlv,
    author = "Cacciapaglia, Giacomo and Caliri, Rosy and Deandrea, Aldo and Fuks, Benjamin and Goodsell, Mark and Hadlik, Jan and Kunkel, Manuel and Porod, Werner",
    title = "{Composite top partners in exotic colour representations}",
    eprint = "2605.04143",
    archivePrefix = "arXiv",
    primaryClass = "hep-ph",
    month = "5",
    year = "2026"
}

@article{Cornell:2022nky,
    author = "Cornell, Alan S. and Deandrea, Aldo and Flacke, Thomas and Fuks, Benjamin and Mason, Lara",
    title = "{Top partners and scalar dark matter: A nonminimal reappraisal}",
    eprint = "2209.13093",
    archivePrefix = "arXiv",
    primaryClass = "hep-ph",
    reportNumber = "KIAS-A22010",
    doi = "10.1103/PhysRevD.107.075004",
    journal = "Phys. Rev. D",
    volume = "107",
    number = "7",
    pages = "075004",
    year = "2023"
}

@article{Araz:2020stn,
    author = "Araz, Jack Y. and Fuks, Benjamin",
    title = "{Implementation of the ATLAS-SUSY-2018-31 analysis in the MadAnalysis 5 framework (sbottoms with multi-bottoms and missing transverse energy; 139 fb${}^{-1}$)}",
    doi = "10.1142/S0217732321410108",
    journal = "Mod. Phys. Lett. A",
    volume = "36",
    number = "01",
    pages = "2141010",
    year = "2021"
}

@article{Darme:2020hxc,
    author = "Darm{\'e}, Luc and Fuks, Benjamin",
    title = "{Implementation of the CMS-TOP-18-003 analysis in the MadAnalysis 5 framework (four top quarks with at least two leptons; 137 fb$^{-1}$)}",
    doi = "10.1142/S021773232141008X",
    journal = "Mod. Phys. Lett. A",
    volume = "36",
    number = "01",
    pages = "2141008",
    year = "2021"
}

@article{Conte:2021xtt,
    author = "Conte, Eric and Ducrocq, Robin",
    title = "{Implementation of the CMS-EXO-19-002 search in the MadAnalysis 5 framework (physics beyond the Standard Model with multilepton final states; 137 fb$^{-1}$)}",
    doi = "10.1142/S0217732321410121",
    journal = "Mod. Phys. Lett. A",
    volume = "36",
    number = "01",
    pages = "2141012",
    year = "2021"
}

@article{Mrowietz:2020ztq,
    author = "Mrowietz, Malte and Bein, Sam and Sonneveld, Jory",
    title = "{Implementation of the CMS-SUS-19-006 analysis in the MadAnalysis 5 framework (supersymmetry with large hadronic activity and missing transverse energy; 137 fb$^{-1}$)}",
    doi = "10.1142/S0217732321410078",
    journal = "Mod. Phys. Lett. A",
    volume = "36",
    number = "01",
    pages = "2141007",
    year = "2021"
}

@data{DVN/NW3NPG_2021,
author = {Ambrogi, Federico},
publisher = {Open Data @ UCLouvain},
title = {{Implementation of a search for squarks and gluinos in the multi-jet + missing energy channel (139 fb$^{-1}$; 13 TeV; ATLAS-CONF-2019-040)}},
year = {2021},
version = {V1},
doi = {10.14428/DVN/NW3NPG},
url = {https://doi.org/10.14428/DVN/NW3NPG}
}

@data{DVN/GBDC91_2021,
author = {Ambrogi, Federico and Sonneveld, Jory},
publisher = {Open Data @ UCLouvain},
title = {{Implementation of a search for supersymmetry in the multi-jet + missing energy channel (35.9 fb$^{-1}$; 13 TeV; CMS-SUS-16-033)}},
year = {2021},
version = {V1},
doi = {10.14428/DVN/GBDC91},
url = {https://doi.org/10.14428/DVN/GBDC91}
}

@article{Banerjee:2016wls,
    author = "Banerjee, Shankha and Barducci, Daniele and B{\'e}langer, Genevi{\`e}ve and Delaunay, C{\'e}dric",
    title = "{Implications of a High-Mass Diphoton Resonance for Heavy Quark Searches}",
    eprint = "1606.09013",
    archivePrefix = "arXiv",
    primaryClass = "hep-ph",
    reportNumber = "LAPTH-033-16",
    doi = "10.1007/JHEP11(2016)154",
    journal = "JHEP",
    volume = "11",
    pages = "154",
    year = "2016"
}

@article{Wang:2020ips,
    author = "Wang, Daohan and Wu, Lei and Zhang, Mengchao",
    title = "{Hunting for top partner with a new signature at the LHC}",
    eprint = "2007.09722",
    archivePrefix = "arXiv",
    primaryClass = "hep-ph",
    doi = "10.1103/PhysRevD.103.115017",
    journal = "Phys. Rev. D",
    volume = "103",
    number = "11",
    pages = "115017",
    year = "2021"
}

@article{Corcella:2021mdl,
    author = "Corcella, Gennaro and Costantini, Antonio and Ghezzi, Margherita and Panizzi, Luca and Pruna, Giovanni Marco and {\v{S}}alko, Jakub",
    title = "{Vector-like quarks decaying into singly and doubly charged bosons at LHC}",
    eprint = "2107.07426",
    archivePrefix = "arXiv",
    primaryClass = "hep-ph",
    doi = "10.1007/JHEP10(2021)108",
    journal = "JHEP",
    volume = "10",
    pages = "108",
    year = "2021"
}

@article{Dasgupta:2021fzw,
    author = "Dasgupta, Sayan and Pramanick, Rohan and Ray, Tirtha Sankar",
    title = "{Broad toplike vector quarks at LHC and HL-LHC}",
    eprint = "2112.03742",
    archivePrefix = "arXiv",
    primaryClass = "hep-ph",
    doi = "10.1103/PhysRevD.105.035032",
    journal = "Phys. Rev. D",
    volume = "105",
    number = "3",
    pages = "035032",
    year = "2022"
}

@article{Dermisek:2021zjd,
    author = "Dermisek, Radovan and Lunghi, Enrico and Mcginnis, Navin and Shin, Seodong",
    title = "{Tau-jet signatures of vectorlike quark decays to heavy charged and neutral Higgs bosons}",
    eprint = "2105.10790",
    archivePrefix = "arXiv",
    primaryClass = "hep-ph",
    doi = "10.1007/JHEP08(2021)159",
    journal = "JHEP",
    volume = "08",
    pages = "159",
    year = "2021"
}

@article{Belyaev:2022shr,
    author = "Belyaev, Alexander and Deandrea, Aldo and Moretti, Stefano and Panizzi, Luca and Ross, Douglas A. and Thongyoi, Nakorn",
    title = "{Fermionic portal to vector dark matter from a new gauge sector}",
    eprint = "2204.03510",
    archivePrefix = "arXiv",
    primaryClass = "hep-ph",
    doi = "10.1103/PhysRevD.108.095001",
    journal = "Phys. Rev. D",
    volume = "108",
    number = "9",
    pages = "095001",
    year = "2023"
}

@article{Aguilar-Saavedra:2019ghg,
    author = "Aguilar-Saavedra, J. A. and Alonso-Gonz{\'a}lez, J. and Merlo, L. and No, J. M.",
    title = "{Exotic vectorlike quark phenomenology in the minimal linear {\ensuremath{\sigma}} model}",
    eprint = "1911.10202",
    archivePrefix = "arXiv",
    primaryClass = "hep-ph",
    reportNumber = "IFT-UAM/CSIC-19-148, FTUAM-19-20",
    doi = "10.1103/PhysRevD.101.035015",
    journal = "Phys. Rev. D",
    volume = "101",
    number = "3",
    pages = "035015",
    year = "2020"
}

@article{Dermisek:2020gbr,
    author = "Dermisek, Radovan and Lunghi, Enrico and McGinnis, Navin and Shin, Seodong",
    title = "{Signals with six bottom quarks for charged and neutral Higgs bosons}",
    eprint = "2005.07222",
    archivePrefix = "arXiv",
    primaryClass = "hep-ph",
    doi = "10.1007/JHEP07(2020)241",
    journal = "JHEP",
    volume = "07",
    pages = "241",
    year = "2020"
}

@article{Bhardwaj:2022nko,
    author = "Bhardwaj, Akanksha and Mandal, Tanumoy and Mitra, Subhadip and Neeraj, Cyrin",
    title = "{Roadmap to explore vectorlike quarks decaying to a new scalar or pseudoscalar}",
    eprint = "2203.13753",
    archivePrefix = "arXiv",
    primaryClass = "hep-ph",
    doi = "10.1103/PhysRevD.106.095014",
    journal = "Phys. Rev. D",
    volume = "106",
    number = "9",
    pages = "095014",
    year = "2022"
}

@article{Bhardwaj:2022wfz,
    author = "Bhardwaj, Akanksha and Bhide, Kartik and Mandal, Tanumoy and Mitra, Subhadip and Neeraj, Cyrin",
    title = "{Discovery prospects of a vectorlike top partner decaying to a singlet boson}",
    eprint = "2204.09005",
    archivePrefix = "arXiv",
    primaryClass = "hep-ph",
    doi = "10.1103/PhysRevD.106.075024",
    journal = "Phys. Rev. D",
    volume = "106",
    number = "7",
    pages = "075024",
    year = "2022"
}

@article{Verma:2022nyd,
    author = "Verma, Shivam and Biswas, Sanjoy and Chatterjee, Anirban and Ganguly, Joy",
    title = "{Exploring maverick top partner decays at the LHC}",
    eprint = "2209.13888",
    archivePrefix = "arXiv",
    primaryClass = "hep-ph",
    doi = "10.1103/PhysRevD.107.115024",
    journal = "Phys. Rev. D",
    volume = "107",
    number = "11",
    pages = "115024",
    year = "2023"
}

@article{Ghosh:2022rta,
    author = "Ghosh, Anupam and Konar, Partha and Roshan, Rishav",
    title = "{Top-philic dark matter in a hybrid KSVZ axion framework}",
    eprint = "2207.00487",
    archivePrefix = "arXiv",
    primaryClass = "hep-ph",
    doi = "10.1007/JHEP12(2022)167",
    journal = "JHEP",
    volume = "12",
    pages = "167",
    year = "2022"
}

@article{Bardhan:2022sif,
    author = "Bardhan, Jai and Mandal, Tanumoy and Mitra, Subhadip and Neeraj, Cyrin",
    title = "{Machine learning-enhanced search for a vectorlike singlet B quark decaying to a singlet scalar or pseudoscalar}",
    eprint = "2212.02442",
    archivePrefix = "arXiv",
    primaryClass = "hep-ph",
    doi = "10.1103/PhysRevD.107.115001",
    journal = "Phys. Rev. D",
    volume = "107",
    number = "11",
    pages = "115001",
    year = "2023"
}

@article{Chala:2017xgc,
    author = "Chala, Mikael",
    title = "{Direct bounds on heavy toplike quarks with standard and exotic decays}",
    eprint = "1705.03013",
    archivePrefix = "arXiv",
    primaryClass = "hep-ph",
    doi = "10.1103/PhysRevD.96.015028",
    journal = "Phys. Rev. D",
    volume = "96",
    number = "1",
    pages = "015028",
    year = "2017"
}

@article{Kim:2018mks,
    author = "Kim, Jeong Han and Lewis, Ian M.",
    title = "{Loop Induced Single Top Partner Production and Decay at the LHC}",
    eprint = "1803.06351",
    archivePrefix = "arXiv",
    primaryClass = "hep-ph",
    doi = "10.1007/JHEP05(2018)095",
    journal = "JHEP",
    volume = "05",
    pages = "095",
    year = "2018"
}

@article{Han:2018hcu,
    author = "Han, Huayong and Huang, Li and Ma, Teng and Shu, Jing and Tait, Tim M. P. and Wu, Yongcheng",
    title = "{Six Top Messages of New Physics at the LHC}",
    eprint = "1812.11286",
    archivePrefix = "arXiv",
    primaryClass = "hep-ph",
    doi = "10.1007/JHEP10(2019)008",
    journal = "JHEP",
    volume = "10",
    pages = "008",
    year = "2019"
}

@article{Arhrib:2024nbj,
    author = "Arhrib, Abdesslam and Benbrik, Rachid and Berrouj, Mbark and Boukidi, Mohammed and Manaut, Bouzid",
    title = "{Search for charged Higgs bosons through vectorlike top quark pair production at the LHC}",
    eprint = "2407.01348",
    archivePrefix = "arXiv",
    primaryClass = "hep-ph",
    doi = "10.1103/PhysRevD.111.095026",
    journal = "Phys. Rev. D",
    volume = "111",
    number = "9",
    pages = "095026",
    year = "2025"
}

@article{Qureshi:2024naw,
    author = "Qureshi, Umar Sohail and Gurrola, Alfredo and Fl{\'o}rez, Andres and Rodriguez, Cristian",
    title = "{Probing light scalars and vector-like quarks at the high-luminosity LHC}",
    eprint = "2410.17854",
    archivePrefix = "arXiv",
    primaryClass = "hep-ph",
    doi = "10.1140/epjc/s10052-025-14085-1",
    journal = "Eur. Phys. J. C",
    volume = "85",
    number = "4",
    pages = "379",
    year = "2025"
}

@article{Arhrib:2024mbq,
    author = "Arhrib, Abdesslam and Benbrik, Rachid and Boukidi, Mohammed and Moretti, Stefano",
    title = "{Large hadron collider signatures of exotic vector-like quarks within the 2-Higgs doublet model type-II}",
    eprint = "2409.20104",
    archivePrefix = "arXiv",
    primaryClass = "hep-ph",
    doi = "10.1088/1361-6471/ae09be",
    journal = "J. Phys. G",
    volume = "52",
    number = "10",
    pages = "105002",
    year = "2025"
}

@article{Arhrib:2026coy,
    author = "Arhrib, A. and Benbrik, R. and Boukidi, M. and Ech-chaouy, M. and Moretti, S. and Kahime, K. and Salime, K. and Yan, Q. S.",
    title = "{VLQBounds: Confronting Vector-Like Quark Models with LHC Searches}",
    eprint = "2605.20848",
    archivePrefix = "arXiv",
    primaryClass = "hep-ph",
    reportNumber = "IFJPAN-IV-2026-9",
    month = "5",
    year = "2026"
}

@phdthesis{Kunkel:2025qld,
    author = "Kunkel, Manuel",
    title = "{Collider Phenomenology of Composite Higgs Models}",
    doi = "10.25972/OPUS-39149",
    school = "U. Wurzburg (main)",
    year = "2025"
}

@article{Aguilar-Saavedra:2013qpa,
    author = "Aguilar-Saavedra, J. A. and Benbrik, R. and Heinemeyer, S. and P{\'e}rez-Victoria, M.",
    title = "{Handbook of vectorlike quarks: Mixing and single production}",
    eprint = "1306.0572",
    archivePrefix = "arXiv",
    primaryClass = "hep-ph",
    doi = "10.1103/PhysRevD.88.094010",
    journal = "Phys. Rev. D",
    volume = "88",
    number = "9",
    pages = "094010",
    year = "2013"
}

@dataset{zenodo,
    author = "Flacke, Thomas and Kunkel, Manuel",
    title = "Constraint data base for pair production of vector-like quarks",
    year = "2026",
    publisher = "Zenodo",
    doi = "10.5281/zenodo.22142862",
    url = {https://doi.org/10.5281/zenodo.22142862}
}

\end{document}